\RequirePackage{fix-cm}
\documentclass[a4paper,english,aps,prd,twocolumn,10pt,a4paper,superscriptaddress,nofootinbib,nobibnotes,altaffillsymbol,showpacs,showkeys,floatfix]{revtex4-1}

\newcommand{\skipthis}[1]{}

\newcommand{\Tr}{{\text{Tr}}}

\usepackage[T1]{fontenc}
\usepackage[utf8]{inputenc}
\usepackage{babel}

\usepackage{amsmath}
\usepackage{amssymb}
\usepackage{mathrsfs}
\usepackage{esint} 
\usepackage{slashed} 
\usepackage{bbm} 

\usepackage{graphicx}

\usepackage{tabularx}

\usepackage[unicode=true,
 bookmarks=true,bookmarksnumbered=false,bookmarksopen=true,bookmarksopenlevel=2,
 breaklinks=true,pdfborder={0 0 0},backref=false,colorlinks=true]
 {hyperref}

\hypersetup{pdftitle={Cooling Stochastic Quantization with colored noise},
 pdfauthor={Jan M.~Pawlowski, Ion-Olimpiu Stamatescu, Felix P.G.~Ziegler},
 linkcolor=BlueViolet,anchorcolor=BlueViolet,citecolor=BlueViolet,filecolor=BlueViolet,urlcolor=BlueViolet}

\allowdisplaybreaks
\usepackage{color}
\usepackage[usenames,dvipsnames]{xcolor}

\def\eq#1{(\ref{#1})}

\def\s0#1#2{\mbox{\small{$ \frac{#1}{#2} $}}}
\def\0#1#2{\frac{#1}{#2}}

\newcommand{\be}{\begin{equation}}
\newcommand{\ee}{\end{equation}}
\newcommand{\bear}{\begin{eqnarray}}
\newcommand{\eear}{\end{eqnarray}}
\newcommand{\kappaprime}{\kappa^{\prime}}
\newcommand{\lambdaprime}{\lambda^{\prime}}
\newcommand{\latcutoff}{{s_{\Lambda}}}
\newcommand{\latcutoffmax}{s_{\Lambda, \text{\tiny{max}}}}

\newcommand{\RGTime}{t}
\newcommand{\MaxRGTime}{t_{\text{max}}}

\begin{document}

\title{Cooling Stochastic Quantization with colored noise}

\author{Jan M.~Pawlowski}

\affiliation{Institut für Theoretische Physik, Universität Heidelberg, Philosophenweg
16, 69120 Heidelberg, Germany}

\affiliation{ExtreMe Matter Institute EMMI, GSI, Planckstraße 1, D-64291 Darmstadt,
Germany}

\author{Ion-Olimpiu Stamatescu}

\affiliation{Institut für Theoretische Physik, Universität Heidelberg, Philosophenweg
16, 69120 Heidelberg, Germany}

\author{Felix P.G.~Ziegler}

\affiliation{Institut für Theoretische Physik, Universität Heidelberg, Philosophenweg
16, 69120 Heidelberg, Germany}


\begin{abstract}

  Smoothing of field configurations is highly important for precision
  calculations of physical quantities on the lattice. We present a
  cooling method based on Stochastic Quantization with a built-in UV
  momentum cutoff. The latter is implemented via a UV-regularized,
  hence colored, noise term.

  Our method is tested in a two-dimensional scalar field theory. We
  show, that UV modes can be removed systematically without altering
  the physics content of the theory. The approach has an
  interpretation in terms of the non-perturbative (Wilsonian)
  renormalization group that facilitates the physics interpretation of
  the cutoff procedure. It also can be used to define the maximal
  colored cooling applicable without changing the theory. 
\end{abstract}
\maketitle


\section{Introduction}
\label{sec:introduction}

Lattice field theory is a powerful non-perturbative approach that
allows for \textit{ab initio} calculations of realistic quantum field
theories. It has been applied to many areas of physics ranging from
condensed matter systems to quantum gravity. In nuclear physics
applications range from the details of the hadronic spectrum to the
numerical study of topological properties and configurations in
Yang-Mills theory and Quantum Chromodynamics (QCD). A further task for
lattice field theory and other non-perturbative approaches is to map
out the phase structure of strongly correlated systems. However,
lattice simulations for finite density QCD are hampered by the sign
problem. This situation has triggered a plethora of work in this
direction, for a recent review see \cite{Aarts:2015tyj}.

Amongst the contenders for beating the sign problem
Stochastic Quantization based on a complex Langevin equation (CLE) is a very
promising candidate \cite{Aarts:2014fsa}. The complexification
however is non-trivial, see e.g.~\cite{Aarts:2012ft} for a brief account
and \cite{Aarts:2017vrv} for a state of the art. If applied to gauge
theories it has to be equipped with cooling algorithms such as gauge
cooling \cite{Seiler:2012wz} in order to even achieve
convergence. The CLE has been applied 
to explore the phase diagram of QCD and that of related models 
\cite{Sinclair:2016nbg, Nagata:2016mmh, Langelage:2013paa, Aarts:2016qrv}.

Cooling algorithms, see e.g.\ \cite{Teper:1985rb}, are set-up to
eliminate configurations that carry large ultraviolet
fluctuations. They are based on the assumption that physics scales can
be safely separated from the ultraviolet scales where cooling is
applied. Then, cooling simply improves the signal-to-noise ratio
without altering the physics under investigation.  For example, this
works very well for observables such as the action density or the
topological charge density.  However, if cooling is not stopped it
produces classical configurations for large cooling times. Hence, the
crucial question is that after a well-defined stopping time. This
stopping time can be related to the physical scale of the cooled
theory of interest, see e.g.\ \cite{GarciaPerez:1998ru}. The same
intricacy also is present for the recently put forward gradient flow
cooling \cite{Luscher:2010iy, Narayanan:2006rf, GonzalezArroyo:2012fx,
  Datta:2015bzm, Capponi:2015ucc}, for further applications see
\cite{Kagimura:2015via, Bornyakov:2015eaa}.  A comparison of the two
approaches can be found in \cite{Bonati:2014tqa}.

In the present work we suggest to use the Stochastic Quantization
approach cooled with colored noise. This combines a Langevin equation
(LE) with the gradient flow. To that end we first notice that the
Langevin equation without noise simply is the gradient flow. Hence,
removing the noise for high momentum modes above a UV cutoff scale
$\Lambda$ leaves us with a gradient flow for these modes. Then, the
related colored noise Langevin evolution completely removes the
momentum modes with $p^2 >\Lambda^2$. In summary, a Langevin equation
with such a colored noise introduces a UV momentum cutoff $\Lambda$
to the path integral. By varying the cutoff $\Lambda$ we interpolate
between the full quantum evolution characterized by the LE with
Gaussian white noise ($\Lambda\to \infty$) and the classical evolution
characterized by the gradient flow ($\Lambda=0)$. This approach is closely 
related to the concept of stochastic regularization
\cite{Bern:1985dk, Bern:1985he}.

Here, our approach is put to work in a scalar theory and numerical
results are presented in two dimensions. Colored noise is also related
to standard Kadanoff block spin steps \cite{Kadanoff:1966wm}, as well
as to the realization of the latter within the functional
renormalization group, for reviews see e.g.\
\cite{Berges:2000ew,Polonyi:2001se,Pawlowski:2005xe,%
  Delamotte:2007pf,Rosten:2010vm,2012RvMP...84..299M}. We show that a
large regime of ultraviolet fluctuations can be removed without
altering the physics content of the theory. Hence, cooling with
colored noise can significantly reduce the numerical costs of lattice
simulations done within Stochastic Quantization. Such a procedure could 
in principle also be applied to CLE simulations.

The paper is organized as follows. In section \ref{sec:SQ-SR} we
recall basic concepts of Stochastic Quantization and introduce
stochastic regularization.  The lattice field theory formulation 
of the LE with colored noise is described in section
\ref{sec:LQFT-CN}. In section \ref{sec:model} we review real scalar
field theory on the lattice providing a suitable model for our
numerical studies. This is followed by the discussion of the relation between colored
noise and the renormalization group in section \ref{sec:RG-study}.  In
section \ref{sec:Numerical-Results} we discuss numerical results from
simulations with colored noise. The approach is put to work in section
\ref{sec:CNC}, where also its relation to the functional
renormalization group is discussed and utilized. We finish with our
conclusions in section \ref{sec:conclusions}.


\section{Stochastic quantization with colored noise}
\label{sec:SQ-SR}
In this section we briefly review the main concepts of Stochastic
Quantization and stochastic regularization within the example of a
Euclidean real scalar field theory with the action $S = S[\phi]$.

Stochastic Quantization is based on the fact that a Euclidean quantum
field theory can be described by a classical statistical mechanical
system in thermal equilibrium with a heat reservoir
\cite{Parisi:1980ys, Damgaard:1987rr}.  This is formulated in terms of
a stochastic process with a stationary distribution $\exp(-S[\phi]) / Z$, where
\begin{equation}
Z = \int \mathcal{D}\phi\, \exp(-S[\phi])
\label{eq:partition-function}
\end{equation} 
denotes the partition function.  The stochastic process evolves the
fields according to the corresponding Langevin equation in a Langevin-time $\tau$
  \begin{equation}
    \frac{\partial \phi(x,\tau)}{\partial \tau} = -
    \frac{\delta S}{\delta \phi(x,\tau)} + \eta(x,\tau)\, .
   \label{eq:SQ_Langevin}
  \end{equation}
  Here, $\phi(x,\tau)$ denotes the $\tau$-dependent scalar field and
  $\eta(x, \tau)$ is the white noise field representing the quantum
  fluctuations.  With vanishing noise the solution of the
  Langevin evolution converges to a solution of the classical
  equations of motion.  The white noise is characterized by Gaussian
  distributed random numbers with 
 \begin{align}\nonumber 
  \langle \eta(x,\tau) \rangle &= 0\,, \\[2ex]
  \langle \eta(x,\tau)\, \eta(y,\tau') \rangle &= 2\, \delta^{(d)}(x-y)\, 
\delta(\tau-\tau')\,.
  \label{eq:noise-gaussian-relations}
 \end{align}
In the limit
 $\tau \to \infty$ thermal equilibrium is reached and the equal
 Langevin-time correlation functions of the statistical mechanical
 system converge to the Green's functions of the Euclidean quantum
 field theory.  The real Langevin evolution \eq{eq:SQ_Langevin} can be
 applied as an updating algorithm in lattice simulations to sample
 field configurations from the Boltzmann distribution.

 For a given Langevin equation there is an associated Fokker-Planck
 equation.  The latter describes the Langevin-time evolution of a
 probability distribution function $P(\phi, \tau)$ and reads
\begin{align}
  &\frac{\partial P(\phi, \tau)}{\partial \tau} = \nonumber \\[2ex]
  &\int \textrm{d}^{d}{x} \frac{\delta}{\delta \phi(x, \tau)} \left(\frac{
    \delta S}{\delta \phi(x, \tau)} + \frac{\delta}{\delta 
    \phi(x, \tau)} \right) P(\phi, \tau)\, .
 \label{eq:SQ_Fokker_Planck}
\end{align}
One can verify that the Boltzmann distribution $\exp(-S[\phi])$ is
therefore the stationary distribution of \eq{eq:SQ_Fokker_Planck} with
$\partial_\tau P(\phi,\tau)=0$. More generally, if the action is real
and positive semi-definite a stationary distribution of the
Fokker-Planck equation exists which equals $\exp(-S[\phi])$ and the
solution converges exponentially fast \cite{Damgaard:1987rr,
  Aarts:2013uxa}. In summary, Stochastic Quantization provides an alternative to
the standard quantization approach based on the path integral
formalism. 

In the Langevin formulation the noise containing the quantum
fluctuations can be regularized in the ultraviolet by introducing a
cutoff parameter $\Lambda$ \cite{Bern:1985he}. The altered
stochastic process in terms of the Langevin equation with a colored
noise kernel then reads
\begin{equation}
  \frac{\partial \phi(x,\tau)}{\partial \tau} = - 
  \frac{\delta S}{\delta \phi(x, \tau)} + r_\Lambda(\Delta_x) \,\eta(x,\tau)\, ,
  \label{eq:kerneled-LE}
\end{equation}
where the dimensionless regularization function $r_\Lambda(\Delta_x)$
is a function of the ratio $\Delta_x/\Lambda^2$ of the Laplace
operator and the square of the cutoff $\Lambda$. Using a short-hand
notation for the functional derivatives, see
app.~\ref{sec:app-CN-FPE}, the associated Fokker-Planck equation is
\begin{align}
  \frac{\partial P(\phi, \tau)}{\partial \tau}  =
  \int \mathrm{d}^dx \frac{\delta}{\delta \phi_x} \left(\frac{\delta S}{\delta \phi_x} 
 + \,r_\Lambda^2(\Delta_x)\, \frac{\delta}{\delta \phi_x}\right) P(\phi, \tau)\,.
  \label{eq:kerneled-FPE}
\end{align}
Note, that with $r_\Lambda(\Delta_x) \to 1$ in the limit
$\Lambda \to \infty$ the full quantum theory is recovered. For a detailed
derivation of the Fokker-Planck equation from the Langevin equation
with a noise kernel see Appendix \ref{sec:app-CN-FPE}.  Note, that the
regularization function can be chosen in different ways.  A simple and
intuitive choice of the regularization function is a sharp
cutoff in momentum space
\begin{equation}
  r_{\Lambda}(p^2) = \theta(\Lambda^2 - p^2)\, .
  \label{eq:sharp-cutoff-continuum}
\end{equation}
Using \eq{eq:sharp-cutoff-continuum} in the Fokker-Planck equation
\eq{eq:kerneled-FPE} allows for a simple relation of Stochastic
Quantization with colored noise with functional renormalization group
equations, for reviews see
\cite{Berges:2000ew,Polonyi:2001se,Pawlowski:2005xe,%
Delamotte:2007pf,Rosten:2010vm,2012RvMP...84..299M}.
A solution of the fixed point equation $\partial_\tau P = 0$ in
momentum space is given by
\begin{align}\label{eq:prob-Seff}
  P_\Lambda(\phi,\tau)= \exp\left( -S -\Delta S_\Lambda\right)\,, 
\end{align}
with the cutoff term  
\begin{align}\label{eq:delta-S-sharpcutoff}
  \Delta S_\Lambda[\phi]=\012 \int_p \phi_p \,\Lambda^2 
  \left(\0{1}{ r_{\Lambda}(p^2) }-1 \right)    \phi_{-p}\,.
\end{align} 
Inserting (\ref{eq:prob-Seff}) with (\ref{eq:delta-S-sharpcutoff}) into
(\ref{eq:kerneled-FPE}) we are led to the fixed point equation
\begin{align}\label{eq:FPL}
  \left[ \Bigl(1 - r_\Lambda(p^2)\Bigr) \0{\delta S}{\delta \phi_p} - 
 r_\Lambda(p^2) \0{\delta \Delta S_\Lambda}{\delta \phi_p} \right] P_\Lambda(\phi,\tau)=0\,. 
\end{align}
Both terms in the square brackets in \eq{eq:FPL} vanish for
$p^2<\Lambda^2$ as they are proportional to $1-r_\Lambda(p^2)$. Note
that in the second term this comes from
$r_\Lambda (1/r_\Lambda -1)= 1-r_\Lambda$. In turn, for
$p^2>\Lambda^2$ the measure $P(\phi,\tau)$ vanishes and hence
\eq{eq:FPL} is satisfied for all fields and momenta. In summary, this
entails that the ultraviolet modes satisfy the classical equations of
motion and no quantum effects are taken into account. For more details on 
the connection between the kerneled
Fokker-Planck equation and the functional renormalization group see
Appendix {\ref{sec:app-SR-FRG}}.

\begin{figure}[t]
\includegraphics[width=\columnwidth]{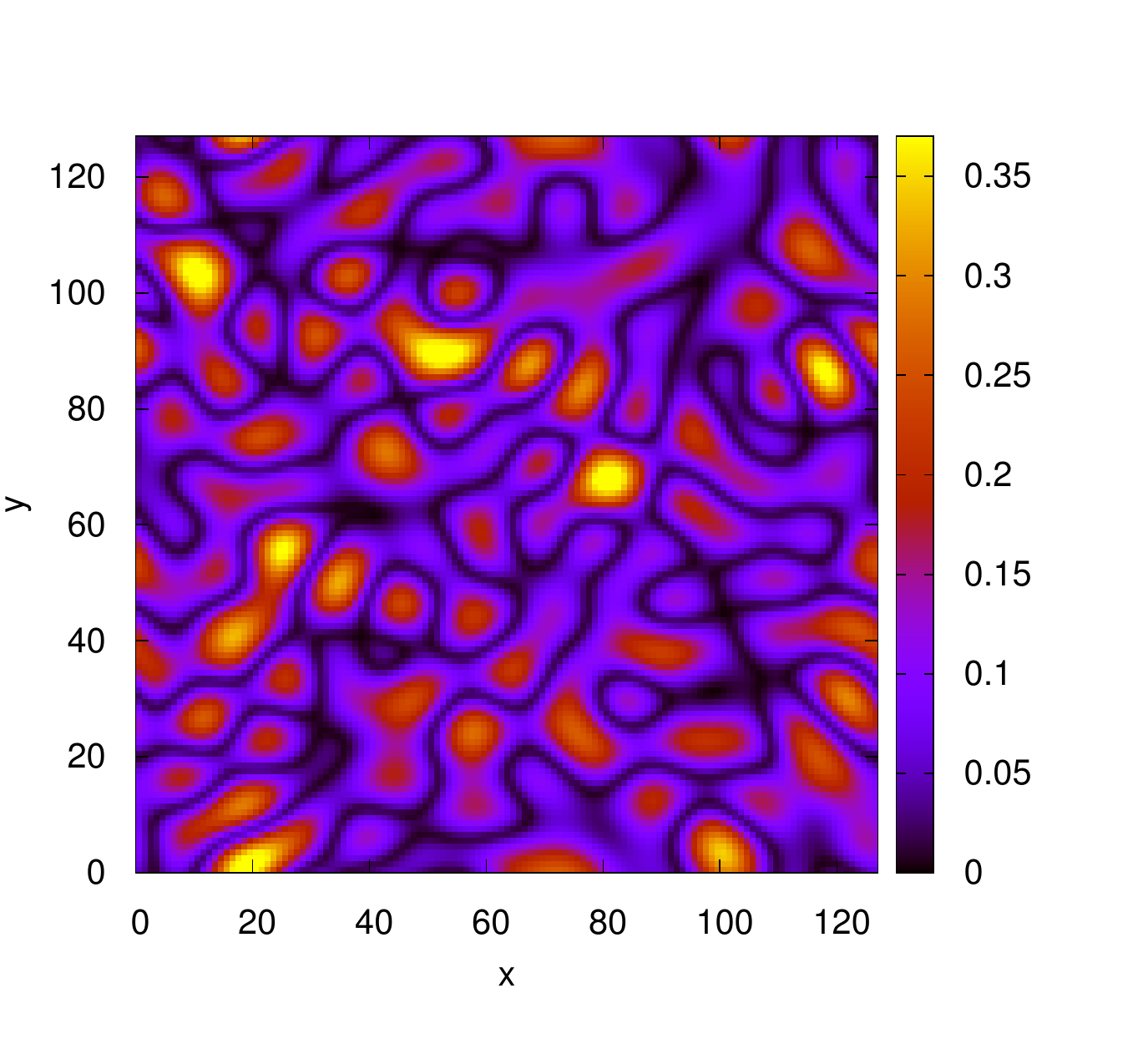}
\caption{Heat map plot of the absolute value of colored noise on a 128
  x 128 lattice for momentum cutoff
  $\latcutoff = 8$ (see Sec.\ \ref{sec:LQFT-CN}
  for the definition of the lattice cutoff). In contrast to the white
  noise picture described in (\ref{eq:noise-gaussian-relations})
  non-delta like spatially correlated structures extending over
  several lattice spacings are visible. The arrangement of the
  structures however appears to be random for we require the
  UV-modified stochastic process to remain Markovian.}
\label{fig:ColNoisePlane}
\end{figure}

The regularization function \eq{eq:sharp-cutoff-continuum} defines the colored noise field 
\begin{equation}
\eta_{\text{col}}(p, \tau) :=  \eta(p, \tau)\, \theta(\Lambda^2 - p^2)\, ,
\label{eq:col-noise-momspace}
\end{equation}
with the space-time representation  
\begin{align}
  \eta_{\text{col}}(x, \tau) = \frac{1}{(2 \pi)^d} \int\, \mathrm{d}^d p\,  
  \eta_{\text{col}}(p, \tau) \, \mathrm{e}^{-ip \cdot x}\, .
\label{eq:col-noise-space-time}
\end{align}
This leads us to the Langevin equation with colored noise 
\begin{align}
  \frac{\partial \phi(x,\tau)}{\partial \tau} =
 - \frac{\delta S}{\delta \phi(x, \tau)} + \eta_{\text{col}}(x,\tau)\, ,
\label{eq:cont-LE-CN}
\end{align}
which is used throughout the work. A visualization of the colored
noise $\eta_{\text{col}}(x, \tau)$ in \eq{eq:col-noise-space-time} with
the sharp cutoff \eq{eq:col-noise-momspace} is illustrated in
Fig.~\ref{fig:ColNoisePlane}.


\section{Lattice QFT with colored noise}
\label{sec:LQFT-CN}
In this section we present the implementation of our method for
lattice simulations of Euclidean quantum field theories. We consider
finite isotropic space-time lattices with lattice spacing $a$ and $N$
lattice points in each direction. Hence, the physical volume is
$\Omega = (aN)^d$.  Then, the allowed lattice momenta on the dual
momentum lattice are given by
\begin{equation}
p_{\mu} = \frac{2 \pi}{a\, N}\, n_{\mu} \,,\qquad {\rm with}  \qquad  \mu = 1, \hdots, d\,,
\label{eq:lat-momenta}
\end{equation}
where $n_{\mu} = -N / 2 + 1, \hdots, N / 2$.
In the thermodynamic limit $N \to \infty$ the $d$-dimensional 
Brillouin zone is given by the interval $(-\pi / a, \pi / a]^d$. 

\begin{figure}[t]
\includegraphics[width=.91\columnwidth]{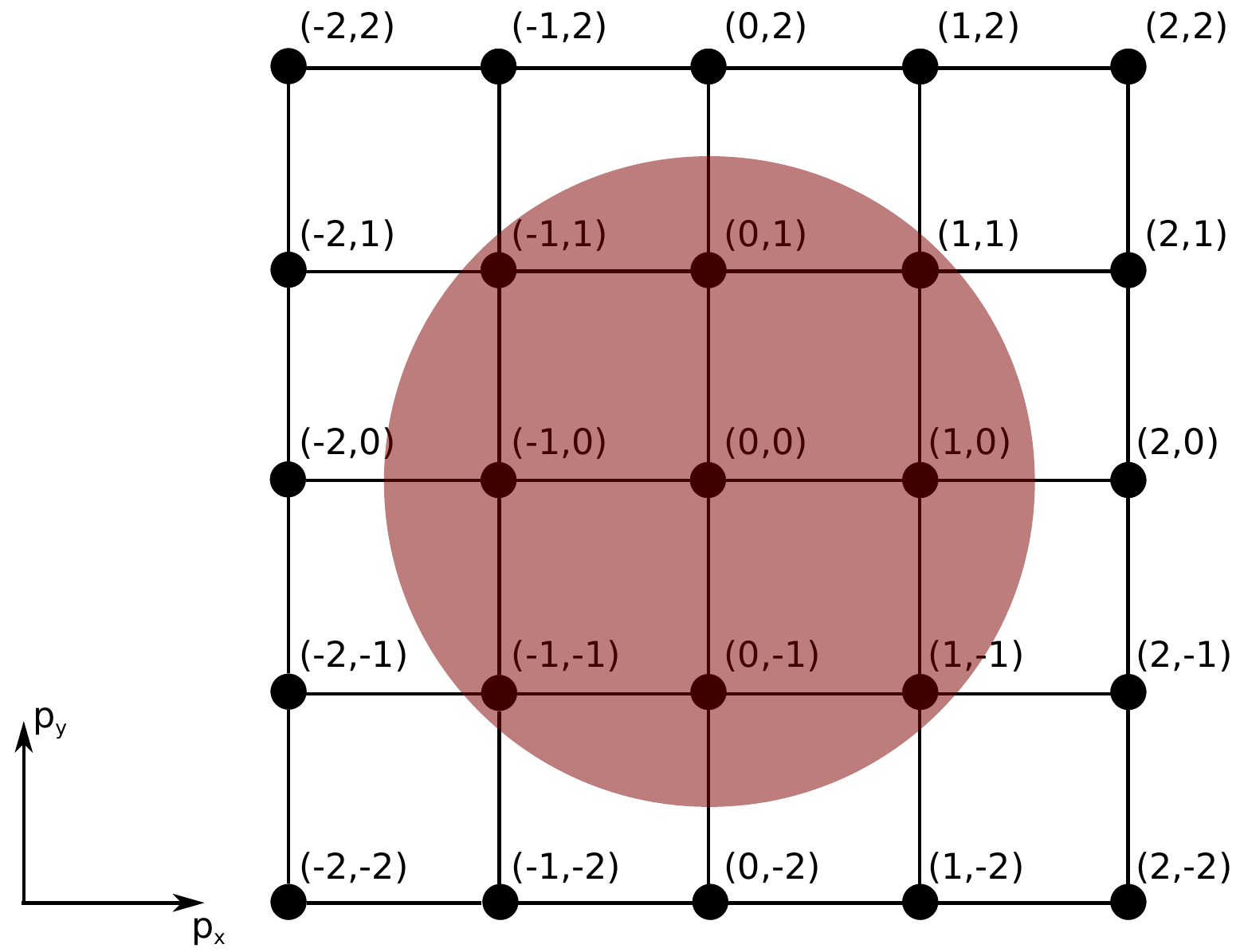}
\caption{Schematic illustration of the dual lattice in $d = 2$ 
         describing our implementation of the cutoff.
         The radius of the red shaded circle 
         corresponds to the cutoff $\Lambda$.
         Noise modes outside of the circle have zero weight in 
         the Langevin evolution. Colored noise only receives finite contributions
         for momentum modes from inside the red circle. 
         }
\label{fig:schematic-lat-mom}
\end{figure}

\begin{figure}[t]
\includegraphics[width=1\columnwidth]{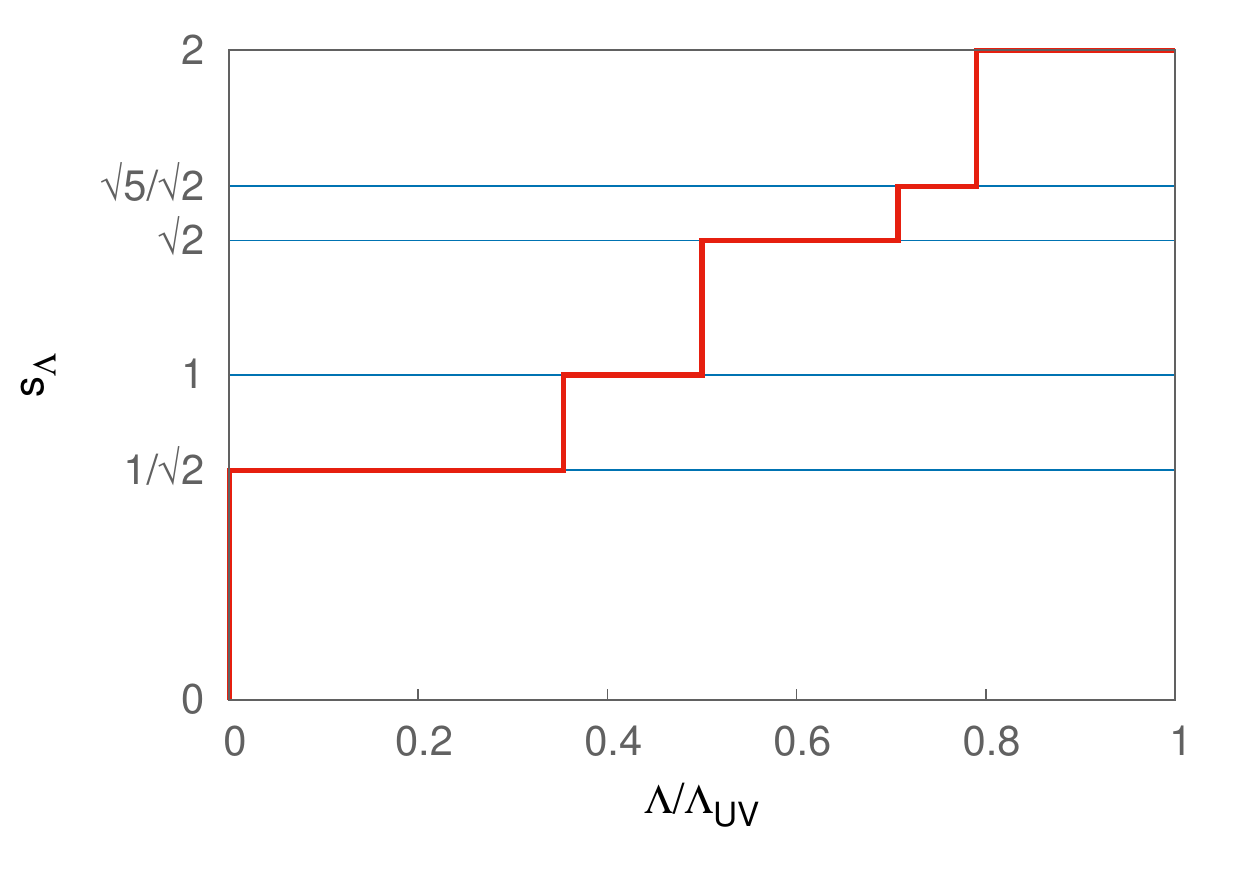}
\caption{$\latcutoff(\Lambda)$ for $N = 4$ in two dimensions. The
  integer values of $\latcutoff$ characterize the standard cubic
  momentum lattices. The latter will be used in the further analysis.}
\label{fig:sLambda-definition}
\end{figure}

In lattice simulations using the Langevin equation with colored noise 
we work with the sharp regulator (\ref{eq:sharp-cutoff-continuum})
introduced in the previous section. 
Similarly as in the continuum, 
colored noise is generated by 
cutting off the noise modes on the momentum
lattice followed by a discrete Fourier transformation
back to the real space lattice
which leads to
\begin{equation}
  \eta_{\text{col}}(x, \tau) = \frac{1}{\Omega} \, \sum_{p} \, 
\mathrm{e}^{i p \cdot x}\,  r_\Lambda(p^2)\, \eta(p, \tau)\, .
\label{eq:lat-CN-momentumspace}
\end{equation}
The discretized Langevin equation with colored noise
thus reads
\begin{align}
\phi(x,\tau_{n+1}) =
\phi(x,\tau_n) - \frac{\delta S}{\delta \phi(x, \tau_n)}\, \Delta \tau +
 \sqrt{\Delta \tau}\,  \eta_{\text{col}}(x, \tau_n)
\label{eq:lat-LE-CN}
\end{align}
with the Langevin time step $\Delta \tau$.  In our implementation we
retain noise modes with $p^2 \leq \Lambda^2$ and remove larger modes,
see Fig.~\ref{fig:schematic-lat-mom}.  Modes are being removed as the
decreasing cutoff $\Lambda$ sweeps over the discrete lattice
momenta. Note that the lattice theory only changes at the discrete
values $\Lambda=\latcutoff \pi/a$ with
\begin{align}\label{eq:sLambda}
\latcutoff= \max \left(\01{\sqrt{d }} \sqrt{n_\mu^2 }\right)\quad {\rm with} 
\quad p^2(n_\mu) \leq \Lambda^2\,. 
\end{align}
For the $\Lambda$-dependence see Fig.~\ref{fig:sLambda-definition}.
For these values the cutoff $\Lambda$ sweeps over the discrete
momentum values, see Fig.~\ref{fig:schematic-lat-mom} for a 
two-dimensional dual lattice. We also notice that
integer values of $\latcutoff$ indicate a standard cubic momentum
lattice of non-zero quantum fluctuations.  Moreover
$\latcutoff = N / 2$ corresponds to the standard Langevin evolution
with Gaussian white noise. For $\latcutoff = 0$ only the zero-momentum
mode contributes to the colored noise.  For the simulation with the
gradient flow we use the Langevin equation with the noise term set to
zero.

For the further discussion it is useful to split the full field in
momentum space in a classical and quantum contribution,
\begin{align} 
\phi(p)=\phi_{\text{\tiny{cl}}}(p)+\delta\phi_{\text{\tiny{qu}}}(p)\,,
\label{eq:cl-qm-lat-field}
\end{align} 
with 
\begin{align}
  \delta\phi_{\text{\tiny{qu}}}(p) =0 \quad {\rm for} \quad 
  p^2 > \latcutoff^2\,.
  \label{eq:quantum-lattice}
\end{align}
Note that the field $\delta\phi_{\text{\tiny{qu}}}$, that carries the
quantum fluctuations, lives on the momentum lattice defined by
$p^2 \leq \latcutoff^2$. Henceforth we call this generically smaller lattice the {\it
  quantum} lattice. In turn, the classical field
$\phi_{\text{\tiny{cl}}}$ lives on the full momentum lattice
which we therefore call the {\it classical} lattice. In
position space this translates into a fine classical lattice and a
coarser quantum lattice.


\section{Scalar field theory}
\label{sec:model}
\subsection{Lattice formulation}
Scalar field theories on the lattice have been investigated in
numerous works over the recent decades and their applications range over
a broad spectrum of topics involving particle, statistical and
condensed matter physics. Here, we consider a Euclidean real
single-component scalar field theory in $d$ dimensions with lattice
action
\begin{align}
S = &\sum_{x} 
a^d \left[\frac{1}{2}\sum_{\mu=1}^d\frac{(
\phi_0(x+a\hat\mu)-\phi_0(x))^2}{a^2} \right. \nonumber \\
&\hspace*{1.2cm}\left. + \frac{m_0^2}{2}\phi_0^2 +
 \frac{g_0}{4!}\phi_0^4\right]\, ,
\label{eq:lattice_action} 
\end{align}
where $\hat \mu$ denotes the unit vector in $\mu$-direction. 
The subscript $0$ indicates bare quantities, 
i.e.\ the bare mass $m_0$, the bare coupling $g_0$ and 
the bare field $\phi_0$ in the action. For numerical simulations the action 
is cast in the following dimensionless form 
\begin{multline}
S = \sum_{x}\left[-2\kappa 
    \sum_{\mu=1}^d\phi(x)\phi(x+\hat \mu)\right.\\
    \left. +\, \vphantom{\sum_{\mu = 1}^d} (1 - 2 \lambda) \, 
    \phi(x)^2 +\lambda \, \phi(x)^4 \right] \, .
 \label{eq:lattice_action_kappa_lambda}
\end{multline}
The parameter $\kappa$ is the so-called hopping parameter 
and $\lambda$ describes the quartic coupling of the theory.
Note that here, the parameters $\kappa$ and $\lambda$ are positive. 
They are related to the bare mass, bare coupling and 
the lattice spacing in the following way 
\begin{align}
 a^{\frac{d-2}{2}}\phi_0 &= (2\kappa)^{1/2}\, \phi\, , \nonumber \\
 (am_0)^2 &= \frac{1-2\lambda}{\kappa} -2d\, , \nonumber \\
 a^{-d + 4}\, g_0 &= \frac{6\lambda}{\kappa^2}\, ,
 \label{eq:kappa-lambda-params}
\end{align}
where we have introduced the dimensionless field $\phi\, .$
The white noise Langevin update step ($\tau_{n} \to \tau_{n + 1}$) 
of a field variable at lattice point $x$ is given by
\begin{align}
\phi(x,\tau_{n+1}) = 
\phi(x,\tau_n) + K[\phi(x, \tau_n)] \Delta \tau + \sqrt{ \Delta \tau}\,  \eta(x, \tau_n)\, ,
\label{eq:lat-LE-update}
\end{align}
where the drift term $K[\phi(x)]= - \delta S[\phi]/\delta\phi(x)$ explicitly reads
\begin{align}
K[\phi(x)] &= 2 \kappa \sum_{\mu = 1}^d 
              [\phi(x + \hat \mu) + \phi(x - \hat \mu)] \nonumber \\
           & \hspace*{10pt} + 2 \phi(x)\, (2 \lambda\, (1-\phi(x)^2) - 1)\, .
\label{eq:lat-drift}
\end{align}
The process (\ref{eq:lat-LE-update}) can be solved iteratively by using an explicit 
Euler-Maruyama discretization scheme. Higher order Runge-Kutta schemes 
are possible as well and are discussed in \cite{Damgaard:1987rr, Batrouni:1985}.

\begin{figure*}[t]
    \centering
     \includegraphics[width=.65\columnwidth]{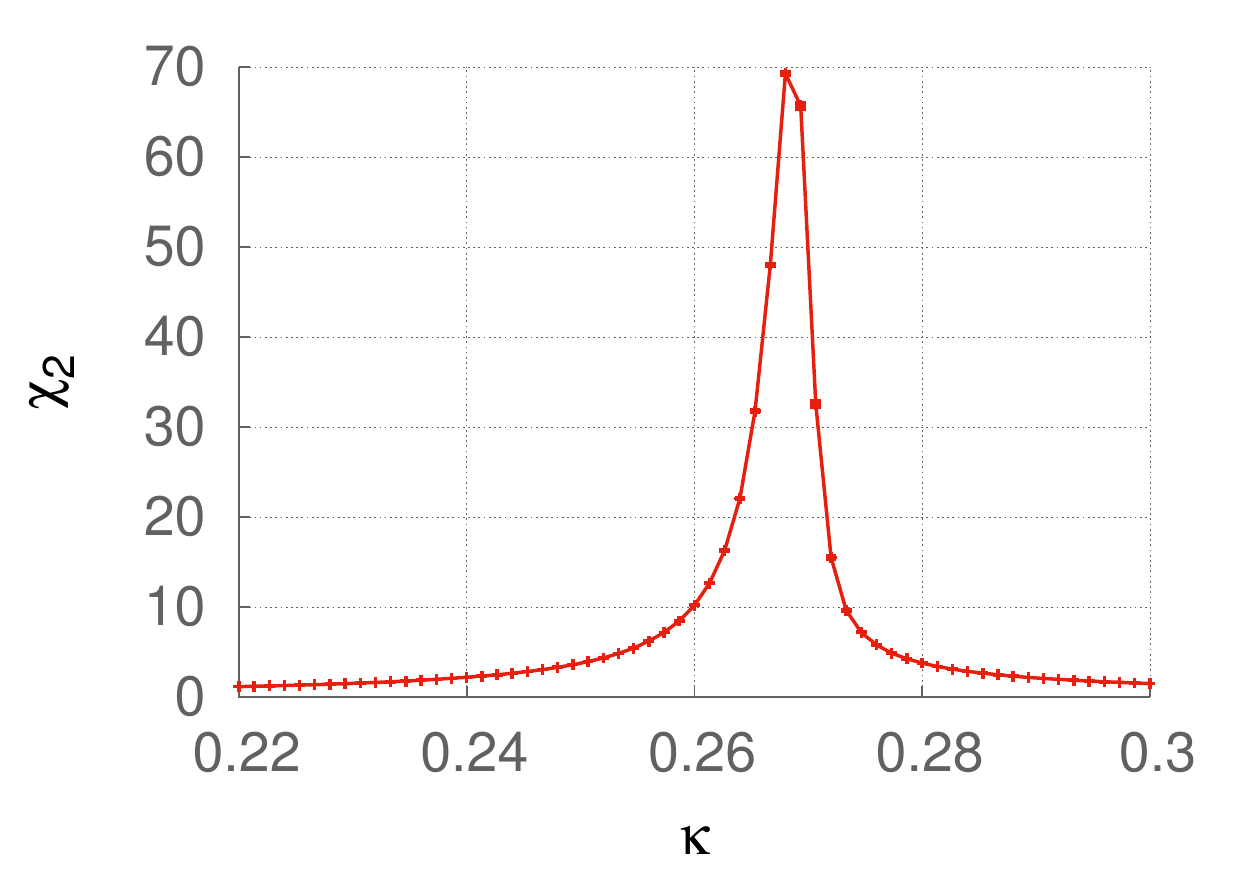}
     ~
     \includegraphics[width=.65\columnwidth]{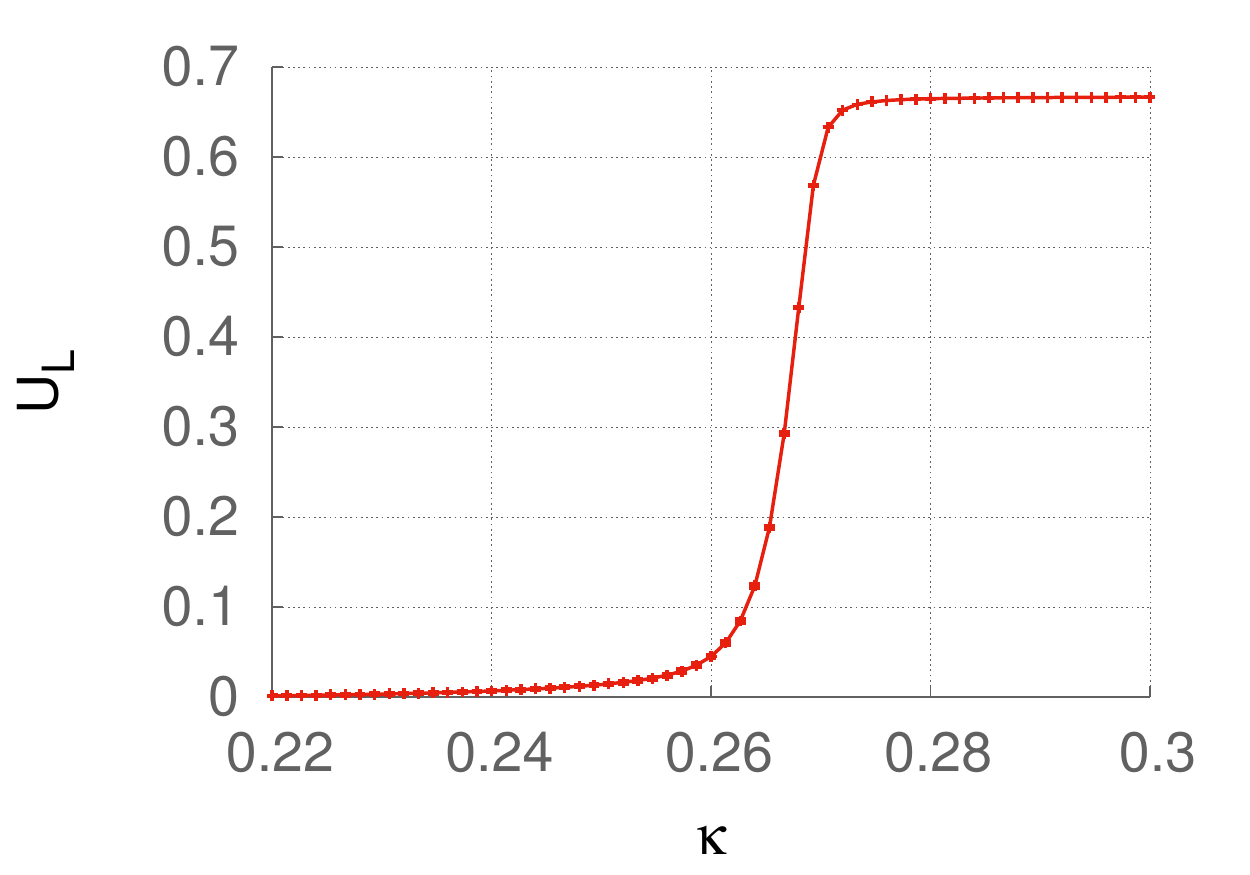}
     ~
     \includegraphics[width=.65\columnwidth]{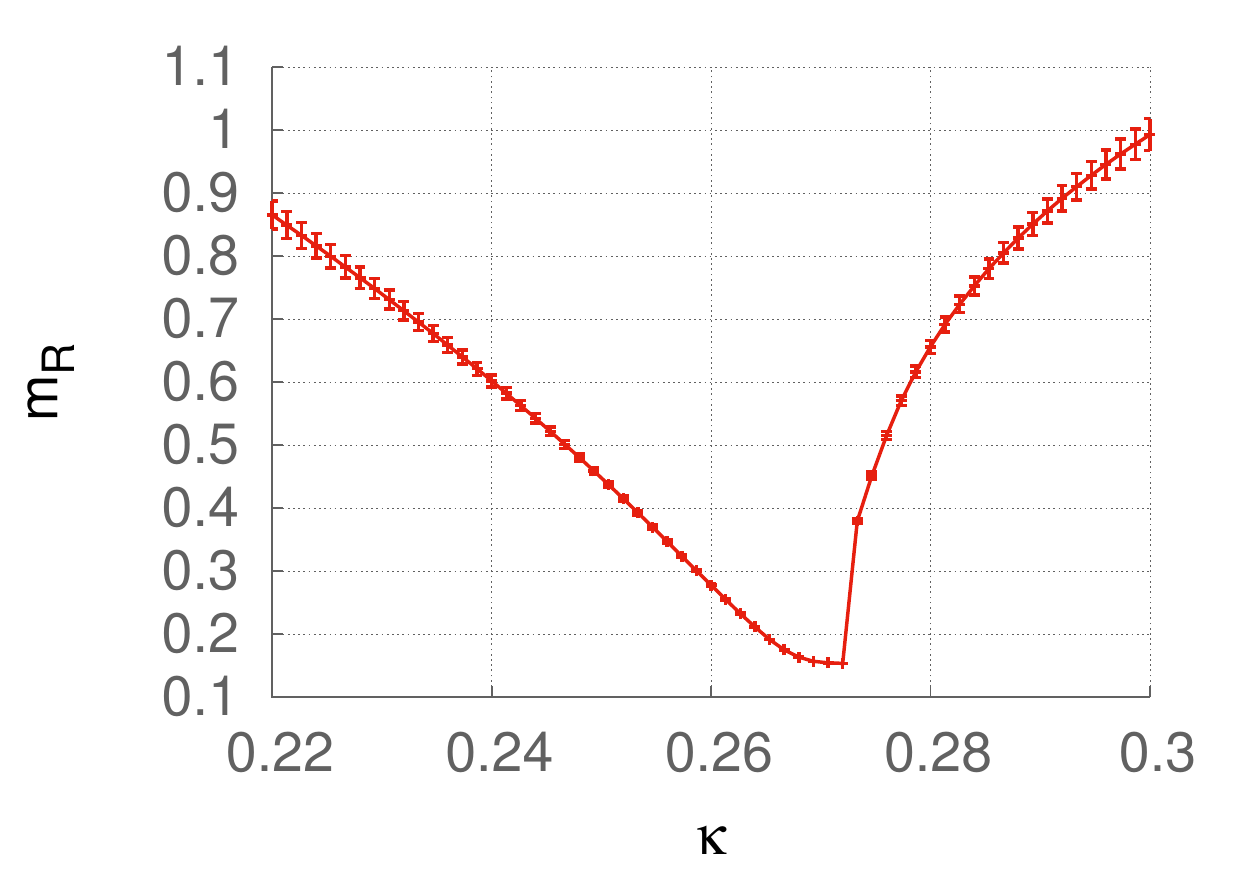}     
     \caption{The phase diagram of $O(1)$ scalar field theory 
              on the lattice characterized by three of the main 
              observables, here at (fixed) coupling $\lambda = 0.02$ 
              for a $32 \times 32$ lattice. 
              (Left) The plot on the left-hand side shows the 
              susceptibility as a function of the hopping parameter 
              $\kappa$. 
              (Middle) The central figure depicts the Binder cumulant. 
              The critical value is given by $\kappa_c \approx 0.27$. 
              The range for $\kappa <= \kappa_c$ describes the symmetric phase. 
              For values of $\kappa$ larger than the critical point the theory 
              is in the phase with spontaneously broken symmetry. 
              (Right) The figure on the right-hand side shows 
              the renormalized mass $m_R$ in lattice units. 
              The results were produced using the 
              Langevin equation with white noise taking $10^6$ measurements 
              in equilibrium $\tau \to \infty$. The data points are 
              connected by lines to guide the eye.}
\label{fig:Scalar-Phase-diagram}
\end{figure*}
Let us consider the case $d \geq 2$. If the action contains no
explicit symmetry breaking term for each value of $\lambda$, there
exists a critical value of the hopping parameter $\kappa_c$ at which
the system undergoes a second order phase transition. The
$\mathbb{Z}_2$ symmetry of the system becomes spontaneously broken
above the critical point. The phase transition for the case of $d = 2$
is illustrated in Fig.\@ \ref{fig:Scalar-Phase-diagram}.  Classically,
the broken phase is characterized by a negative mass term
$(am_0)^2 < 0$, leading to two degenerated minima in the
potential. Within the dimensionless formulation these minima are at
$\pm \phi_{\text{\tiny{min}}}$ with
\begin{equation}
 \phi_{\text{\tiny{min}}}\,  =
  \left[\frac{1}{2 \lambda}(2d \kappa - 1 + 2 \lambda)\right]^{1/2}\, .
\label{eq:class-min}
\end{equation}
The critical value for the hopping parameter in the classical theory 
can be determined by requiring the mass term to vanish, leading to 
\begin{equation}
\kappa_c(\lambda) = \frac{1 - 2 \lambda}{2 d}\, .
\label{eq:class-crit-kappa}
\end{equation}


\subsection{Observables}
We now discuss some of the main observables to explain 
the properties of the theory. Those are useful 
in the analysis of the effects of colored noise. 
The vacuum expectation value of the field also called 
the magnetization reads
\begin{equation}
 \langle M \rangle := \left\langle \frac{1}{\Omega} \sum_{x} \phi(x) \right\rangle\, .
 \label{eq:obs_lat_magnetization}
\end{equation}
It is zero in the symmetric phase of the 
theory and takes a finite value in the broken phase.
Note that $\Omega$ is given by the number of lattice points 
since we consider the dimensionless formulation. 
The connected two-point correlation is defined as
\begin{align}
 G_c(x,y) = \langle \phi(x) \phi(y) \rangle_c 
 \equiv \langle \phi(x) \phi(y) \rangle - \langle \phi(x) \rangle \langle \phi(y) \rangle\, .
 \label{eq:obs_conn_correlator}
\end{align}
From this we obtain the two-point correlation function of time slices
by evaluating the spatial Fourier transform of $G_c(x, y)$ at
vanishing spatial momentum
\begin{align}
 G_c(t) = \frac{1}{V}\, \sum_{\vec x} G_c(x, 0)\,.
 \label{eq:obs_2point_tempcor}
\end{align}
It measures the decay of correlations over the time extent of the lattice.
The mass is related to the inverse correlation length.
Moreover, (\ref{eq:obs_2point_tempcor}) is related to the connected susceptibility by 
\begin{align}
 \chi_2 = V \sum_t  G_c(t)  = \Omega\, \left( \langle M^2 \rangle - \langle M \rangle^2 \right)\,.
 \label{eq:obs_2point_susceptibility}
\end{align}
Hence, the susceptibility is the (d-dimensional) Fourier transform of the correlator
(\ref{eq:obs_conn_correlator}) evaluated at zero momentum.
The susceptibility measures the Gaussian fluctuations 
of the magnetization.
The fourth-order cumulant or Binder cumulant \cite{Binder} 
quantifies the curtosis of the fluctuations. 
It can be used to study phase transitions 
and to determine critical exponents. The Binder cumulant reads
\begin{equation}
 U_L = 1 - \frac{1}{3} \frac{\langle M^4 \rangle}{\langle M^2 \rangle^2}\, .
 \label{eq:obs_Binder}
\end{equation}
It vanishes in the symmetric phase and assumes 
the value $2 /3$ in the phase with broken symmetry. The second moment is defined by
\begin{equation}
\mu_2 := d\, V \sum_{t} \, t^2\, G(t)_c\, .
\label{eq:obs_second_moment}
\end{equation}
From (\ref{eq:obs_2point_susceptibility}) and 
(\ref{eq:obs_second_moment}) the renormalized 
mass can be computed according to
\begin{equation}
m_R^2 = \frac{2\, d\, \chi_2}{\mu_2}\, .
\label{obs_ren_mass}
\end{equation}
This is derived in more detail in Appendix \ref{sec:app-obs}. 
In Fig.~\ref{fig:Scalar-Phase-diagram} the behaviour of the connected
susceptibility, the Binder cumulant and the renormalized mass 
as a function of $\kappa$ for constant $\lambda = 0.02$ are 
shown across the phase transition. 
\begin{figure*}[t]
    \includegraphics[width=.9\columnwidth]{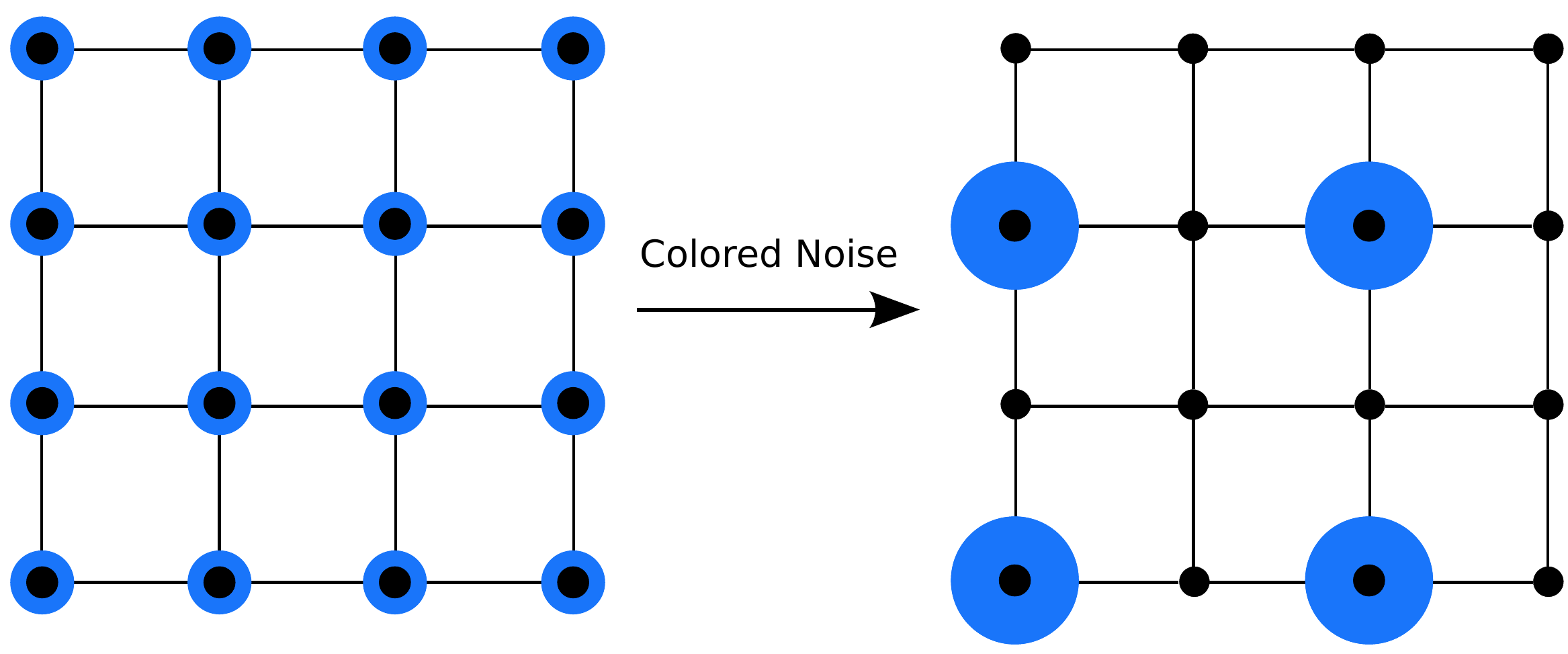}
    ~
    \hspace*{0.5cm}
\includegraphics[width=.9\columnwidth]{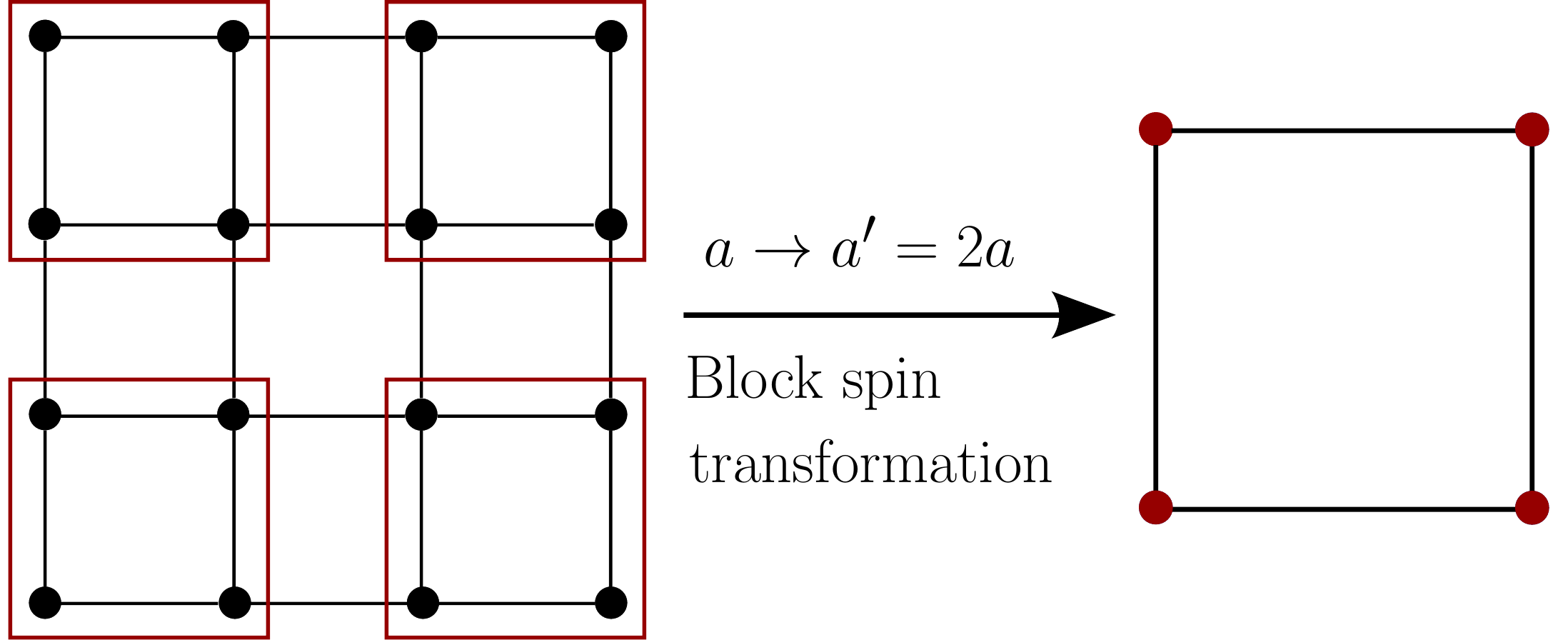}
    \\[2ex]
\caption{(Left) Illustration of the classical and 
         the quantum lattice in two-dimensional real space (blue figure). 
         The two lattices agree for the Langevin equation with white noise
         where $\latcutoff = \latcutoffmax$, see the 
         left lattice sketch. Using a sharp regulator only field modes 
         with $|p| \leq \latcutoff$ receive a 
         non-vanishing contribution from the colored noise term. 
         This leads to a coarser quantum lattice (blue circles) 
         compared to its classical
         counterpart (black points) which is unaffected by the cutoff.\\ 
         (Right) A typical block spin transformation 
         in two dimensions is illustrated (red figure). 
         A possible choice of the transformation is to 
         define the blocked field variables by averages over the four fields 
         inside the red squares. This leads to a coarser lattice with double the lattice spacing 
         and a quarter of the original lattice points.
         The illustrations point out the analogy between colored 
         noise coarsening the quantum lattice and standard 
         block spin transformations.
         }
\label{fig:sketch_block_spin}
\end{figure*}
\section{Colored Noise and the Renormalization Group}
\label{sec:RG-study}

Colored noise introduces a UV cutoff $\latcutoff$. The change of the
theory with an infinitesimal change of the cutoff is governed by the
renormalization group. In terms of our lattice setup colored noise
leads to the separation into the classical and the quantum lattice
(\ref{eq:cl-qm-lat-field}).  The momentum space quantum lattice
(\ref{eq:quantum-lattice}) contains only field modes with
$|p| \leq \latcutoff$. Those receive a non-zero contribution to
fluctuations from the colored noise term in the Langevin equation
(\ref{eq:lat-LE-CN}). The remaining contribution encoded in the drift
term is purely classical and applies to all field modes. 
Let $\latcutoffmax$ denote the maximum cutoff 
on the lattice with $N^d$ points.
In general, for a given cutoff $\latcutoff \leq \latcutoffmax$ the
quantum lattice in momentum space has less points than the classical
lattice.  In the limit $\tau \to \infty$ the field modes with
$|p| > \latcutoff$ assume their classical value according to the
limit of the gradient flow. The fewer points of the quantum momentum
lattice translate into a coarser quantum real space lattice as
compared to the classical real space lattice, see
Fig.~\ref{fig:sketch_block_spin}. 
With this in mind we
study the relation of the colored noise Langevin evolution
to the renormalization group in more detail. We investigate 
if the effect of the 
cutoff $\latcutoff$ may be compensated by
varying the lattice spacing $a$, 
thus tuning the coarseness of the quantum real space lattice (\ref{eq:quantum-lattice}). To this end we compare a simulation with white noise  at $\latcutoff = \latcutoffmax$ 
on a coarse lattice with spacing $a_1$ with a colored noise 
simulation at cutoff $\latcutoff < \latcutoffmax$
on a fine lattice with spacing $a_2 < a_1$.

Our procedure is to introduce 
scale factors for the following parameters
\begin{equation} 
 a \to a^{\prime} = s^{-1} a 
 \, , \ N \to N^{\prime} = s N
 \, , p \to p^{\prime} = s p\, ,
 \label{eq:scale-factors}
 \end{equation}
where $\ s \geq 1$ and $a$, $a^{\prime}$
are the original coarse and the fine lattice spacing.
Correspondingly, the lattice size $N$ as well as 
the lattice momenta $p$ are transformed.
The physical volume $(a N)^d$ is thereby kept constant.
The cutoff is transformed according to
\begin{equation}
 \latcutoff(a, N)
 \to \latcutoff^{\prime}(a^{\prime}, N^{\prime}) = s^{-1} \latcutoff(a^{\prime}, N^{\prime})\,.
 \label{eq:transformed-plambda}
\end{equation}
To give an explicit example of our transformation logic we consider
the case $N = 4$, $s = 2$.  Let the cutoff on the coarse lattice be
$\latcutoff(a, N = 4) \equiv \latcutoffmax(a, 4) = 2$.  This
corresponds to the white noise case.  The transformed cutoff
reads
$\latcutoff^{\prime}(s^{-1} a, s N) = 2 = \latcutoffmax(a / 2\,, 8)\,
/ \, 2$.  This corresponds to a colored noise simulation at half the
maximum cutoff on the finer lattice.

 The above scaling transformations result in a change of the
 parameters $\kappa$ and $\lambda$ in the scalar theory introduced in
 Sec.\ \ref{sec:model}.  From now on we explicitly consider the two-dimensional 
 theory. To derive the tree-level renormalization group
 equations for the parameters $\kappa$ and $\lambda$ we fix the bare
 parameters $m_0,\,g_0$, see (\ref{eq:kappa-lambda-params}) in Sec.\
 \ref{sec:model}.  The first expression of (\ref{eq:scale-factors})
 yields
\begin{align}
(am_0)^2 &\to  s^{-2}\,(am_0)^2\,, \nonumber \\
a^2g_0 &\to s^{-2}a^2g_0\, .
\label{eq:a-ren}
\end{align}
Next we use the definition (\ref{eq:kappa-lambda-params}) in
(\ref{eq:a-ren}) leading to 
\begin{align}
s^{-2} \left[\frac{1 - 2\lambda}{\kappa} - 4\right] &= \frac{1 -
 2\lambda^{\prime}}{\kappa^{\prime}} - 4 \nonumber \\
s^{-2}\frac{6 \lambda}{\kappa^2} &= \frac{6 \lambda^{\prime}}{{\kappa^{\prime}}^2}\, .
\label{eq:tree-level-trafo-1}
\end{align}
These equations can be solved for
$\kappaprime$ and $\lambdaprime$ used in the colored noise simulation
with cutoff $\latcutoff^{\prime}$. 
We remark, that the equations (\ref{eq:tree-level-trafo-1}) 
are akin with standard block-spinning equations.
Under a complete block spin transformation the partition function is
invariant. This requires an adjustment of the couplings of the theory
which completes the renormalization group step.
The right-hand side of Fig.~\ref{fig:sketch_block_spin} shows a typical block spin
transformation on a two-dimensional lattice. Field variables are 
organized into blocks by local averaging which reduces the number of lattice 
points and renders the lattice coarser. The physical volume thereby remains fixed.
For the correlation length this entails 
\begin{equation}
2\, \xi(N^{\text{\tiny{B}}} = N/2, \kappa^{\text{\tiny{B}}}, \lambda^{\text{\tiny{B}}}) = \xi(N, \kappa, \lambda) \,,
\label{eq:block-spin-trafo}
\end{equation} 
where $N^{\text{\tiny{B}}}, \kappa^{\text{\tiny{B}}}, \lambda^{\text{\tiny{B}}}$
are the number of lattice points and the adjusted couplings on the blocked lattice.
Our procedure is therefore analogous to block spinning since decreasing 
the cutoff $\latcutoff$ generates the local averaging and the coarsening
of the quantum lattice. 


\section{Numerical Results}
\label{sec:Numerical-Results}
In this section, we present numerical results for the scalar theory in
two dimensions. All simulations in this work have been carried out using
the sharp regulator function defined in (\ref{eq:sharp-cutoff-continuum})
and a fixed Langevin time step $\Delta\tau = 10^{-2}$.  For a comparison of
different regularization choices see App.~\ref{sec:app-reg-function}. In the first 
part of this section we study the effect of the sliding 
cutoff scale $\latcutoff$ by means of the 
observables introduced in Sec.~\ref{sec:model}. 
Our simulation with maximal $\latcutoff$
(white noise) reproduces the results in \cite{De:2005ny}.
In the second part we focus on the relation between colored 
noise and the real space renormalization group.

\subsection{Colored Noise: incomplete blocking}
\label{sec:CN-Effects}

\begin{figure}[t]
\includegraphics[width=\columnwidth]{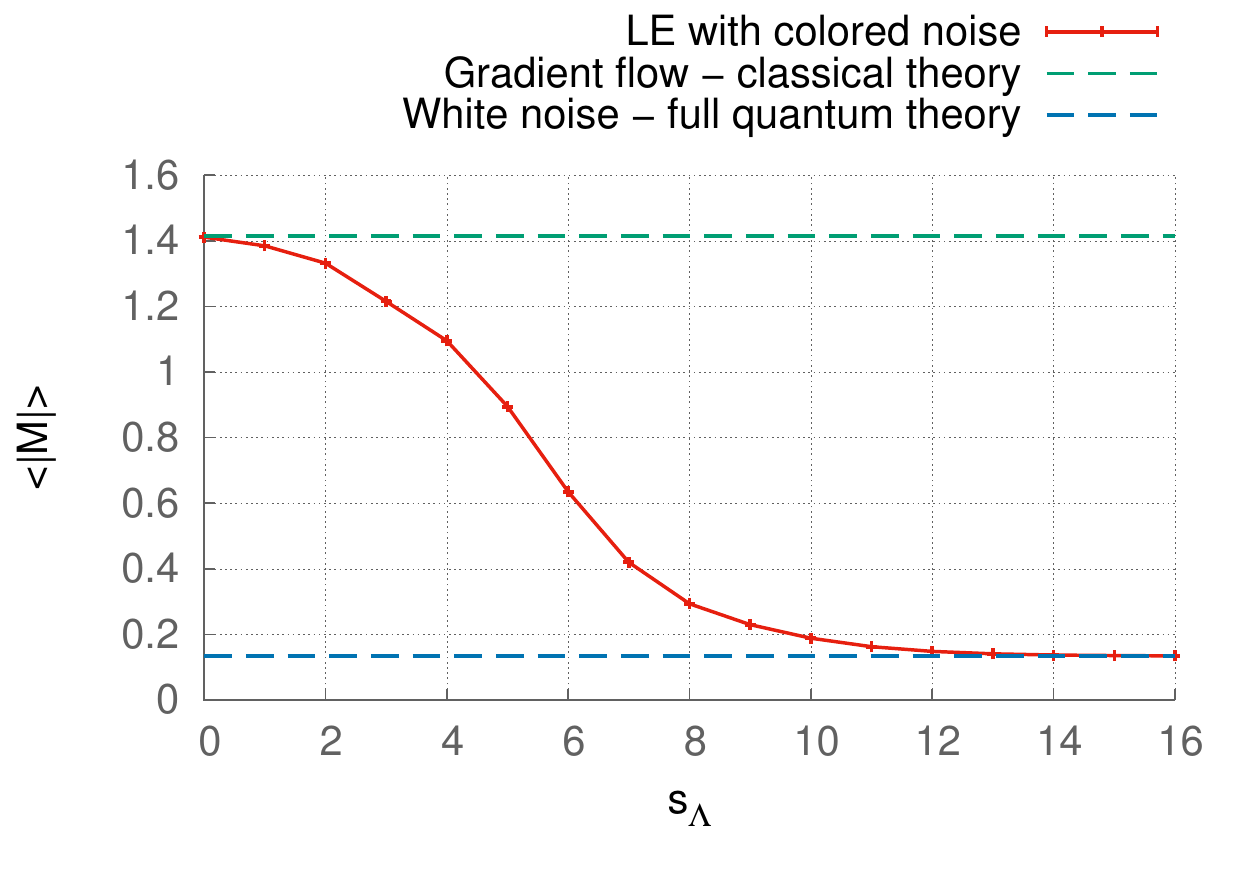}
\caption{Magnetization $\langle |M|\rangle$ as a function of the
  cutoff $\latcutoff$ for $\kappa = 0.26$, $\lambda = 0.02$ and
  $N = 32$.  For these external parameters $(\kappa, \lambda)$ the
  minima of the classical potential are located at
  $\pm \phi_{\text{\tiny{min}}}$ with
  $\phi_{\text{\tiny{min}}} =\sqrt{2}$ as given by
  (\ref{eq:class-min}). We find a consistent interpolation
  between the classical and the full quantum theory using our colored
  noise setup.}
\label{fig:WF-CN-WN-Benchmark}
\end{figure}

A first check of our colored noise approach is shown in Fig.~\ref{fig:WF-CN-WN-Benchmark}.
The expectation value of the absolute magnetization measured on a $32 \times 32 $ 
lattice is plotted as a function of the cutoff $\latcutoff$. 
Here, for the parameter choice ($\kappa = 0.26$, $\lambda = 0.02$) 
the classical theory is in the broken phase 
and the full quantum theory is in the symmetric phase. 
The white noise result ($\latcutoff=16$) is indicated by the blue dashed line.
We find that colored noise (red data) allows for a consistent interpolation 
between the full quantum theory and the classical theory.

\begin{figure*}
\includegraphics[width=0.65\columnwidth]{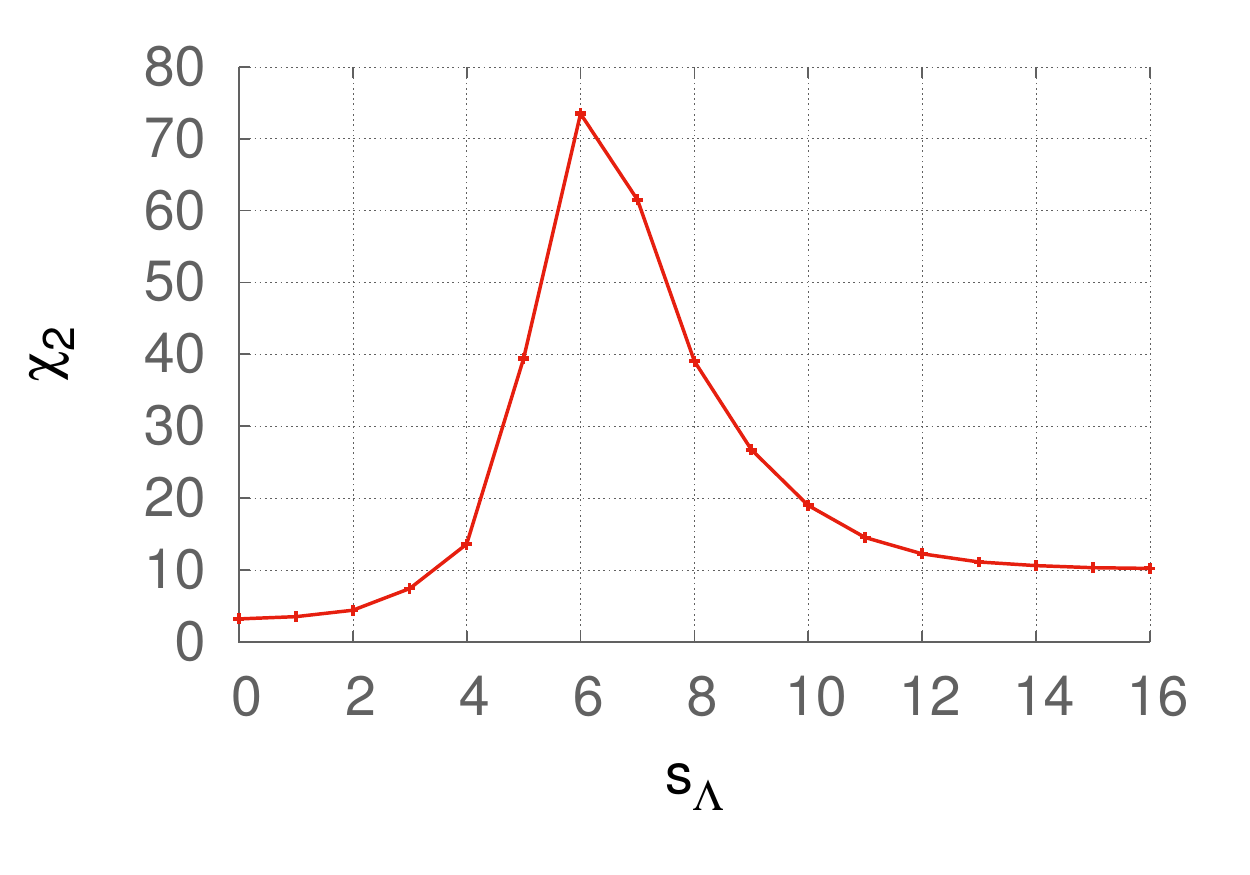}
~
\includegraphics[width=0.65\columnwidth]{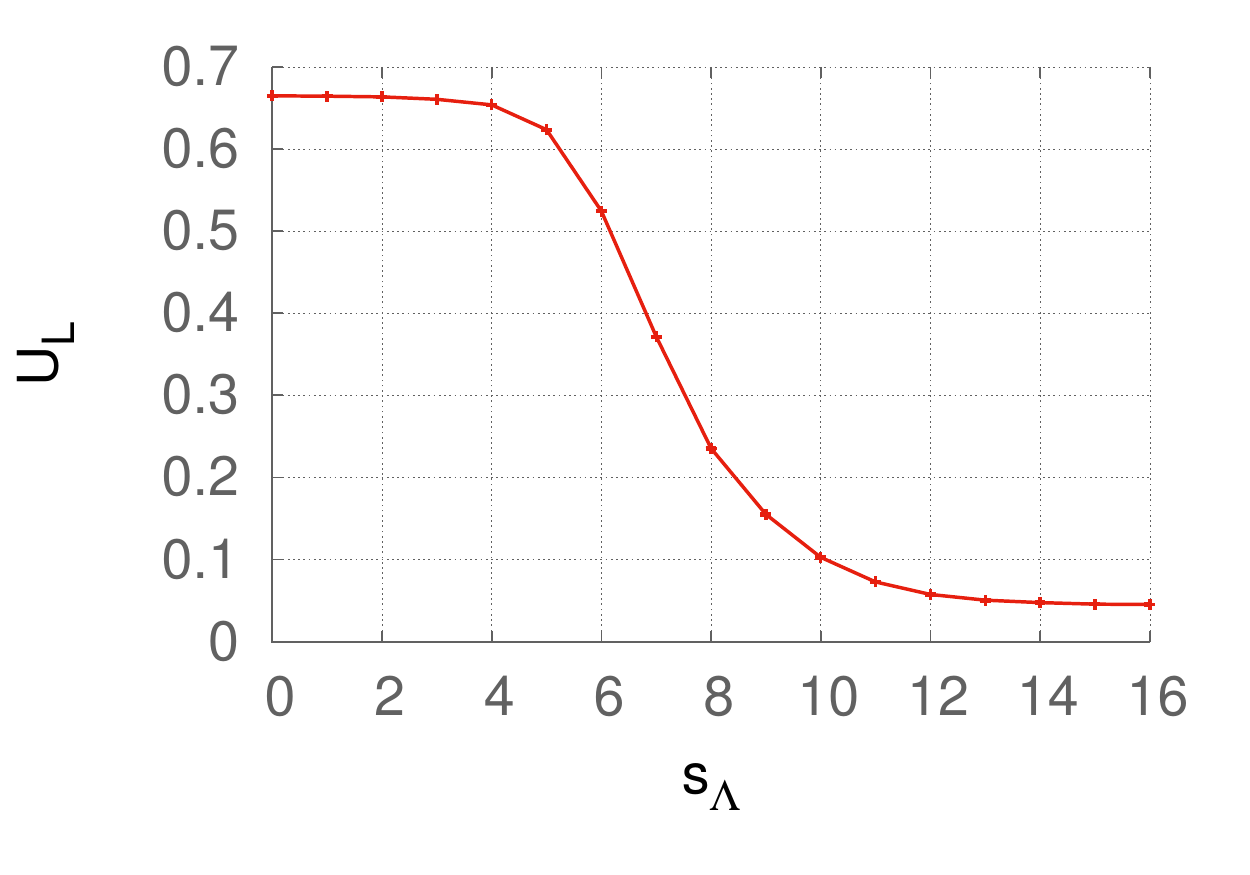}
~
\includegraphics[width=0.65\columnwidth]{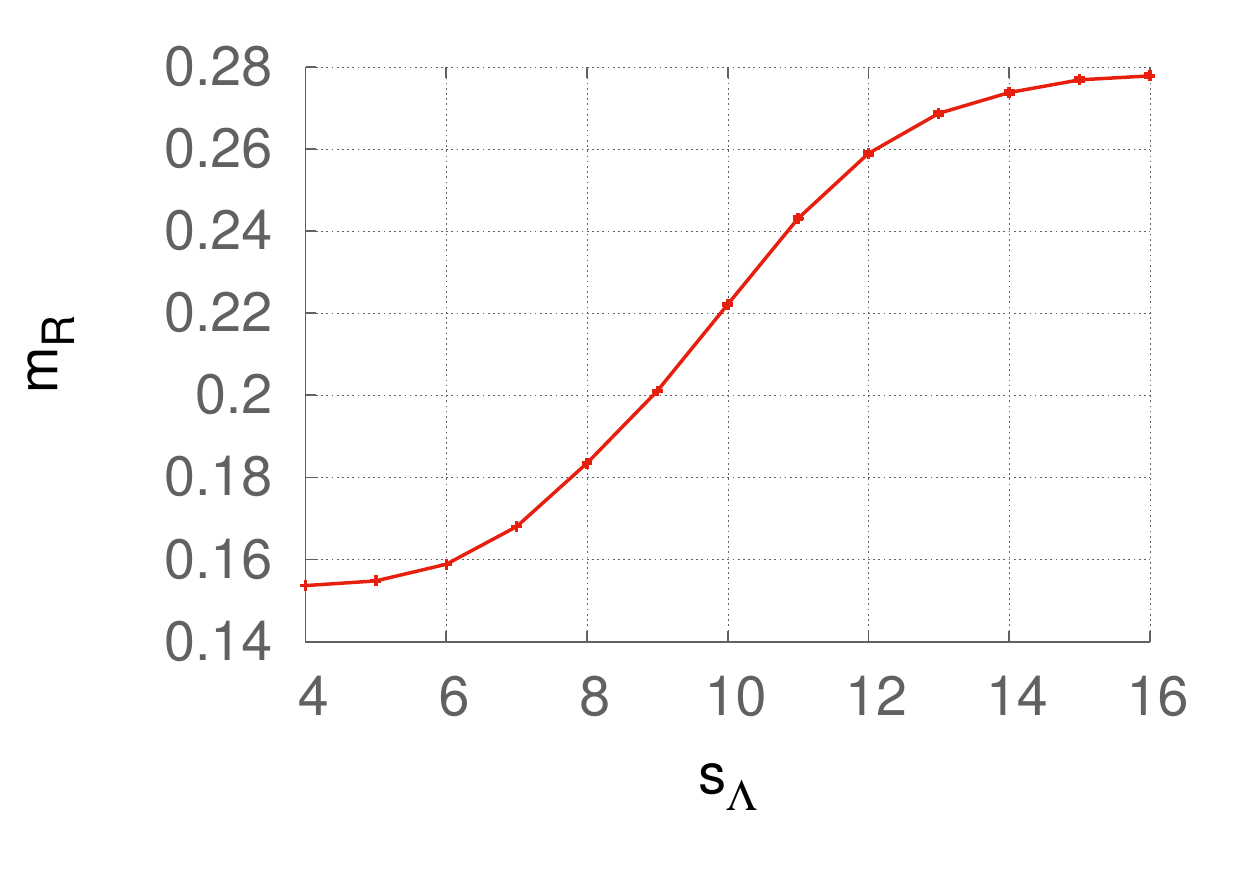}
\caption{The phase transition of the theory is shown by means of its
  characteristic observables as a function of the cutoff
  $\latcutoff$. The parameters $N = 32, \kappa = 0.26$ and
  $\lambda = 0.02$ are fixed.  (Left) The susceptibility shows a peak
  indicating that the quantum theory \textit{moves} from the symmetric
  to the broken phase where the classical theory lives as ultraviolet
  modes are being removed successively.  (Middle) The Binder cumulant
  confirms this effect of colored noise.  (Right) The renormalized
  mass $m_R$ smoothly decreases towards the phase transition as the
  cutoff is lowered from $\latcutoff = 16$ to $\latcutoff = 4$. 
  For $\latcutoff < 4$, $m_R$ as defined in \eq{obs_ren_mass} 
  shows a delayed phase transition from the symmetric to the broken phase.
  Moreover, below $\latcutoff = 4$ the sharp cutoff imprints artifacts
  on the correlation function.
  For a detailed illustration of this 
  behaviour see also Appendix \ref{sec:app-reg-function}.}
\label{fig:obs-CN-shells-N-32}
\end{figure*}

Next the interpolation between the two phases is investigated further
by considering the fluctuation content of the theory. Thereto, we
analyze the susceptibility, the Binder cumulant and the renormalized
mass shown in Fig.~\ref{fig:obs-CN-shells-N-32}. The parameters are
the same as for Fig.~\ref{fig:WF-CN-WN-Benchmark}. 
Cutting off ultraviolet modes gradually moves the susceptibility
(left) and the Binder cumulant (middle) across the phase transition.
This confirms the effects of colored noise observed 
in Fig.~\ref{fig:WF-CN-WN-Benchmark}. 
The right plot in Fig.~\ref{fig:obs-CN-shells-N-32} 
shows the renormalized mass calculated
from the second moment and the connected susceptibility. The mass
decreases with lowering the cutoff which means that the correlation
length (in lattice units) increases.  Beyond the critical point we
expect the renormalized mass to increase again. However, for
$\latcutoff < 4$ the sharp regulator induces oscillations in the
correlation function of time slices. Then, $m_R$ 
as defined in (\ref{obs_ren_mass}) shows a delayed transition
from the symmetric to the broken phase. This problem can be 
resolved with the application of a smooth regulator function.  
In Appendix~\ref{sec:app-reg-function} we analyze the
behaviour of the correlation function of time slices comparing two
different regularization choices. From this we can draw conclusions
on the behaviour of $m_R$ for any $\latcutoff$.
In summary
we find that the susceptibility, the Binder cumulant and the
renormalized mass represent quantities that are sensitive to the
application of colored noise if all bare parameters
($\kappa, \lambda$) and the lattice size $N$ are kept fixed.

\begin{figure}[h]
\includegraphics[width=\columnwidth]{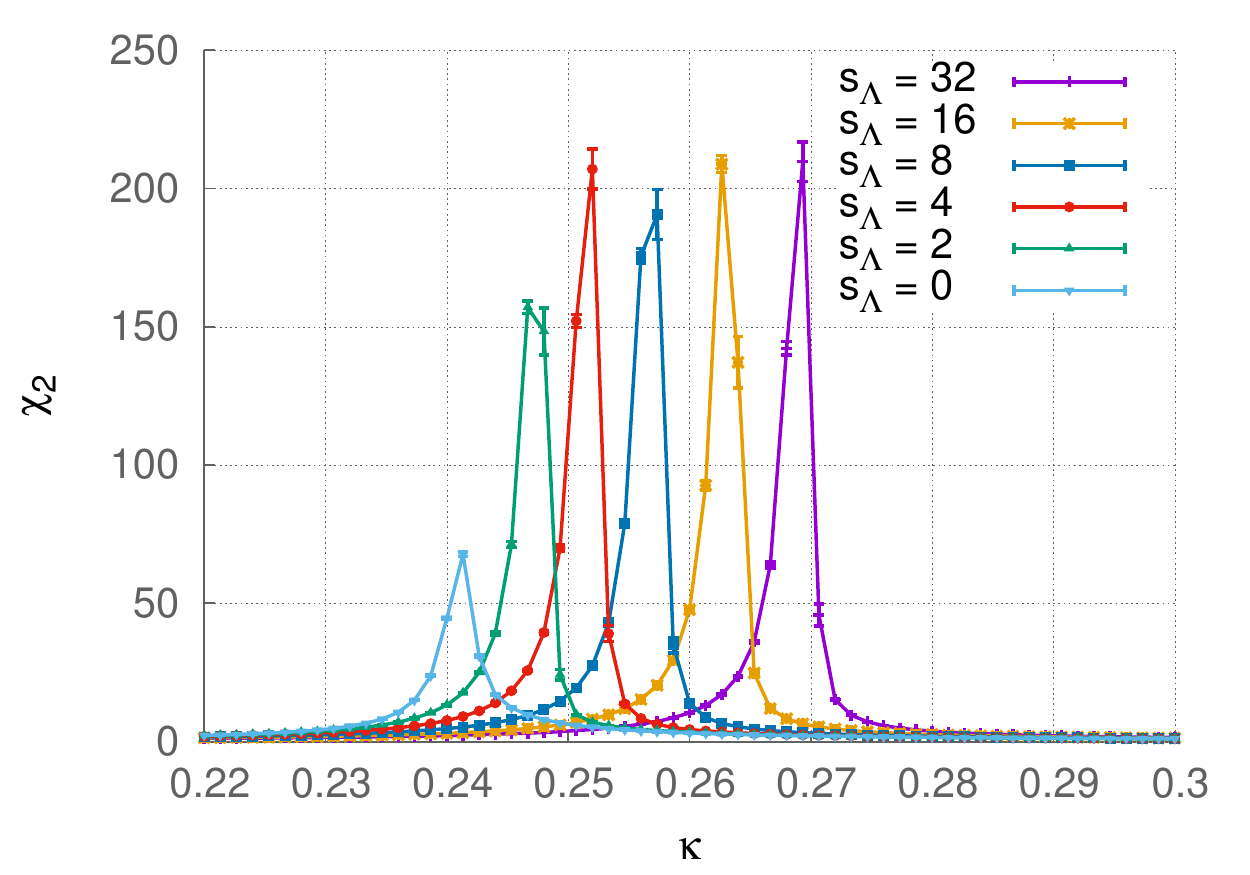}
\caption{The susceptibility $\chi_2$ as a function 
         of $\kappa$ for $N = 64$ and fixed $\lambda = 0.02$ 
         for different cutoffs $\latcutoff$. As the cutoff is 
         lowered the peak of the susceptibility is being
         shifted towards smaller values of $\kappa$ and 
         approaches the classical critical value 
         $\kappa_c^{(\text{cl})} = 0.24$. For the lowest cutoffs
         the peak height shrinks. Note that the white
         noise result is represented by the violet 
         curve for $s_{\Lambda, \text{\tiny{max}}}=32$.}
\label{fig:susceptibility-CN-kappa-N-64}
\end{figure}

To continue our analysis we investigate the susceptibility as a
function of $\kappa$ for different cutoffs $\latcutoff$ as shown in 
Fig.~\ref{fig:susceptibility-CN-kappa-N-64}.  Here, the results were
produced on a $64 \times 64$ lattice where $\lambda = 0.02$ is kept
fixed.  The violet curve depicts the white noise result. Our observations are: The peak position corresponding
to $\kappa_c$ is successively shifted towards lower values of $\kappa$
with decreasing cutoff. This is consistent with the previous results
in this section. Colored noise removes quantum fluctuations rendering
the theory more classical.  In the limit of the pure gradient flow the
peak of the susceptibility would lie directly on the tree-level value
of $\kappa_c^{(\text{cl})} = 0.24$ according to (\ref{eq:class-crit-kappa}).
The stepwise UV
regularized theory shows its critical behaviour in ranges  
of $\kappa$ where the full quantum theory 
($\latcutoff = \latcutoffmax$) lives in the symmetric phase.

Moreover, in the situations studied here the peak height shrinks when
lowering the cutoff below $\latcutoff = 4$. This suggests that the momentum
fluctuations below this characteristic momentum scale given by
$\latcutoff = 4$ are relevant for the physics observed. Removing these momenta 
with a lower cutoff therefore modifies the theory. In turn, the momentum
fluctuations above this scale are physically irrelevant.

\subsection{Colored noise: complete blocking at tree level}
\label{sec:CN-complete-blocking-tree-level}
\begin{figure}[t]
\includegraphics[width=\columnwidth]{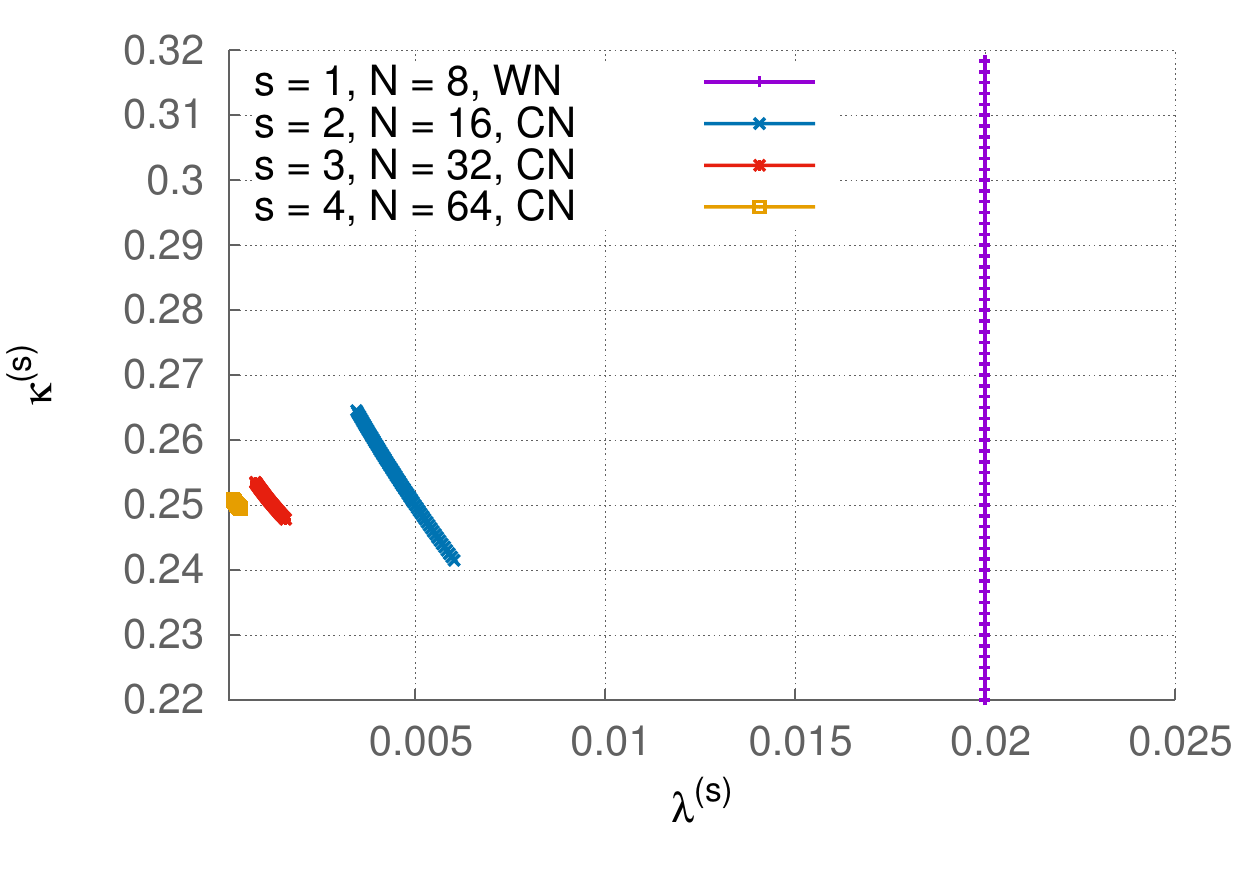}
\caption{Transformed values for $\kappa^{(s)}$ and $\lambda^{(s)}$ used in 
  simulations. The violet line represents the parameter range used for
  the $s=1$ white noise simulation.}
\label{fig:kappa-lambda-vals}
\end{figure}

In this section we relate the effects of colored 
noise to the real space renormalization group. 
Following the procedure outlined in Sec.~\ref{sec:RG-study}
we set up the white noise reference simulation $(s = 1, 
\latcutoff = \latcutoffmax)$ on an $8 \times 8$ lattice. 
The parameters are chosen 
to be $\lambda=\lambda^{(1)}=0.02$ and $0.22 \leq \kappa = \kappa^{(1)} \leq 0.32$
with lattice spacing $a=a^{(1)}$. This determines the 
full quantum theory we want to compare our colored noise results
with.
We proceed by setting $s = 2, 4, 8$ and perform 
colored noise simulations with finer lattice spacings $a^{(s)}$ 
on $N^{(s=2)}\times N^{(s=2)} = 16 \times 16$,
$N^{(s=4)} \times N^{(s=4)}= 32 \times 32$ 
and $N^{(s=8)} \times N^{(s=8)} = 64 \times 64\,$ lattices
at the corresponding cutoffs $s^{-1}\, \latcutoffmax(a^{(s)}, N^{(s)})$.
Table \ref{table:RG-data}
summarizes the lattice spacings $a^{(s)}$ and cutoffs $\latcutoff$ 
used in the simulations.
Accordingly, the transformed parameters are determined from 
(\ref{eq:tree-level-trafo-1}).
The resulting parameters 
$(\kappa^{(s)},\,\lambda^{(s)})$ are plotted in 
Fig.~\ref{fig:kappa-lambda-vals}. 
\begin{table}[t]
\begin{minipage}{\columnwidth}
\def\arraystretch{2}
\begin{tabular}{c|c|c|c}
 $N$ & lattice spacing & $\latcutoff $ & $\sqrt{d}\, (2 \pi / a^{\tiny{(s)}} N)\, \latcutoff$ \\
 \hline
 $8$ & $a^{(s=1)} = a$ & $\latcutoffmax = 4$ & $\sqrt{2}\frac{\pi}{a}$ \\
 \hline
 $16$ & $a^{(s=2)}$ = $a/2$ & $\latcutoffmax/2 = 4$ & $\sqrt{8}\frac{\pi}{a}$ \\
 \hline
 $32$ & $a^{(s=4)}$ = $a/4$ & $\latcutoffmax/4 = 4$ & $\sqrt{32}\frac{\pi}{a}$ \\
\hline
 $64$ & $a^{(s=8)}$ = $a/8$ & $\latcutoffmax/8 = 4$ & $\sqrt{128}\frac{\pi}{a}$ \\
\end{tabular}
\caption{Lattice scales for the RG transformation (\ref{eq:a-ren}) in two dimensions.
         The last column shows the lattice cutoff with reinstated lattice spacing.}
\label{table:RG-data}
\end{minipage} 
\end{table}

\begin{figure}[t]
\includegraphics[width=\columnwidth]{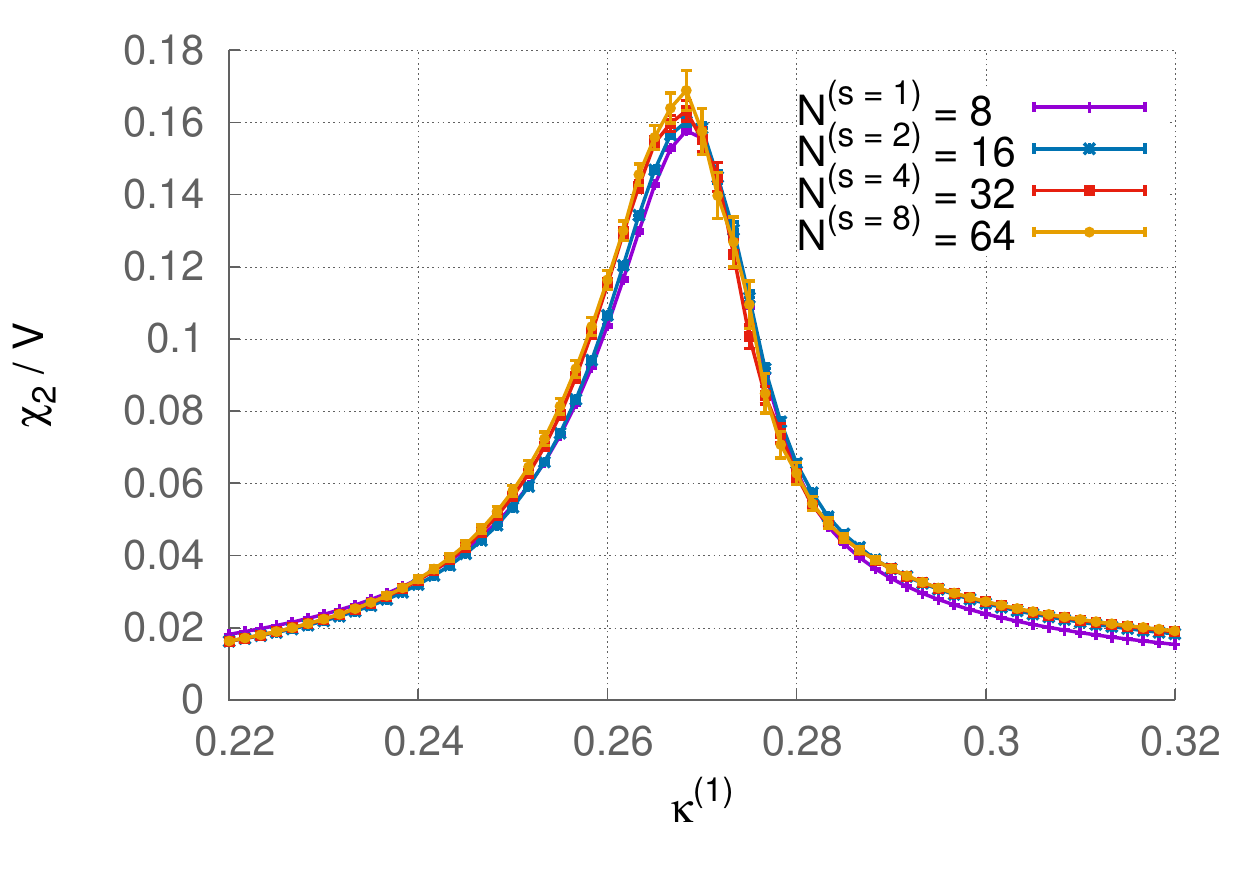}
\caption{The connected two-point susceptibility rescaled by the volume
  as a function of $\kappa^{(1)}$ for different cutoff scales $\latcutoff$.
  The violet curve represents the white
  noise reference result on the smallest and coarsest lattice. All
  of the remaining curves for $s=2,4,8$ were produced by 
  means of colored noise with couplings adjusted according to the 
  RG transformation (\ref{eq:tree-level-trafo-1}).}
\label{fig:susc-kappa-RG-check}
\end{figure}

\begin{figure}
\includegraphics[width=\columnwidth]{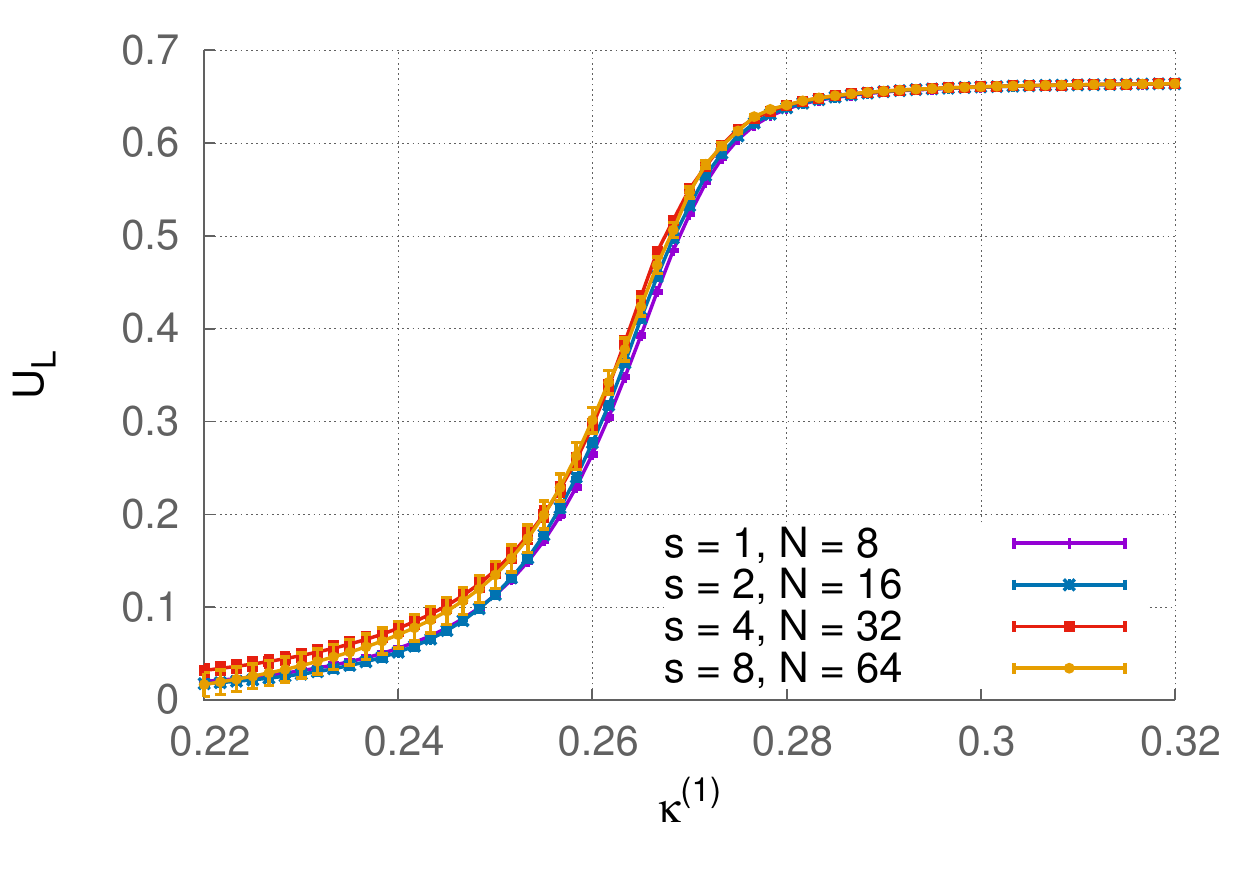}
\caption{The Binder cumulant $U_L$ as a function of $\kappa^{(1)}$ 
         for different cutoff scales $\latcutoff$. The violet curve
         represents the white noise result.}
\label{fig:Binder-kappa-RG-check}
\end{figure}

\begin{figure}
  \includegraphics[width=\columnwidth]{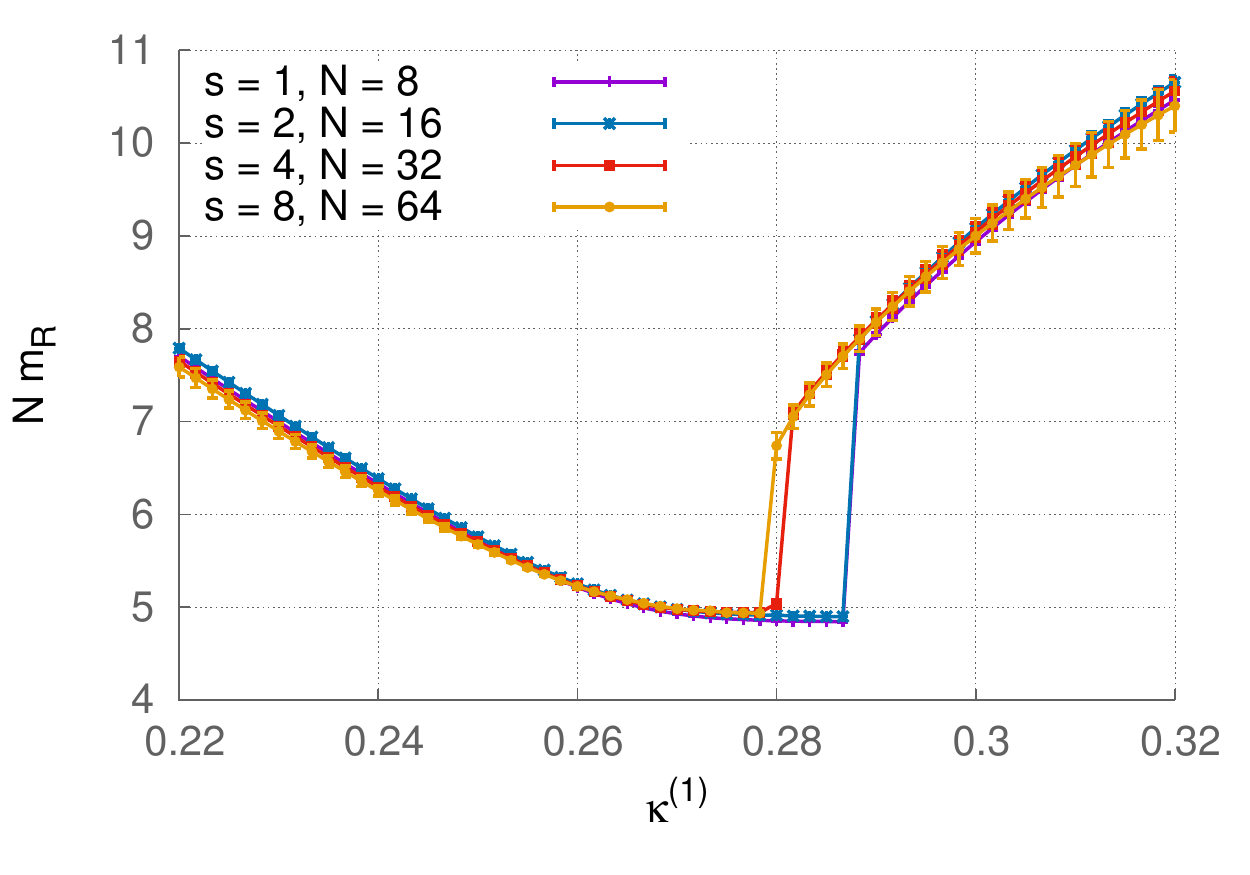}
  \caption{The rescaled renormalized mass $Nm_R$ as a function of $\kappa^{(1)}$ 
           for different cutoff scales $\latcutoff$. The violet curve
           represents the white noise result.}
\label{fig:mR-kappa-RG-check}
\end{figure}

Fig.~\ref{fig:susc-kappa-RG-check} shows the Gaussian fluctuations 
by means of the volume rescaled connected susceptibility
$\chi_2 / V$ plotted as a 
function of the untransformed hopping parameter 
$\kappa^{(1)}$ at fixed $\lambda^{(1)} = 0.02$.
That is we consider
$(\chi_2 /V)[\kappa^{(s)}(\kappa^{(1)}, \lambda^{(1)}),
\lambda^{(s)}(\kappa^{(1)}, \lambda^{(1)})]$. 
Analogously, the Binder cumulant $U_L$ as well 
as the rescaled renormalized mass $Nm_R$
are presented in Fig.~\ref{fig:Binder-kappa-RG-check} 
and Fig.~\ref{fig:mR-kappa-RG-check}.
The violet curve 
represents the full quantum theory 
produced on the $8 \times 8$ lattice with white noise.
We find that the colored noise results 
(blue for $N^{(s=2)} = 16$, red for $N^{(s=4)} = 32$ and 
dark yellow for $N^{(s=8)} = 64$) are 
in close agreement with the results for the full theory.
This meets the expectations of our construction. Although
the classical lattices for different $s$ do not coincide
in size, the quantum lattices are the same due to the rescaling
(\ref{eq:a-ren}).
However, a few deviations are clearly visible.

\begin{figure}
  \includegraphics[width=\columnwidth]{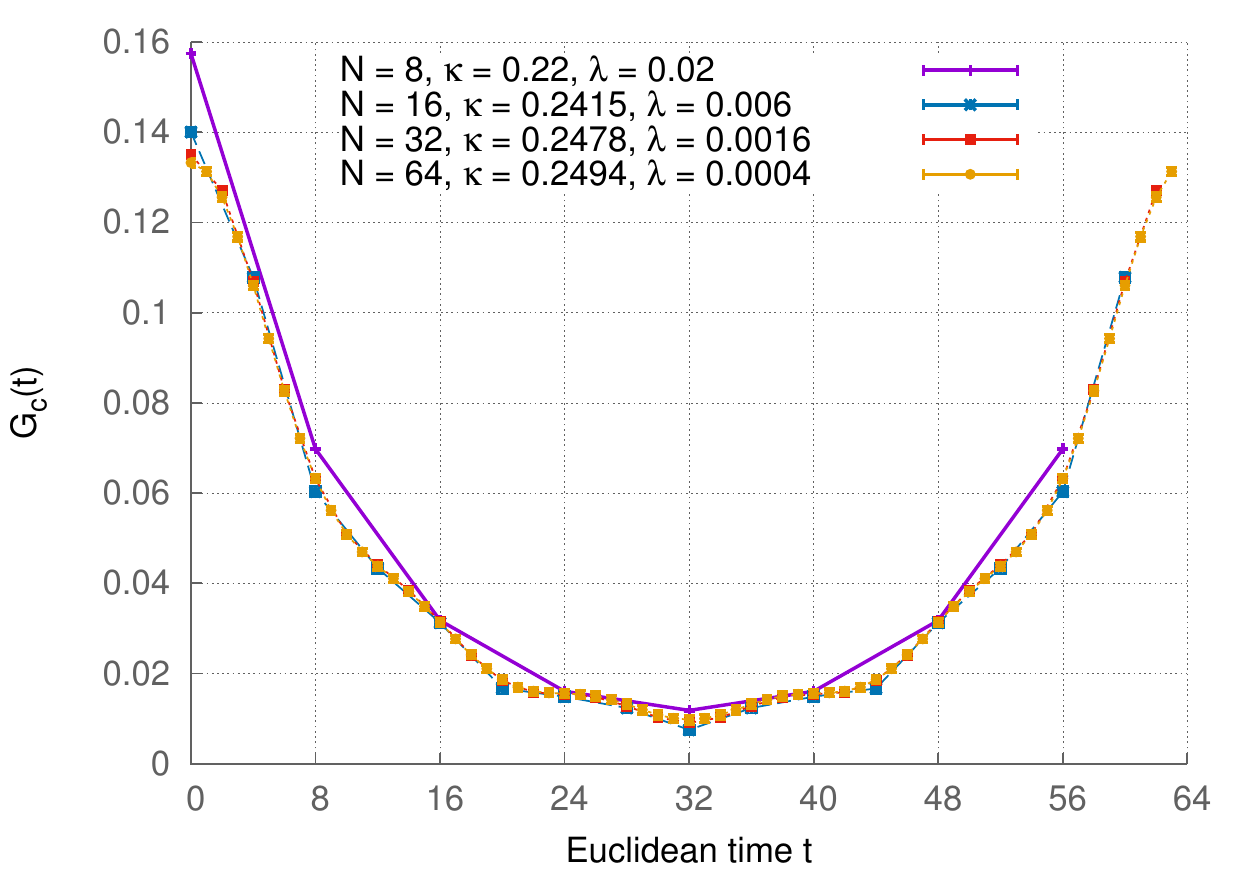}
  \caption{The correlation function of time slices $G_c(t)$ for different
    lattice sizes $N^{(s)}$ and transformed couplings 
    $(\kappa^{(s)}, \lambda^{(s)})$.
    The violet curve represents the white noise result.}
\label{fig:Correlator-RG-check}
\end{figure}

The correlation function of time slices 
for the choice $\kappa^{(1)} = 0.22, \lambda^{(1)} = 0.02$
and the transformed parameters $\kappa^{(s)}, \lambda^{(s)}$
thereof are shown in Fig.~\ref{fig:Correlator-RG-check}. 
The sharp regulator affects 
the colored noise correlators at small Euclidean 
times and causes oscillations for larger times as 
already mentioned in Sec.~\ref{sec:Numerical-Results}.
However, the results seem to agree well if 
we rescale the Euclidean time axis for the $s = 1, 2, 4$ 
cases to the time extent of the $N^{(s=8)} = 64$ lattice. 
The shape of the correlator hints also the behaviour of 
the correlation length regardless of the artifacts from  
the sharp cutoff. In agreement with (\ref{eq:block-spin-trafo}) 
we find that the correlation lengths in lattice 
units fulfill $\xi(N^{(s)}, \kappa^{(s)}, \lambda^{(s)}) 
\approx s\, \xi( N, \kappa^{(1)}, \lambda^{(1)})$. 
The correlation length increases which is 
consistent with the requirement $a \to a^{(s)} = a/s$.
This is moreover in agreement with the concept 
of the block spin transformation (here in a kind of
reverted sense)
as discussed in Sec.~\ref{sec:RG-study}. 
Accordingly, from investigating the renormalized mass in 
Fig.~\ref{fig:mR-kappa-RG-check} we find that 
$m_R(N, \kappa^{(1)}, \lambda^{(1)}) \approx s\, 
m_R(N ^{(s)}, \kappa^{(s)}, \lambda^{(s)})$. 

\begin{figure}
  \includegraphics[width=\columnwidth]{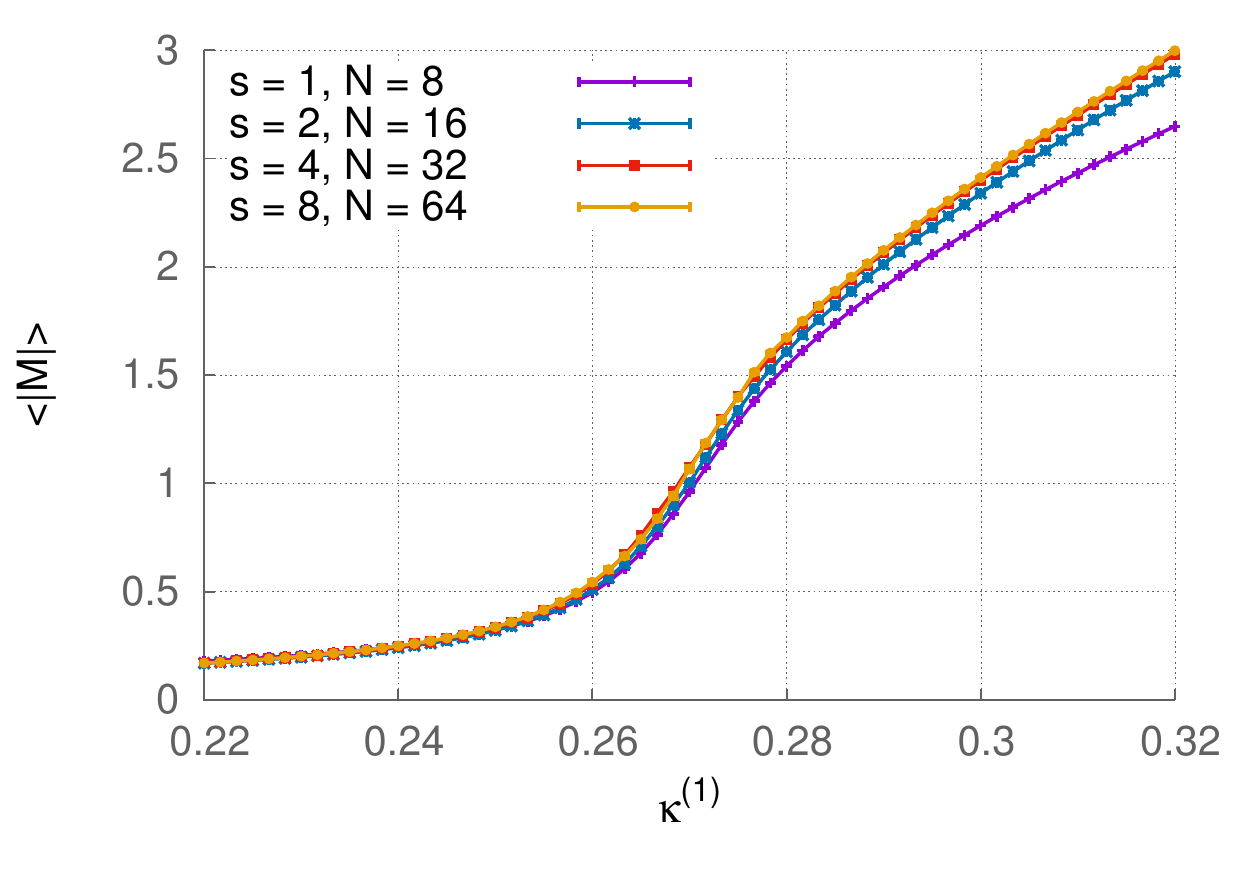}
  \caption{The magnetization as a function of $\kappa^{(1)}$ 
           for different cutoff scales $\latcutoff$. The violet
           curve represents the white noise result.}
\label{fig:magnetization-kappa-RG-check}
\end{figure}

In Fig.~\ref{fig:magnetization-kappa-RG-check} 
the order parameter $\langle |M| \rangle$ 
is plotted as a function of $\kappa^{(1)}$. 
The colored noise results seem to converge with increasing 
lattice size to the dark-yellow curve for $N^{(s=8)} = 64$. 

There are several error sources that have to be taken into account.
The deviations in the critical region are influenced by critical
slowing down, see Fig.~\ref{fig:susc-kappa-RG-check} and
Fig.~\ref{fig:mR-kappa-RG-check}. The latter poses a hard issue for a
local updating algorithm such as the Langevin equation.  Furthermore,
finite size effects are a possible error source for the mismatch of
our data in the critical region.  Those are also clearly visible in
the order parameter in Fig.~\ref{fig:magnetization-kappa-RG-check}.
Moreover, the deviations from the full quantum theory observed in our
data indicate that our compensation procedure might be incomplete.
One reason is that our naive renormalization group transformation is
based on the tree-level relations (\ref{eq:tree-level-trafo-1}). With
increasing $s$ the deviation from the tree-level relations should
increase as well due to the running of $m_0^2$ and $g_0$ affecting
$\kappa$ and $\lambda$.  A further reason is that the number of
blocking steps is limited on a finite lattice.  Here, only for the
first RG step our procedure seems to yield correct results.
\begin{figure}
\includegraphics[width=\columnwidth]{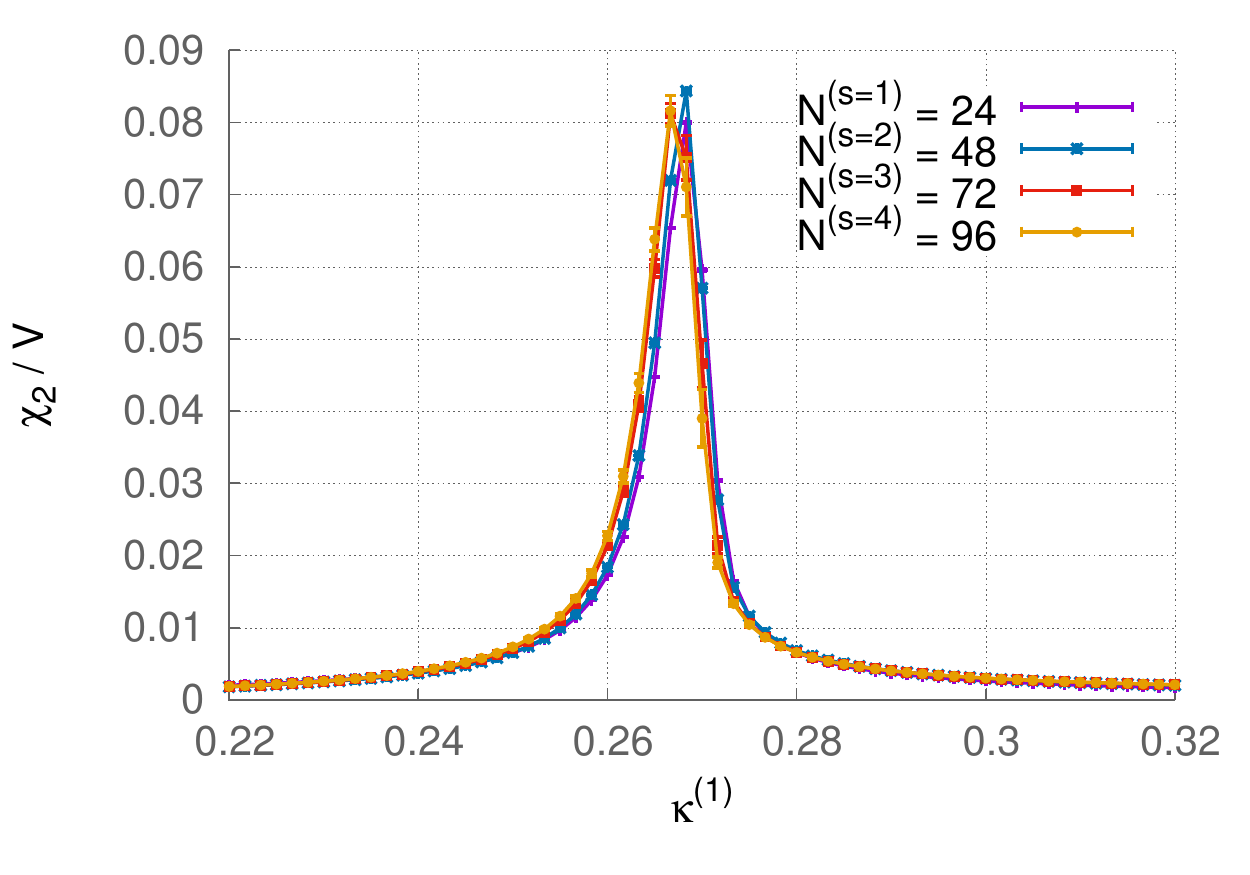}
\caption{The susceptibility rescaled by the volume
  as a function of $\kappa^{(1)}$ for different cutoffs 
  $\latcutoff$ on large lattices.
  The violet curve represents the white
  noise result.}
\label{fig:susc-kappa-RG-check-large}
\end{figure}
\begin{figure}
\includegraphics[width=\columnwidth]{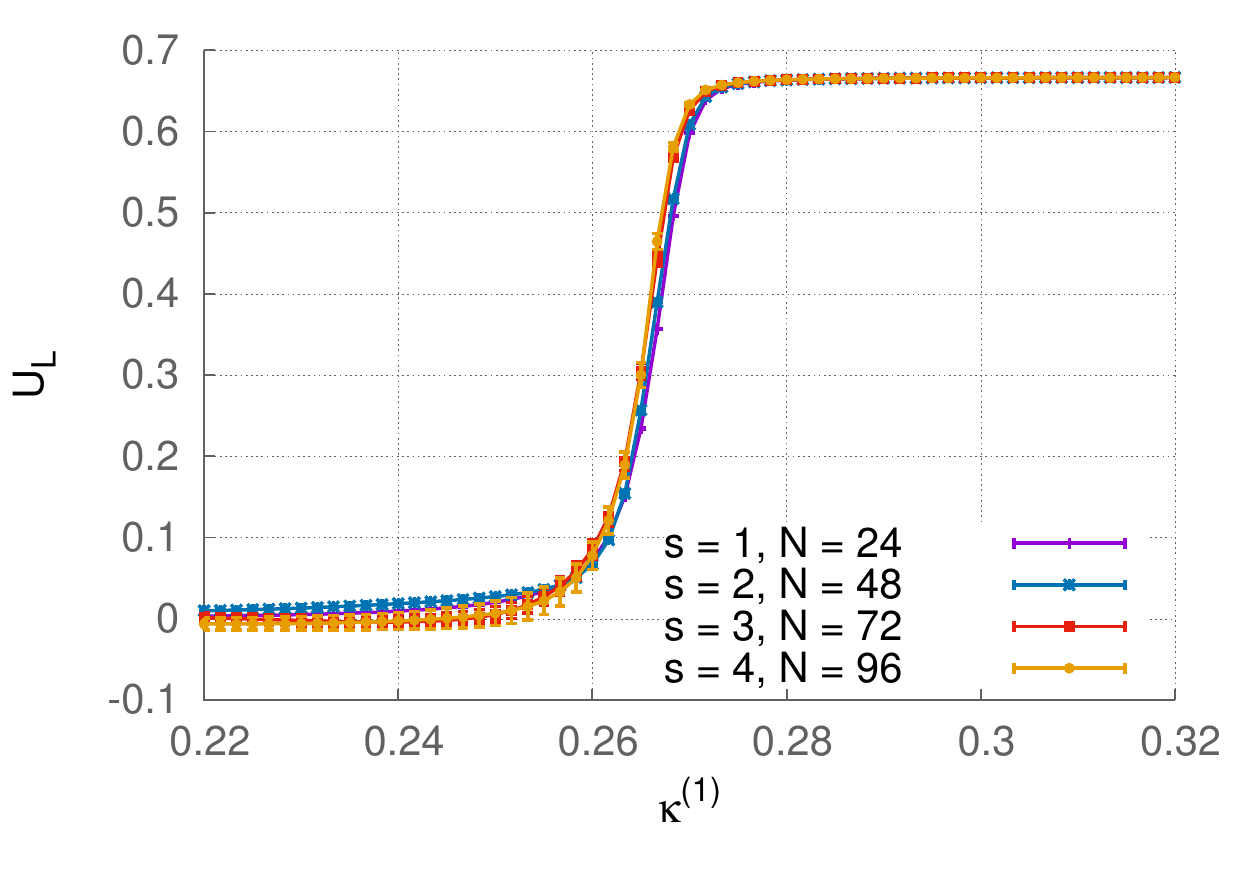}
\caption{The Binder cumulant as a function of $\kappa^{(1)}$
         for different cutoff scales $\latcutoff$.
         The violet curve
         represents the white noise result.}
\label{fig:Binder-kappa-RG-check-large}
\end{figure}
\begin{figure}
\includegraphics[width=\columnwidth]{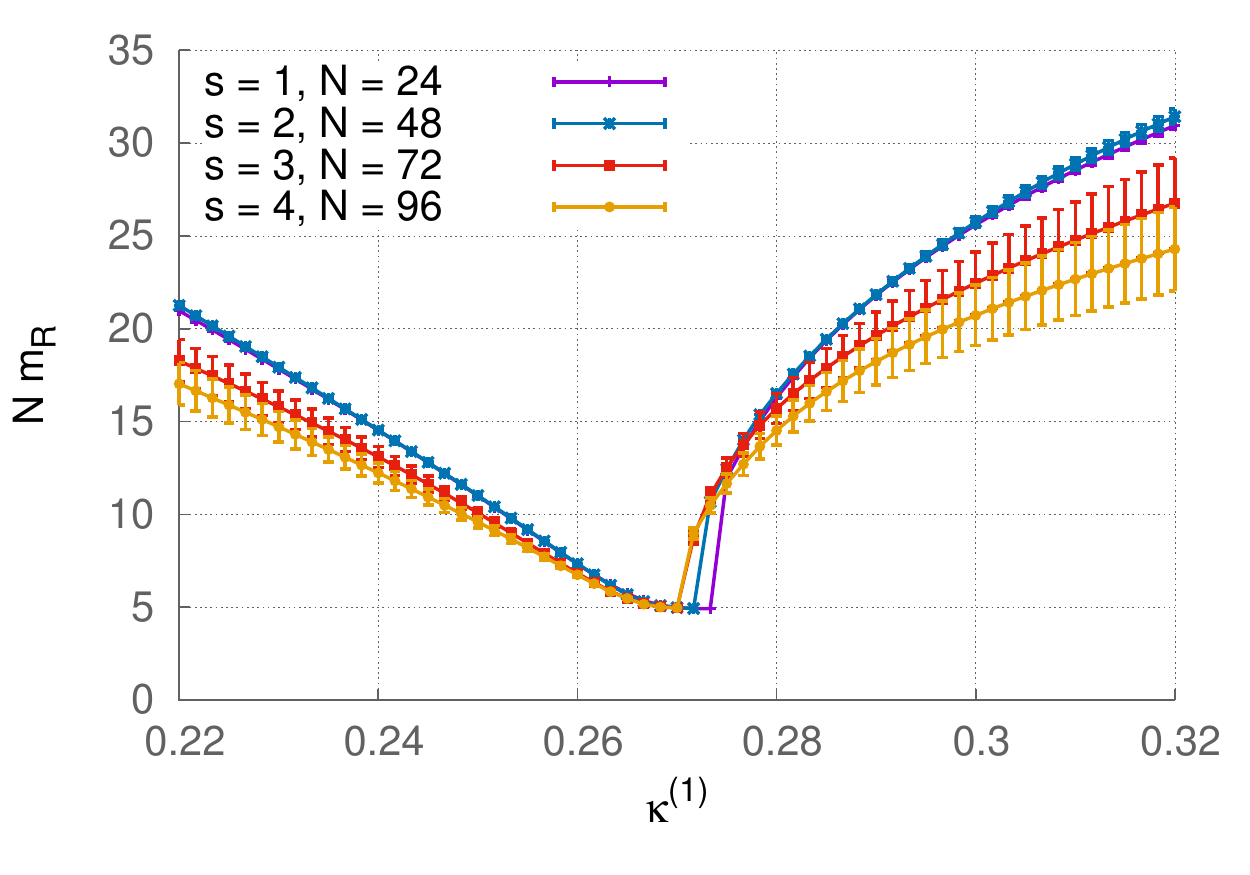}
\caption{The rescaled renormalized mass $N m_R$ as a function of
         $\kappa^{(1)}$ for different
         cutoff scales $\latcutoff$. 
         The violet curve represents the white noise result.}
\label{fig:mR-kappa-RG-check-large}
\end{figure}
\begin{figure}
\includegraphics[width=\columnwidth]{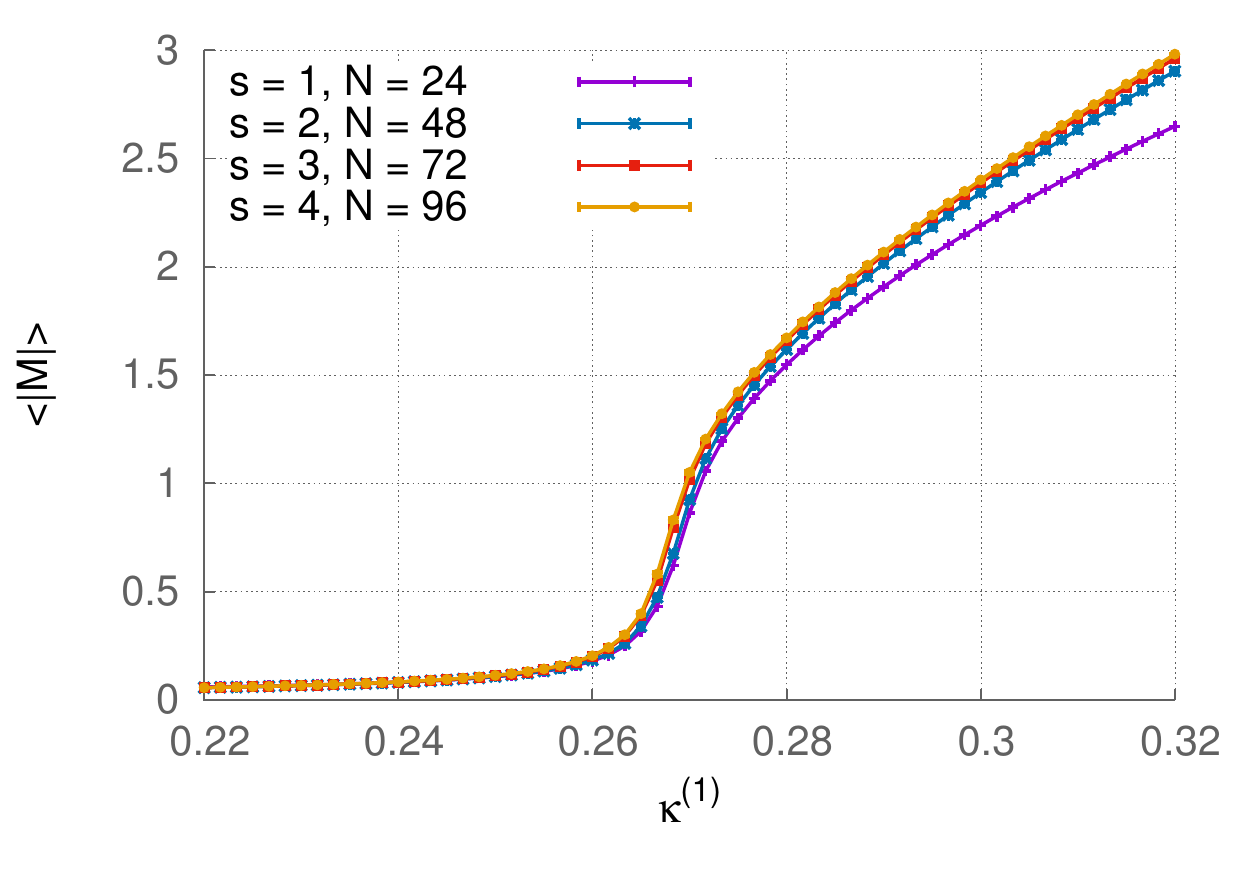}
\caption{The magnetization as a function of $\kappa^{(1)}$
         for different cutoff scales $\latcutoff$. The violet curve
         represents the white noise result.}
\label{fig:magnetization-kappa-RG-check-large}
\end{figure}

To cope with the finite size effects, 
we repeat our analysis considering larger lattices. 
We proceed analogously as before
but in contrast to the discussion above 
we carry out the $s=1$ simulation using 
white noise on a larger $24 \times 24$ lattice 
and set the scale factors for the colored
noise simulations to $s=2,3,4$. The parameter set 
for the full theory is again given by $0 \leq \kappa^{(1)} \leq 0.22$ and 
$\lambda^{(1)} = 0.02$. 
Note, that the lattice sizes for the simulations with 
colored noise at half, third and quarter the maximum cutoff 
are now $N^{(s=2)}\times N^{(s=2)} = 48 \times 48$,  
$N^{(s=3)}\times N^{(s=3)} = 72 \times 72$ and 
$N^{(s=4)}\times N^{(s=4)} = 96 \times 96$. 

From the susceptibility shown in 
Fig.~\ref{fig:susc-kappa-RG-check-large} and 
the Binder cumulant in Fig.~\ref{fig:Binder-kappa-RG-check-large}
we find that by halving the lattice spacing 
the results from the $N^{(s=1)} = 24$ and the $N^{(s=2)} = 48$ 
simulations are in good agreement.
In the critical regime the results for 
larger lattices however deviate  
from the white noise reference result. 

The renormalized mass in Fig.~\ref{fig:mR-kappa-RG-check-large} 
shows that the $N^{(s=1)}=24$ and the $N^{(s=2)}=48$ data 
agree well over the whole range in the hopping parameter 
in spite of the deviation caused by the (remaining) 
finite size effect around the critical point. 
The larger lattices however, indicate that the masses 
differ from the white noise result. 
We remark that the simulations are plagued by a 
bad signal-to-noise ratio, visible in the correlator
for parameters sufficiently
far in the symmetric or broken phase respectively.
The magnetization for the larger lattices in 
Fig.~\ref{fig:magnetization-kappa-RG-check-large}  
shows an analogous behaviour as observed on the small lattices.
We conclude, that except for the renormalized mass our renormalization
procedure gives the same result on large and small lattices.

The results in this section have been produced from
$10^6$ measurements of time slice configurations
for each lattice size. Between two
measurements we have performed 100 subsequent Langevin update sweeps
corresponding to a Langevin time $\tau = 1$
without measurement to reduce the autocorrelation of the
observables. After a standard data blocking check we find that
this is not enough, especially in the case of the 
large and fine lattices.
The data is more severely correlated. For example
at $\kappa^{(1)} = 0.26$ for an $N=96$ lattice a block must have a
minimum length of 5000 which we have used for a standard blocked Jackknife error
analysis. 

\newpage

\section{Cooling with colored noise - applications}
\label{sec:CNC}

In the previous sections we have shown that the cutoff can be
decreased stepwise without changing the physics content of the theory
if the cutoff is still sufficiently large. The complementary
Wilsonian picture is that of integrating out degrees of freedom: with
colored noise the path integral measure $P_{\Lambda}[\phi]$ only
involves modes with $p^2 \leq \Lambda^2$. Accordingly, let us consider
the colored stochastic process  \eq{eq:kerneled-LE}, \eq{eq:kerneled-FPE} with 
$S\to S_{{\rm eff},\Lambda}$, where the latter already contains the
quantum effects of fields with $p^2 > \Lambda^2$, 
\begin{align}  
e^{ -S_{{\rm eff},\Lambda}[\phi]} = \int d\phi_{p^2>\Lambda^2} \, e^{ -S[\phi]}\,,
\end{align}
This leads us to 
\begin{subequations}\label{eq:WilStoch}
\begin{equation}
  \frac{\partial \phi(x,\tau)}{\partial \tau} = - 
  \frac{\delta S_{{\rm eff},\Lambda}}{\delta \phi(x, \tau)} + r_\Lambda(\Delta_x) \,\eta(x,\tau)\, ,
  \label{eq:kerneled-LEWil}
\end{equation}
with 
\begin{align}
  \frac{\partial P_{{\rm eff},\Lambda}}{\partial \tau}  =
  \int \mathrm{d}^dx \frac{\delta}{\delta \phi_x} \left(\frac{\delta S_{{\rm eff},\Lambda}}{\delta \phi_x} 
 + \,r_\Lambda^2(\Delta_x)\, \frac{\delta}{\delta \phi_x}\right) P_{{\rm eff},\Lambda}\,.
  \label{eq:kerneled-FPEWil}
\end{align}
\end{subequations} 
The stochastic process \eq{eq:WilStoch} gives the full 
correlation functions for momenta $p^2\leq \Lambda^2$. The related generating functional 
is that of the full theory  
\begin{align}\label{eq:WilsonPI} 
Z= \int d\phi_{p^2\leq \Lambda^2} \, e^{ -S_{{\rm eff},\Lambda}[\phi]}= \int d\phi\, e^{-S[\phi]}\,,
\end{align}
with the classical action $S[\phi]$ used in the original Langevin
evolution. The Wilsonian effective action $S_{{\rm eff},\Lambda}$ can
be also understood in terms of an improved or perfect lattice action
if an additional block spinning transformation is applied. 
\begin{figure}[t]
    \includegraphics[width=\columnwidth]{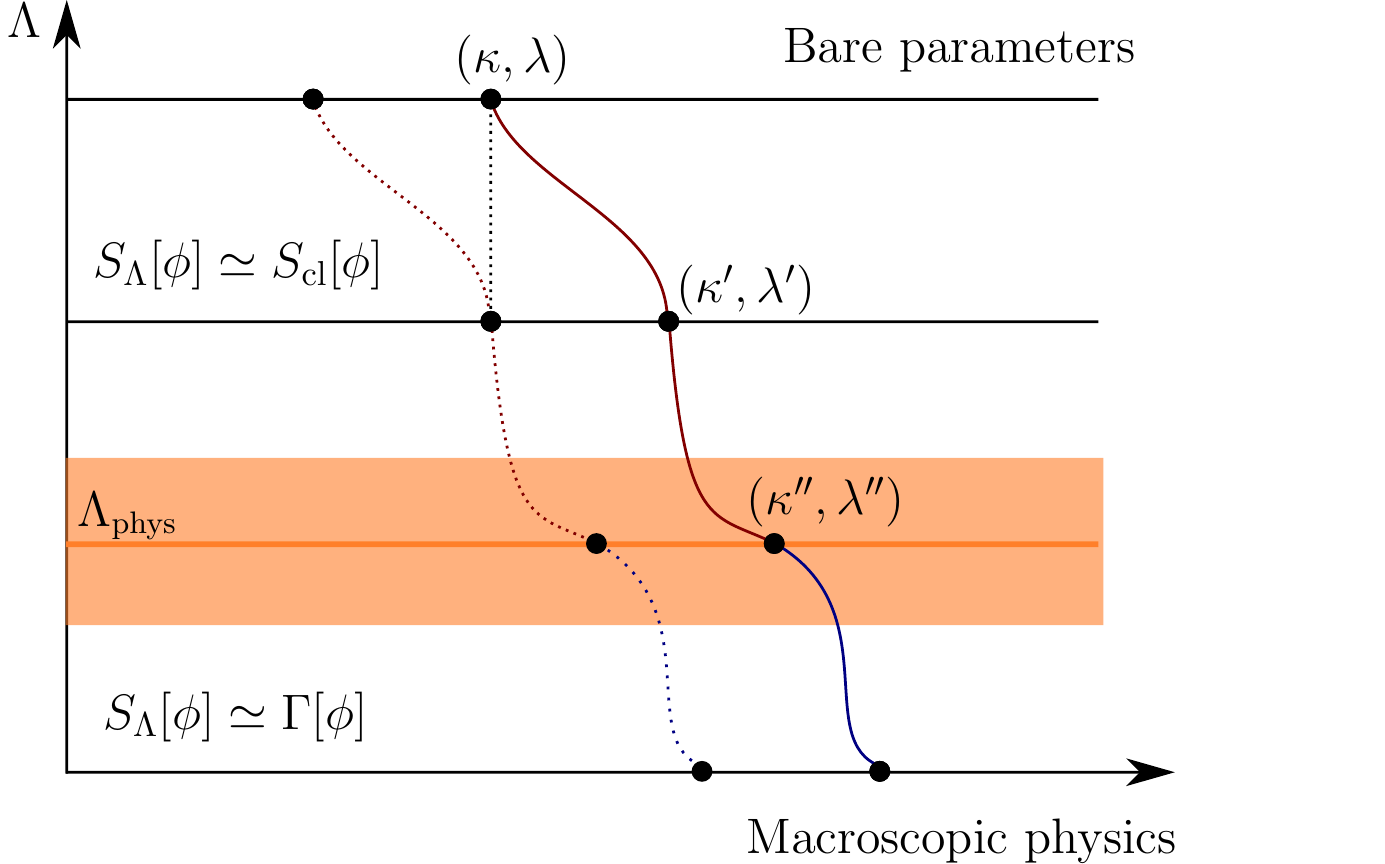}
    \caption{Sketch of colored noise cooling. Each point on a
      horizontal line represents a pair of couplings
      $(\kappa, \lambda)$.  The vertical axis denotes the cutoff scale
      $\Lambda$.  The upper horizontal line depicts the white noise
      limit, and the curves constitute lines of constant physics. The
      couplings of UV-irrelevant operators are also generated during
      the RG flow. The corresponding axes are perpendicular to the
      plane in the plot, and are dropped. \newline For
      $\Lambda \gg \Lambda_{\text{\tiny{phys}}}$ we have
      $S_{\rm eff,\Lambda} = S[\kappa_{\Lambda}, \lambda_{\Lambda}]$,
      see \eq{eq:SUV}, depicted by the dark-red straight and dashed
      lines. Changing $\Lambda$ at fixed couplings effectively changes
      the physics content, see vertical black dashed line and also the
      observables in Fig.~\ref{fig:obs-CN-shells-N-32}. \newline The
      scale $\Lambda_{\text{\tiny{phys}}}$ and the orange band denote
      the bound below which the action in the colored noise simulation
      must be described by the full quantum effective action.     }
\label{fig:CNC_concept_sketch}
\end{figure}

In summary the following picture emerges: if the ultraviolet cutoff is
asymptotically large, lowering the cutoff only changes the bare
couplings $\kappa_\Lambda, \lambda_\Lambda$ in the classical lattice
action to accommodate the RG-running of the theory. Effectively this
defines a scale $ \Lambda_{\text{\tiny{phys}}}$, and for
$\Lambda\gg \Lambda_{\text{\tiny{phys}}}$ the above statement
holds. Higher order operators are suppressed by UV power counting with
powers of $ \Lambda_{\text{\tiny{phys}}}/\Lambda$ and can be safely
dropped. This leads us to
\begin{align}\label{eq:SUV}
  S_{{\rm eff},\Lambda\gg \Lambda_{\text{\tiny{phys}}}}\simeq 
  S[\kappa_{\Lambda}, \lambda_{\Lambda}]\,, 
\end{align}
see also Fig.~\ref{fig:CNC_concept_sketch}.  In turn, for small
cutoffs, $\Lambda \lesssim \Lambda_{\text{\tiny{phys}}}$, physical
fluctuations are removed from the lattice. Then, RG-transformations of
the bare parameters in the classical lattice action do not suffice to
keep the physics constant. Still, the latter can be achieved by RG
transformations leading to improved or perfect actions,
\begin{align}\label{eq:Seff}
  S_{{\rm eff},\Lambda\lesssim \Lambda_{\text{\tiny{phys}}}}\neq  S[\kappa_{\Lambda}, \lambda_{\Lambda}]\,.
\end{align}
This idea is depicted in Fig.~\ref{fig:CNC_concept_sketch}. It also
suggests a systematic way to use the Wilsonian picture in terms of the
(lattice) functional renormalization group (FRG) for improved lattice
computations as well as for effectively determining
$\Lambda_{\text{\tiny{phys}}}$. In contrast to the previous section we shall
consider RG transformations beyond tree-level on lattices of fixed
size and lattice spacing. These transformations are encoded in the
flow equation for the Wilsonian effective action
$S_{\rm eff,\Lambda}$. In the present work we concentrate on the sharp
cutoff, a more general analysis also including smooth cutoffs will be
presented elsewhere. 

For the sharp cutoff $S_{\rm eff,\Lambda}$
satisfies the Wegner-Houghton equation \cite{Wegner:1972ih}. For the
sake of computational convenience we formulate it for the 1PI
effective action, the Legendre transform of $S_{\rm eff,\Lambda}$ (where the
cutoff term is subtracted
\cite{Berges:2000ew,Polonyi:2001se,Pawlowski:2005xe,Delamotte:2007pf,%
  Rosten:2010vm,2012RvMP...84..299M}),
\begin{align}
\label{eq:WH-flow} 
  \partial_\Lambda\Gamma_\Lambda[\bar\phi] = \012 \Tr  \langle 
  \phi(p)\phi(-p)\rangle_c\,\Lambda^2 \partial_\Lambda 
  \left(\0{1}{r_\Lambda(p^2)}-1\right)\,,
\end{align}
where the subscript ${}_c$ stands for the connected part of the
two-point function similarly as introduced in Sec.~\ref{sec:model}.
The trace $\Tr$ stands for the sum over momenta in the Brillouin zone,
and $\bar \phi=\langle \phi\rangle$. In the continuum limit it turns
into the standard momentum integration $\int d^d p/(2 \pi)^d$. In
\eq{eq:WH-flow} a suitable smoothing of the sharp cutoff is assumed
and mandatory on the lattice. The propagator is the inverse of the
second derivative of $\Gamma_\Lambda$ w.r.t.\ the fields,
$\Gamma^{(2)}_\Lambda= \delta^2 /\Gamma_\Lambda\delta\phi^2$, and
hence \eq{eq:WH-flow} is a closed equation for $\Gamma_\Lambda$. In
the continuum it takes the simple form
\begin{align}
\label{eq:WH-flowExpl} 
 \Lambda \partial_\Lambda\Gamma_\Lambda[\bar\phi] = \012 \Tr  
\log \left( \Gamma^{(2)}_\Lambda[\bar\phi]\right)(p^2=\Lambda^2)\,. 
\end{align}
In the UV regime with $\Lambda\gg \Lambda_{\text{\tiny{phys}}}$ the effective
action is given by the classical action, see \eq{eq:SUV}.  Then
the flow equation is a closed equation for $\kappa(t)$ and
$\lambda(t)$ with $t=\log \Lambda/\Lambda_{\rm UV}$, where
$\Lambda_{\rm UV}$ is a normalization scale, typically the initial UV
scale.  In the present case this is the maximal momentum on the
classical lattice. Another convenient definition originates in
$\Lambda/\Lambda_{\rm UV}=s_\Lambda/s_{\Lambda_{\rm UV}}$. Since $s_\Lambda$
is already dimensionless we drop the normalization and use 
\begin{align}\label{eq:flowscale}
t= \log s_\Lambda\,. 
\end{align}
For $\Lambda\lesssim \Lambda_{\text{\tiny{phys}}}$ the simple closed
flows for $\kappa(t)$ and $\lambda(t)$ do not hold anymore, and the
higher operators will be important. By comparing the full flows with
the simplified ones the physical scale $\Lambda_{\text{\tiny{phys}}}$
can be defined as the scale below which the correlation functions
computed from the stochastic processes with either $S$ and
$S_{\rm eff}$ show significant deviations.  Note that this procedure
is less costly than the blocking procedure which involves decreasing
the lattice spacing while simultaneously increasing the number of
lattice points.

A full analysis of this framework goes beyond the scope of the present
work. Here we want to provide some first simple practical computations
that also give indications of the precision needed in fully
quantitative analyses. To that end we approximate the lattice RG
transformations by the functional RG flow equations in the continuum
theory \eq{eq:WH-flowExpl}. In the asymptotic UV regime with
$\Lambda\gg \Lambda_{\text{\tiny{phys}}}$ the effective action $\Gamma_\Lambda$
is given by the classical action, to wit 
\begin{align}\label{eq:GaSclass}
\Gamma_\Lambda[\phi] \simeq \int d^d x \left\{\012 \phi  \left(-\partial^2 +m^2\right) \phi  
+ \0{g }{4!}   \phi^4\right\}\,, 
\end{align}
for $m^2\geq 0$. Taking two and four field derivatives at $\phi=0$ and $p=0$ we are  
lead to the flows
\begin{align}\label{eq:flows}
\partial_\Lambda m = F_m(m,g)\,,\qquad    \partial_\Lambda g = F_{g}(m,g)\,,
\end{align}
for the mass and the coupling. The latter can be converted to flows for the 
dimensionless lattice parameters $\kappa, \lambda$
using the relations (\ref{eq:kappa-lambda-params}).
Note that the flow of $g$ 
runs like $1/\Lambda^2$ for large
$\Lambda$ up to logarithmic corrections. Hence, at leading order only
$m^2$ and $\kappa$ run logarithmically proportional to $\log(\Lambda)$
and $1/\log(\Lambda)$ for large $\Lambda$.  Accordingly, for
$\Lambda\to 1/2 \Lambda$ the mass squared shifts by an amount proportional to 
$\log(2)$. The prefactor can be computed from \eq{eq:flows}. 
In explicit form, the dimensionful continuum flow equations 
for the mass $m$ and the coupling $g$ read
\begin{equation}
\Lambda \, \partial_{\Lambda} m^2 = -\frac{g}{4\pi} \frac{1}{1+m^2 / \Lambda^2}\,,
\label{eq:flow-mass} 
\end{equation}
and 
\begin{equation}
  \Lambda \, \partial_{\Lambda} g = \frac{3}{4\pi}\, \frac{g^2}{\Lambda^2}\,
  \frac{1}{(1+m^2 / \Lambda^2)^2}\,.
\label{eq:flow-cpl} 
\end{equation}
The flow equations are cast into dimensionless form by multiplying
both sides with the fixed square lattice spacing $a^2$.  The
dimensionless cutoff reads $a\, \Lambda$ and the flow time is defined
by $\RGTime:=\log(a\, \Lambda)$.  Using the relations
(\ref{eq:kappa-lambda-params}) leads to the flow equations for the
lattice parameters.
\begin{figure}[t]
\includegraphics[width=\columnwidth]{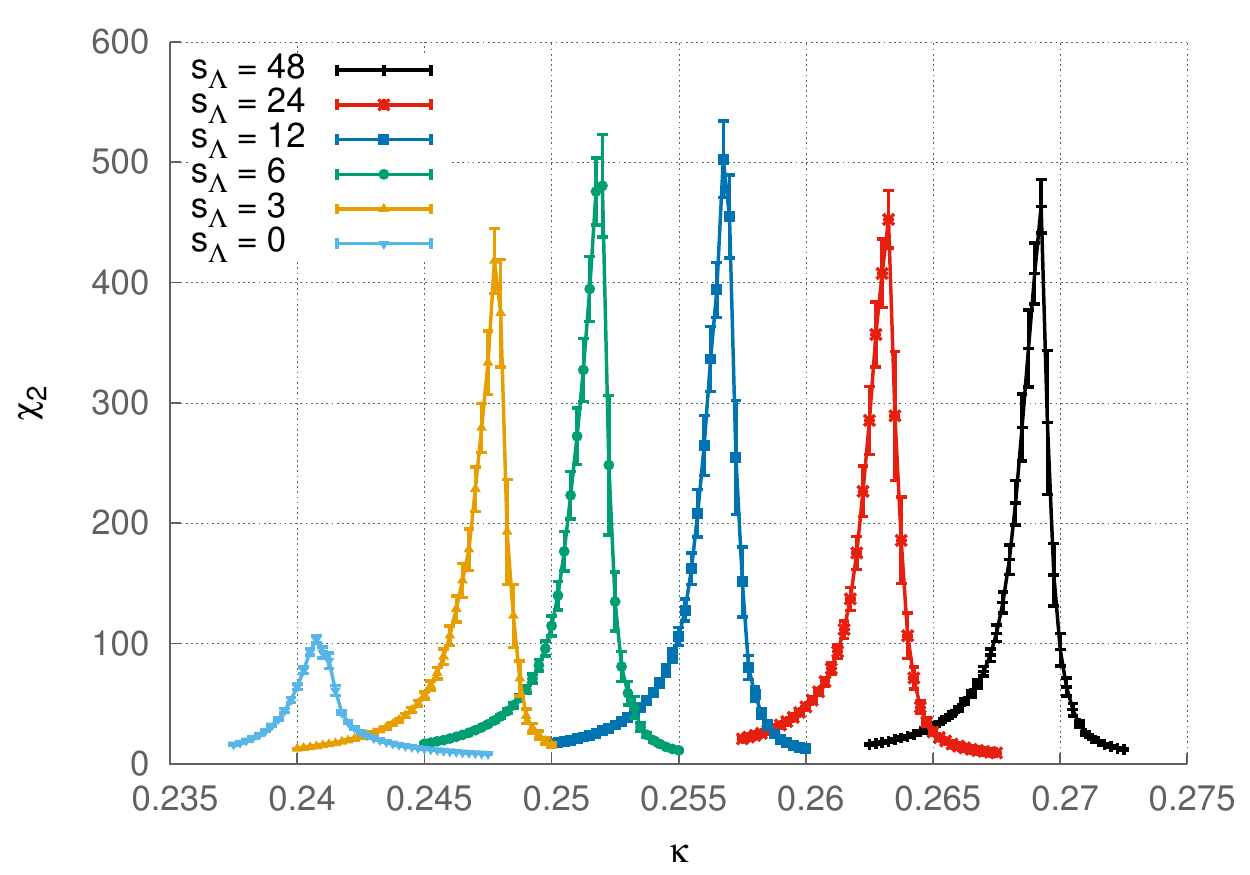}
\caption{Susceptibility $\chi_2$ as a function of the hopping 
parameter $\kappa$ for different cutoffs $\latcutoff$. The coupling 
is fixed to $\lambda = 0.02$. The results shown here 
have been computed using a $96 \times 96$ lattice. The white noise 
result corresponds to $\latcutoff = 48$.}
\label{fig:susceptibility_kappa_N_96}
\end{figure}
%
%
\begin{figure}[t]
\includegraphics[width=\columnwidth]{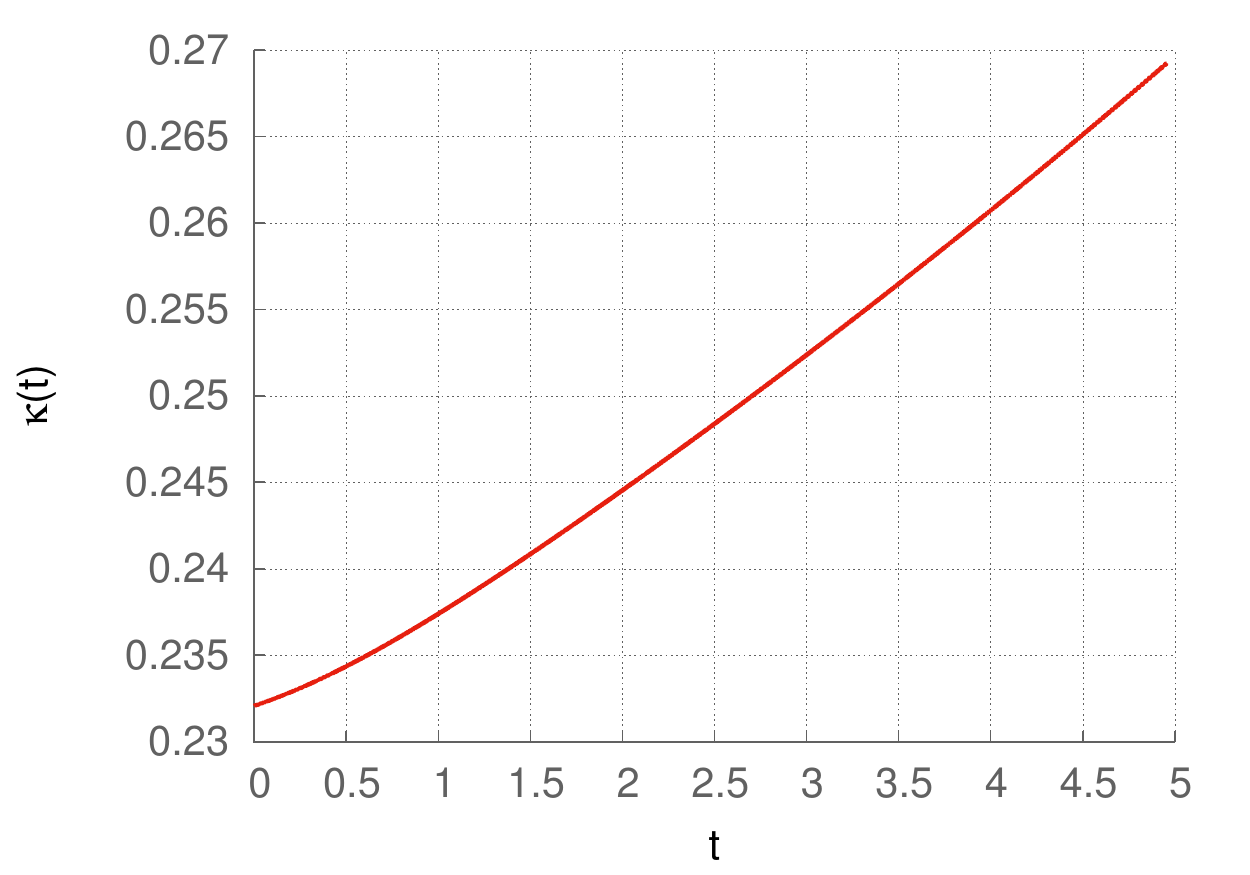}
\caption{Scale dependence of the hopping parameter $\kappa$ 
         from integrating the flow (\ref{eq:kappa-flow}).}
\label{fig:kappa-flow-from-RG}
\end{figure}
%
%
\begin{figure}[t]
\includegraphics[width=\columnwidth]{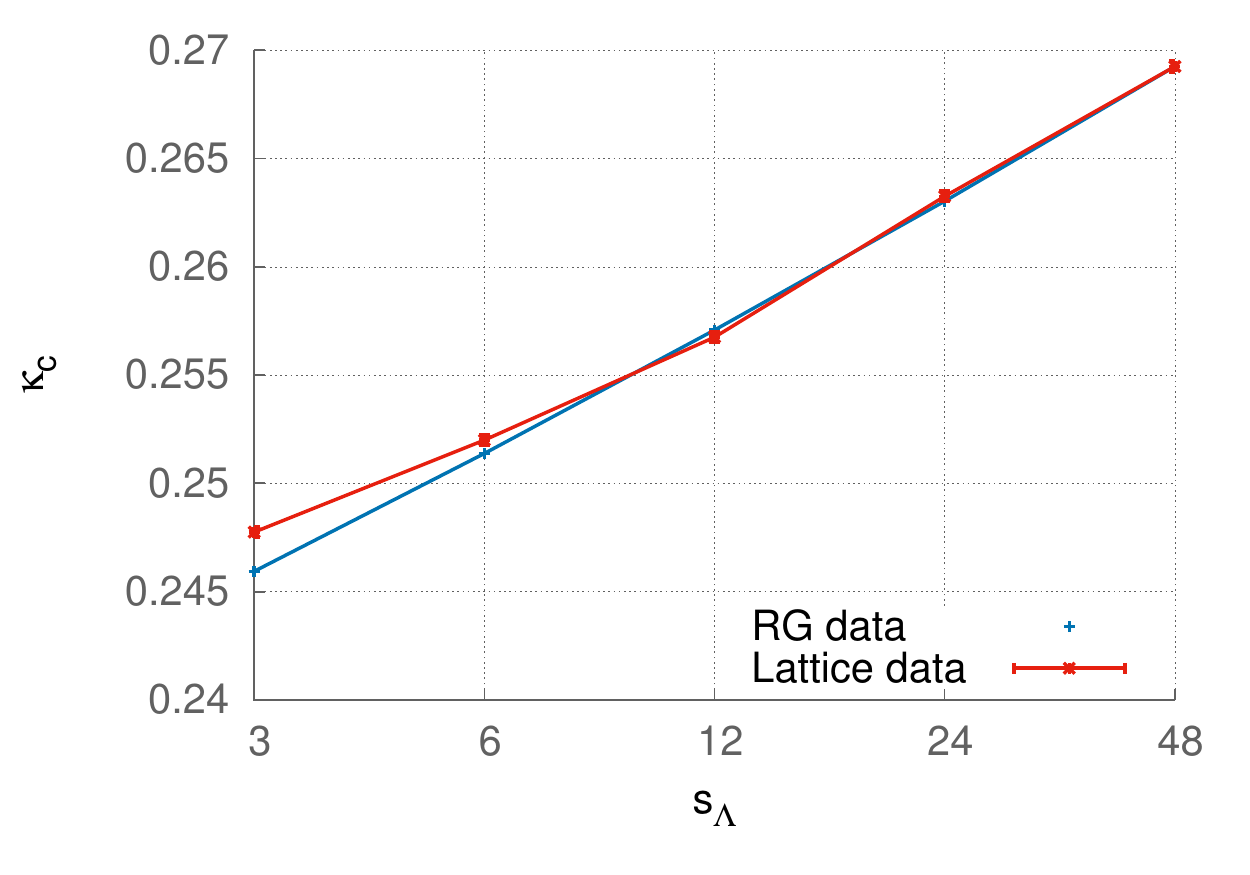}
\caption{Predictions for the critical hopping parameter $\kappa_c$ 
         from the functional RG flow equations (blue) 
         in comparison with colored noise simulations on the lattice (red).}
\label{fig:kappacrit-lattice-RG}
\end{figure}
%
%
\begin{figure}[t]
\includegraphics[width=\columnwidth]{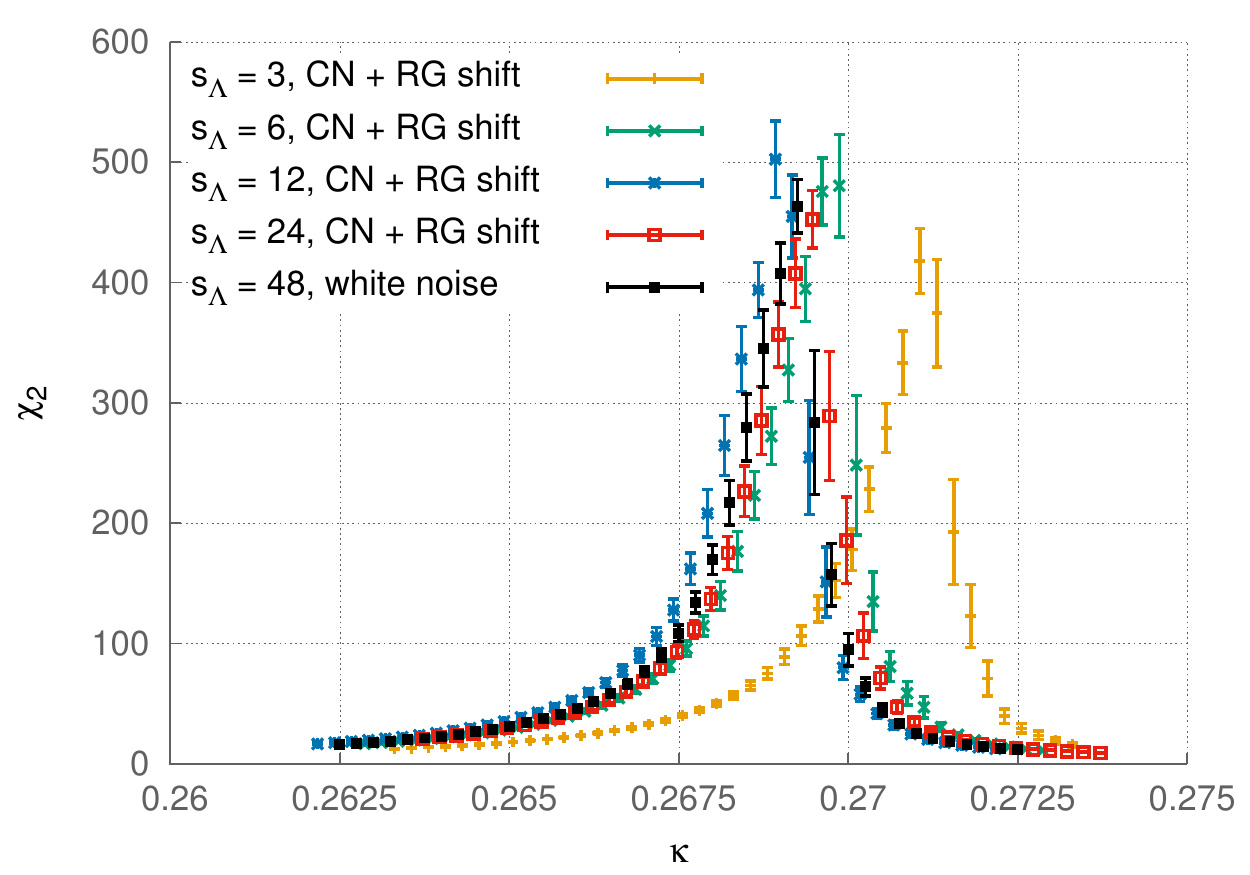}
\caption{Shift of the susceptibility peaks according to the results
for $\kappa$ and $\lambda$ from the flow equations (\ref{eq:kappa-flow}) and (\ref{eq:lambda-flow}).}
\label{fig:kappa-shift-RG-Prediction}
\end{figure}
\begin{align}
\hspace*{-1cm}\partial_{\RGTime}\, \kappa(\RGTime)\, =\, &
\frac{3}{2} \frac{\lambda(\RGTime)}{\pi} \, \kappa(\RGTime)\,
\frac{e^{2 \RGTime}}{1 + 2 \lambda(\RGTime)} \nonumber \\
&\times \frac{ \kappa(\RGTime)\left(e^{2 \RGTime}-4\right)-8
 \lambda(\RGTime)+1}{\left[\kappa(\RGTime) \left(e^{2 \RGTime}
-4\right) -2 \lambda(\RGTime)+1\right]^2}\,, 
\label{eq:kappa-flow}
\end{align}
\begin{align}
\partial_{\RGTime}\, \lambda(\RGTime)\, =\, &\frac{3}{2}\,
 \frac{\lambda(\RGTime)^2}{\pi} \, \frac{e^{2 \RGTime}}{1 + 2 \lambda(\RGTime)} \nonumber \\
&\times \frac{2\, \kappa(\RGTime) \left(e^{2 \RGTime}-4\right) -
10 \lambda(\RGTime)+5}{\left[\kappa(\RGTime)\, \left(e^{2 \RGTime}-4\right)-2 \lambda(\RGTime)+1\right]^2}\, .
\label{eq:lambda-flow}
\end{align}
For a quantitative comparison between the continuum RG and colored
noise cooling on the lattice we consider the peak positions $\kappa_c$
of the susceptibilities for different $\latcutoff$ as shown in
Fig.~\ref{fig:susceptibility_kappa_N_96}.  The data stems from
simulations on a $96 \times 96$ lattice.  The coupling $\lambda=0.02$
is fixed as in the previous sections.  For the comparison we take into
account the data for $\latcutoff = 3, 6, 12, 24, 48$.  The flow
equations (\ref{eq:kappa-flow}) and (\ref{eq:lambda-flow}) are
initialized at the maximum flow time
$\MaxRGTime = \log(a\, \Lambda_{\text{max}})$ using the parameters
$\lambda(\MaxRGTime) = 0.02$ and
$\kappa(\MaxRGTime) = \kappa_{c, \text{WN}}$.  Here,
$\kappa_{c, \text{WN}}$ indicates the critical hopping parameter
obtained from the simulation with white noise ($\latcutoffmax = 48$).
Moreover, the continuum cutoff translates into its lattice counterpart 
with $a\, \Lambda_{\text{max}} = \sqrt{2} \pi\,C$, where $C$ is a
free RG-parameter. The running hopping parameter
$\kappa(\RGTime)$ is depicted in Fig.~\ref{fig:kappa-flow-from-RG}.
To compute the remaining critical hopping parameters $\kappa_c$ from
the RG flow corresponding to lower values of $\latcutoff$ we evaluate
$\kappa(\RGTime)$ at scales $\MaxRGTime - n\,\log(2)$, where
$n = 1, 2, 3, 4$.  The red data points in
Fig.~\ref{fig:kappacrit-lattice-RG} show the critical values
$\kappa_c$ as a function of $\latcutoff$ from the lattice simulations.
The blue points denote the values of $\kappa_c$ obtained from the RG
flow (\ref{eq:kappa-flow}).

We find that at large cutoff scales the 
critical values $\kappa_c$ measured on the lattice
coincide with those calculated from the flow equations. In contrast,
for small cutoff momenta a deviation is visible. This indicates
that at lower momentum scales the stochastic process in terms of the classical 
action fails to describe the full theory. There the classical action
needs to be replaced by an effective action as mentioned above.
We conclude that for the model considered in this work colored noise cooling is 
applicable at scales between the UV and a specific IR scale. 
In the case investigated here this IR scale lies between $\latcutoffmax / 4$
and $\latcutoffmax / 8$. This is also supported by the shifted susceptibility
in Fig.~\ref{fig:kappa-shift-RG-Prediction} .
Here, the peaks have been translated by the 
difference between $\kappa_{c, \text{WN}}$ and the values of $\kappa_c$
from the RG prediction, see Fig.~\ref{fig:kappacrit-lattice-RG}. 
While the agreement between the curves is
quite good up to $\latcutoff = 12$ (blue curve), for lower cutoffs the 
results deviate from the full theory, see the green and yellow curves.

There are a few caveats to mention. Firstly, we work at fixed $\lambda$
in our lattice simulations. When lowering $\latcutoff$, 
$\lambda$ should be adjusted properly. Secondly, we approximate 
RG transformations of the lattice parameters by the continuum 
functional RG.
For a more exact comparison between the RG transformations 
and the lattice results, we need to solve the flow equations (\ref{eq:flow-mass})
and (\ref{eq:flow-cpl}) on the lattice. This however comes with a few technical
complications since the flow is only defined 
at the discrete lattice momenta. 

\section{Conclusions and outlook}
\label{sec:conclusions}
In this work we have investigated lattice theories with Stochastic
Quantization with UV-regularized colored noise.  Cooling the Langevin
evolution by removing field configurations in the UV may be a
promising candidate to optimize lattice simulations of systems with a
clear scale separation between the relevant physics and the asymptotic
UV regime.  There are two possible interpretations of our method. The
first is that the colored noise LE can be applied in the traditional
sense of smoothing out UV-fluctuations. The alternative interpretation
is to sample smooth configurations directly from the UV-regularized
Langevin evolution. 

Here we have exploited the latter interpretation which also
can be connected directly to the renormalization group.  The scale of
the smooth fields is set by using an external cutoff parameter
$\Lambda$. By varying $\Lambda$ the colored noise Langevin
equation interpolates between the full quantum theory accessible in a
standard white noise simulation and the classical theory. 

Our approach has been put to work within a real scalar field theory in
two dimensions using a sharp momentum cutoff. We have shown, that for
sufficiently large cutoff scales $\Lambda$ no relevant physics is
cut off. In Sec.~\ref{sec:CNC} we have analyzed the viability of the
colored noise cooling by sampling configurations with colored noise on
lattices of fixed size.  Thereby the form of the classical action is
kept unchanged. This procedure is only valid for
$\Lambda \gg \Lambda_{\text{\tiny{phys}}}$. In turn, for
$\Lambda \lesssim \Lambda_{\text{\tiny{phys}}}$ deviations grow
large. At this point a description by means of an effective action
might be necessary. Furthermore finite size and volume effects on the
lattice certainly also play a r$\hat {\rm o}$le and prohibit the use
of the continuum approximation for small UV cutoffs. Hence, a refined
analysis may even lower the cooling range.

Even without the refined analysis we have shown that a remarkably
large regime of ultraviolet fluctuations can be removed without
altering the physics content of the theory. The next step is to probe
the maximal colored cooling by identifying the lowest possible cutoff
scale at which the use of the classical action is still
valid. Thereto, we compute the parameters $\kappa$ and $\lambda$ from
the associated RG flows at a desired scale $\Lambda$ and use them in
the lattice simulation. This is current work in progress. 

Moreover, in our ongoing work we use a (Symanzik) improved action
and study the flow of the couplings of operators with 
dimension larger than $O(a^2)$. Further perspectives of 
the method are to explore the effects of regulator functions
different from the sharp cutoff beyond the effects shown in 
Appendix \ref{sec:app-reg-function}.

Applications of the method to SU(N) gauge theories and to finite
density models are also work in progress.  In theories with a complex
action induced e.g.~by a finite chemical potential, the Complex
Langevin evolution might be optimized by colored noise cooling.

\section*{Acknowledgments}
We thank Alexander Rothkopf, Manuel Scherzer and D\'{e}nes Sexty 
for discussions. 
This work is supported by EMMI, the grant ERC-AdG-290623, the BMBF,
grant 05P12VHCTG, and is part of and supported by the DFG
Collaborative Research Centre "SFB 1225 (ISOQUANT)".
I.-O.S.~thankfully acknowledges support from the DFG
under grant STA 283/16-2. F.P.G.Z.~thanks for support 
from the FAIR OCD project.

\newpage

\appendix


\section{Fourier transformation on the Lattice}
\label{sec:app-DFT}
On the lattice the discrete Fourier transformation \
of the field $\phi(x)$ reads
\begin{equation}
\phi(p) = \sum_{x} a^d\, \mathrm{e}^{-i p \cdot x} \phi(x)\, ,
\label{eq:app-lat-FT}
\end{equation}
where the momenta $p$ are elements of the discrete Brillouin zone.
The inverse Fourier transform of
the field $\phi(x)$ is correspondingly given by
\begin{equation}
\phi(x) = \frac{1}{a^d\, N^{\tiny{d}}} \, \sum_{p} \, \mathrm{e}^{i p \cdot x}\, \phi(p)\, ,
\label{eq:app-FT-inverse-lat}
\end{equation}
where the sum runs over all momenta in the Brillouin zone.
In the thermodynamic limit $N \to \infty$ 
the previous equation converges to
\begin{equation}
  \phi(x) = \frac{1}{(2 \pi)^d} \int_{-\pi / a}^{\pi / a}\, \mathrm{d}^{d}p\, \mathrm{e}^{i p \cdot x}\, \phi(p)\, . 
\label{eq:app-inv-lat-FT-therm-limit}
\end{equation}
For the remaining part of this section we work in the thermodynamic limit.

The $O(a^2)$ discretized Euclidean Laplace operator has the form
\begin{equation}
  \Delta_{x, y}  = -\frac{1}{a^2} \sum_{\mu = 1}^d\, (\delta_{x + \hat \mu, y} - 2\, \delta_{x,y} + \delta_{x - \hat \mu,y})\, . 
\label{eq:laplace-lattice}
\end{equation}
Let $\Delta^{-1}_{x, y}$ denote the inverse lattice Laplacian obeying
\begin{equation}
\Delta_{x, y} \Delta^{-1}_{y, z} = \frac{\delta_{x, z}}{a^d}\, .
\label{eq:app-LaplacianFunctionalEquation}
\end{equation} 
Substituting the Fourier transform of the Laplacian according to
(\ref{eq:app-inv-lat-FT-therm-limit}) in the previous equation yields
\begin{align}
  &\quad \ \Delta_{x, y} \left[ \frac{1}{(2 \pi)^d} \int_{-\pi / a}^{\pi / a}\,
    \mathrm{d}^{d}p \, \mathrm{e}^{i p \cdot (y - z)} \,
    \Delta^{-1}(p) \right] \nonumber \\
  &= \frac{1}{(2 \pi)^d} \int_{-\pi
    / a}^{\pi / a}\, \mathrm{d}^{d}p\, \mathrm{e}^{i p \cdot (x - z)}
  \, .
\label{eq:app-LatLaplacianDeriv}
\end{align}
Evaluating this further leads to the lattice Laplacian in momentum space
\begin{equation}
\Delta(p) = \frac{4}{a^2} \sum_{\mu = 1}^d \sin^2\left(\frac{a p_{\mu}}{2}\right)\, .
\label{eq:app-LatLaplacianMS}
\end{equation}
The right-hand side of (\ref{eq:app-LatLaplacianMS}) 
appears in a similar fashion in the
free propagator of a scalar field theory.
It relates the physical momenta to the lattice momenta
(\ref{eq:lat-momenta}) by
\begin{equation}
\tilde{p}_{\mu} := \frac{2}{a} \sin\left(\frac{a\, p_{\mu}}{2}\right).
\label{eq:app-phys-momenta}
\end{equation}


\section{Observables}
\label{sec:app-obs}
In this section we work in lattice units.
Let $V = N_s^{d-1}$ denote the spatial lattice volume and $T = N_t$ the time
extent of the lattice. Similarly as above we work with
$N_s = N = N_t$. The total lattice volume is $\Omega=VT$. 
In the following we derive in more detail a few of the key
observables of a real scalar field theory with the
lattice action given in (\ref{eq:lattice_action_kappa_lambda}). 
We keep our notation close to \cite{montvay1997quantum}.
The connected two-point susceptibility is defined 
as the integrated connected two-point correlation function
(\ref{eq:obs_conn_correlator}). It can be formulated in terms of the magnetization $M$ 
defined in (\ref{eq:obs_lat_magnetization}) using that $\langle \phi(x) \rangle = \langle \phi(0) \rangle = \langle M \rangle$.
\begin{align}
\chi_2 
&:= \sum_x G_c(x, 0) = \sum_x \langle \phi(x) \phi(0) \rangle_c \nonumber \\
&= \sum_x \left( \langle \phi(x) \phi(0) \rangle - \langle \phi(x) \rangle \langle \phi(0) \rangle \right) \nonumber \\
&= \sum_x \left( \frac{1}{\Omega}\, \left \langle \sum_y  \phi(x + y) \phi(y) \right\rangle - \langle M \rangle^2 \right) \nonumber \\
&=  \frac{1}{\Omega}\, \left \langle \sum_{x,y}  \phi(x + y) \phi(y) \right\rangle - \Omega\, \langle M \rangle^2 \nonumber \\
&=  \left \langle \sum_{y} M \phi(y) \right\rangle - \Omega\, \langle M \rangle^2 \nonumber \\
&= \Omega\, ( \langle  M^2 \rangle - \langle  M \rangle^2) = \Omega\, \langle  M^2 \rangle_c\, .
\label{eq:app-susc-coord-space}
\end{align}
In the step from the third to the fourth as well as from the fifth to the sixth equation translation 
invariance on the lattice has been used.
Moreover, we exploited the linearity 
of the (path integral) expectation value.
Alternatively, the connected susceptibility is just
the Fourier transform of the connected correlation function 
with momentum set to zero
\begin{equation}
  \chi_2 \equiv \left. \vphantom{\frac{dummy}{dummy}}\tilde G(p) \right|_{p = 0}\, .
\label{eq:app-chi-2-proj}
\end{equation}
Here, the momentum space correlator for small $p$ has the form
\begin{equation}
\tilde G(p) = \frac{1}{2 \kappa} \frac{Z_R}{m_R^2 + p^2 + O(p^4)}\, .
\label{eq:app-correlator-momentum-space}
\end{equation}
From this, the second moment is determined according to 
\begin{equation}
\mu_2 \equiv -\left.\frac{\partial^2}{\partial p_{\nu}\partial p_{\nu}} \tilde G(p) \right|_{p = 0} \, .
\label{eq:app-mu_2_proj}
\end{equation}
Explicitly, for the susceptibility it holds
\begin{equation}
\chi_2 = \tilde G(0) = \frac{1}{2 \kappa} \frac{Z_R}{m_R^2}\, .
\label{eq:app-chi-2-correlator-momentum}
\end{equation}
The evaluation of (\ref{eq:app-mu_2_proj}) 
for the second moment yields
\begin{equation}
\mu_2 
= \frac{2d}{2 \kappa} \frac{Z_R}{m_R^4}\, .
\label{eq:app-mu-2-explicit}
\end{equation}
Thus, the renormalized mass is given by
\begin{equation}
m_R^2 = 2d \frac{\chi_2}{\mu_2}\, .
\label{eq:app-mR-def}
\end{equation}
Next, we define the time slice as the spatial
average of the field over the lattice at each time $t$ 
\begin{equation}
S(t) = \frac{1}{V} \sum_{\vec x} \phi(\vec x, t)\, .
\label{eq:app-timeslice-def}
\end{equation}
In a similar way as discussed above, we can 
express $\chi_2$ in terms of the integrated correlation function of time slices 
using $\langle S(t) \rangle = \langle S(0) \rangle = \langle M \rangle$.
\begin{align}
&\chi_2 = \frac{1}{\Omega} \sum_{x, y} \langle \phi(x + y) \phi(y) \rangle 
- \Omega \langle M \rangle^2 \nonumber \\
&= \frac{1}{\Omega} \sum_{(t, \vec x) , (t', \vec y)} \langle \phi( \vec x + \vec y, t + t') 
\phi(\vec y, t')  \rangle - \Omega \langle M \rangle^2 \nonumber \\
& = \sum_{t, t'}\, \frac{V}{T} \langle   S(t + t') S(t') \rangle - V \sum_t \langle S(t)\rangle \langle S(0) \rangle \nonumber \\
&= V\, \sum_{t} \langle S(t) S(0) \rangle_c = V\, \sum_t G_c(t)\,.
\label{eq:app-chi-2-ts-corr}
\end{align}
The second moment $\mu_2$ can be expressed in form of time slices 
exploiting $\langle \phi(\vec x, t)\rangle = \langle S(t)\rangle$
as follows
\begin{align}
&\mu_2 = \sum_x x^ 2 G_c(x, 0) = \sum_x x^ 2 \langle \phi(x) \phi(0) \rangle_c \nonumber \\
&= \sum_{\tiny{t, \vec x}} (t^2 + {\vec x}^ 2) \langle \phi(\vec x, t) \phi(0, 0) \rangle_c \nonumber \\
&= d\, \sum_{\tiny{t, \vec x}} t^ 2 \langle \phi(\vec x, t) \phi(0, 0) \rangle_c \nonumber \\
&= d\,\sum_{\tiny{t, \vec x}}\, t^2 \langle \phi(\vec x, t) \phi(0, 0) \rangle - d\, \sum_{\tiny{t, \vec x}} \, t^2 \langle \phi(\vec x, t) \rangle^2 \nonumber \\
&= d\, \sum_{\tiny{t, \vec x}} t^2 \langle \phi(\vec x, t) S(0) \rangle - d\, \sum_{\tiny{t, \vec x}} \, t^2 \langle S(t) \rangle^2 \nonumber \\
&= d\, V\, \sum_{\tiny{t}} t^2 \langle S(t) S(0) \rangle - d\, V \sum_{\tiny{t}} \, t^2 \langle S(t) \rangle^2 \nonumber \\
&= d\, V  \sum_{\tiny{t}} t^2 \langle S(t) S(0) \rangle_c\, .
\label{eq:app-second-moment-time-slice}
\end{align}
In the step from the third to the fourth equation 
we have used the above premise
that there is no distinguished direction on the
lattice.

The corresponding formulae for a scalar field theory in $d = 2$ read
\begin{align}
\chi_2 
&= \frac{1}{N^2} \sum_{x, y} \langle \phi(x + y) \phi(y) \rangle_c 
= N^2 \langle M^2 \rangle_c \nonumber \\
&= \sum_{t, t'} \langle S(t + t') S(t') \rangle_c
=  N \sum_{t} \langle S(t) S(0) \rangle_c\,,
\label{eq:app-chi-2-2D}
\end{align}
where 
\begin{equation}
S(t) = \frac{1}{N} \sum_{x} \phi(x, t) 
\label{eq:app-ts-2D}
\end{equation}
and
\begin{equation}
M = \frac{1}{N^2} \sum_{x} \phi(x)\, .
\label{eq:app-magn-2D}
\end{equation}
For the second moment we find
\begin{equation}
\mu_2 = 2 N \sum_{t} t^2 \langle S(t) S(0) \rangle_c\, .
\label{eq:appmu2-2D}
\end{equation}
Finally, the renormalized mass can be computed from
\begin{equation}
m_R = \left( \frac{4 \chi_2}{\mu_2}\right)^{1/2}\, .
\label{eq:app-mR-2D}
\end{equation}


\section{Spacetime correlation function of colored noise}
\label{sec:app-spacetime-correlator-CN}
First, the spatial Fourier transform of the noise field $\eta(x, \tau)$ is given by
\begin{equation}
\eta(p, \tau) = \int \mathrm{d}^dx\, \eta(x, \tau) \, \mathrm{e}^{-ip \cdot x}\, .
\label{eq:app-noise-ft}
\end{equation}
The white noise correlation function in momentum space 
is obtained by applying the second 
relation from  (\ref{eq:noise-gaussian-relations})
\begin{equation}
\langle \eta(p, \tau) \eta(q, \tau')\rangle 
= 2\, (2 \pi)^d\, \delta^{(d)}(p + q)\, \delta(\tau - \tau')\, .
\label{eq:app-noise-corr-momspace}
\end{equation}
In the continuum colored noise is defined 
by the convolution with the sharp 
regulator function (\ref{eq:sharp-cutoff-continuum})
\begin{equation}
\eta(x, \tau)_{\text{col}} = \frac{1}{(2 \pi)^d} \int\, 
\mathrm{d}^d p\, \eta(p, \tau) \theta(\Lambda^2 - p^2) 
\, \mathrm{e}^{ip \cdot x}\, .
\label{eq:app-CN-cont}
\end{equation}
The correlation function for the colored noise field in $d$ dimensions is derived in the following.
\begin{align}
&\langle \eta_{\text{col}}(x, \tau)\, \eta_{\text{col}}(y, \tau') \rangle \nonumber \\
&= \left\langle \frac{1}{(2 \pi)^{2d}} \int \mathrm{d}^dp \, \mathrm{d}^dq  \, \mathrm{e}^{-ip \cdot x} \mathrm{e}^{-iq \cdot y}\right. \nonumber \\
& \ \ \left. \vphantom{\frac{dummy}{dummy}} 
\times \theta(\Lambda - |p|) \theta(\Lambda - |q|) 
\, \eta(p, \tau) \eta(q, \tau') \right\rangle \\[2ex]
&= \frac{2}{(2 \pi)^{d}} \int \mathrm{d}^dp \, \mathrm{d}^dq  \, \mathrm{e}^{-ip \cdot x} \mathrm{e}^{-iq \cdot y}\, \nonumber \\
& \quad \times \theta(\Lambda - |p|) \theta(\Lambda - |q|) \,
\delta^{(d)}(p + q) \delta(\tau - \tau') \\[2ex]
&= \frac{2}{(2 \pi)^{d}} \, \int \, \mathrm{d}^dp \, \mathrm{e}^{-ip \cdot (x - y )} 
\theta(|\Lambda| - |p|)\, \delta(\tau - \tau')\, \\[2ex]
&= \frac{2}{(2 \pi)^{d}} \int _0^{\Lambda}\, \mathrm{d}|p| |p|^{d-1} 
\int_{\vartheta_{d - 1} = 0}^{2 \pi} \int_{\vartheta_{d - 2} = 0}^{\pi} 
\hdots \int_{\vartheta_{1} = 0}^{\pi} \nonumber \\
& \quad \times \mathrm{e}^{{-i |p| |x - y| \cos(\vartheta_{d-2})}}\, \sin^{d-2}(\vartheta_1) \hdots \sin(\vartheta_{d - 2}) \, \nonumber \\
& \quad \times \mathrm{d}\vartheta_{1} \hdots \mathrm{d}\vartheta_{d - 2}\, \mathrm{d}\vartheta_{d - 1} \, \delta(\tau - \tau') \\[2ex]
&= \frac{2}{(2 \pi)^{d}} \int _0^{\Lambda}\, \mathrm{d}|p| 
|p|^{d-1} \frac{\sin(|p||x - y|)}{|x - y|} \, \int \, \Omega_d \, \delta(\tau - \tau')\\[2ex]
&= \frac{1}{2^{d-2} \pi^{d/2} \, \Gamma(\frac{d}{2})} \int _0^{\Lambda}\, 
\mathrm{d}|p| |p|^{d-1} J_0(|p||x - y|)\, \delta(\tau-\tau')\, .
\label{eq:app-col-noise-correlation}
\end{align}
Here $\Gamma(t) := \int \mathrm{d}y \, \mathrm{e}^{-y} y^{t-1}$ 
denotes the Euler gamma function and 
$J_0(x) \equiv \sin(x) / x$ is a Bessel function of the first kind. 
The Bessel profile is also visible in observables 
such as the correlation function of 
time slices for sufficiently low cutoff 
in numerical simulations. 
This is discussed further 
in Appendix \ref{sec:app-reg-function}.


\section{Aspects of Stochastic Quantization with colored noise}
\label{sec:app-CN-FPE}
\subsection{Fokker-Planck equation}
\label{sec:app-FPE}
In this section, we derive the Fokker-Planck equation (FPE)
with a noise kernel (\ref{eq:kerneled-FPE}) 
which describes the evolution of 
the probability distribution $P(\phi, \tau)$ in fictitious time $\tau$. 
The derivation presented in 
\cite{Damgaard:1987rr, Bern:1985he} is worked out in 
more detail focusing on the important technical steps.
Thereto, we consider a real one-component 
interacting scalar field theory in $d$ dimensions 
whose Euclidean action reads
 \begin{align}
  S = \int \mathrm{d}^dx \, \left[\frac{1}{2} \left( \sum_{\mu = 1}^d (\partial_{\mu} \phi(x))^2 \right) 
\vphantom{\sum_{\mu = 1}^d}+ \frac{m^2}{2} \phi^2(x) + V(\phi)\right]\, .
  \label{eq:app-action_standard}
 \end{align}
 The regularized Langevin equation reads
 \begin{equation}
  \frac{\partial \phi(x,\tau)}{\partial \tau} = - \frac{\delta S}{\delta \phi(x, \tau)} 
  + r_{\Lambda}(\Delta_x) \, \eta(x,\tau)\,,
  \label{eq:app-LEQ_col}
 \end{equation}
where $r_{\Lambda}(\Delta_x)$ denotes the
regularization function which depends on the cutoff parameter $\Lambda$
and the Laplacian $\Delta_x$ with $r_{\Lambda}(\Delta_x) \to 1$ in the limit 
$\Lambda \to \infty$.
The field is evolved in Langevin time according to
\begin{align}\nonumber 
 \phi(x,\tau)  =&\,\int_{x'}\, \int_{-\infty}^{\tau} \mathrm{d}\tau' \, G(x-x', \tau-\tau') \\[2ex]
&\hspace{.3cm}  \times \left[ r_\Lambda(\Delta_x)\, \eta(x,\tau') - \frac{\delta V}{\delta \phi}\phi(x',\tau)\right]\,, 
 \label{eq:app-sol_LEQ_col}
\end{align}
where the Langevin Green's function, see 
\cite{Damgaard:1987rr} for the derivation, is given by
\begin{multline}
 G(x-x', \tau-\tau') = \\
 \theta(\tau - \tau') \int_p \mathrm{e}^{-ip \cdot (x - x')}\mathrm{e}^{-(\tau-\tau')(p^2 + m^2)}\, .
 \label{eq:app-Lan_GF}
\end{multline}
Note, that the lower bound in the fictitious 
time integral is set to $-\infty$ such that at 
finite (positive) Langevin times 
the system is in thermal equilibrium.
Stochastic averages are equivalent to functional 
averages over a probability distribution  $P(\phi, \tau)$.
Moreover, let $F[\phi]$ be an arbitrary functional 
of the field variables. To stress the explicit noise
dependence of the field obtained 
as a solution of the Langevin equation we write $\phi_{\eta}$.
Stochastic averages are written as
\begin{align}
\langle F[\phi_{\eta}]\rangle_{\eta} = &\,
\int \, \mathcal{D}\eta \,F[\phi_{\eta}] \, \mathrm{exp}\left[-\frac{1}{4} \int \mathrm{d}\tau \mathrm{d}^dx \, \eta^2(x,\tau)\right] \nonumber \\[2ex] 
= &\int \mathcal{D}\phi \, F[\phi] \, P(\phi, \tau)\, .
 \label{eq:app-averages}
\end{align}
Before we proceed, we derive some useful identities.
First, it follows from (\ref{eq:app-sol_LEQ_col})
\begin{align}
 \frac{\delta \phi(x, \tau)}{\eta(y, \tau)} 
  &= \theta(0) \int_{x'} r_{\Lambda}(\Delta_{x})\, \delta^{(d)}(x-y) \int_p \mathrm{e}^{-ip \cdot (x-x')} \nonumber \\
  &= \frac{1}{2}\, r_{\Lambda}(\Delta_x)\, \delta^{(d)}(x-y)\,,
   \label{eq:app-phi-eta-deriv}
   \\[2ex] \nonumber
\end{align}
where the convention $\theta(0) = \frac{1}{2}$ is used.
Finally, we note the trivial identity
\begin{equation}
 \left(2 \frac{\delta}{\delta \eta_{y,\tau}} + \eta_{y,\tau} \right) \, \exp\left[-\frac{1}{4} \int_{\tau}\,\int_x\, \eta^2_{x, \tau}\right] 
 = 0\, . 
 \label{eq:app-noise_deriv}
\end{equation}
To derive the FPE we consider the derivative 
with respect to fictitious time $\tau$ 
of the stochastic average given in (\ref{eq:app-averages}).
For simplicity, we drop the subscript $\eta$.
\begin{align}
 &\frac{d}{d\tau} \langle F[\phi]\rangle = \,\left\langle \int_x\, \frac{\delta F[\phi]}{\delta \phi_{x, \tau}} \frac{\partial \phi_{x, \tau}}{\partial \tau}\right\rangle \nonumber \\[2ex]
 &= \left\langle \int_x\, \frac{\delta F[\phi]}{\delta \phi_{x, \tau}}  \left( - \frac{\delta S}{\delta \phi_{x, \tau}} + r_{\Lambda}(\Delta_x) \, \eta_{x,\tau} \right) \right\rangle \nonumber \\[2ex] 
 &= \int \mathcal{D}\eta \left\{ \, \int_x\, \frac{\delta F[\phi]}{\delta \phi_{x, \tau}} 
 \left( - \frac{\delta S}{\delta \phi_{x, \tau}} +  r_{\Lambda}(\Delta_x) \, \eta_{x,\tau} \right)\right. \nonumber \\[2ex]
 & \quad \left. \times \mathrm{exp}\left[-\frac{1}{4} \int_{\tau}\, \int_z\, \eta^2_{z,\tau}\right]\right\} \nonumber \\[2ex] 
 &= \,\int \mathcal{D}\eta \, \int_x\, \frac{\delta F[\phi]}{\delta \phi_{x, \tau}} \nonumber \\
 &\quad \times \left( - 
\frac{\delta S}{\delta \phi_{x, \tau}}  -2 \int_y \, r_{\Lambda}(\Delta_x)\, \delta^{(d)}(x-y) \, \frac{\delta}{\delta \eta_{y,\tau}} \right)  \nonumber \\[2ex]
 & \quad \times \mathrm{exp}\left[-\frac{1}{4} \int \, \mathrm{d}\tau \, \mathrm{d}^dz \, \eta^2_{z,\tau}\right]\, .  
 \end{align}
In the second equation the Langevin equation (\ref{eq:app-LEQ_col}) was inserted. 
The third equation follows by writing the noise average in the functional integral form.
In the fourth equation the identity for the functional derivative with respect to the noise field from (\ref{eq:app-noise_deriv}) was used.
 This can be simplified further as follows
 \begin{widetext}
 \begin{align}
 \frac{d}{d\tau} \langle F[\phi]\rangle &= \int \mathcal{D}\eta \, \mathrm{exp}\left[-\frac{1}{4} \int _{\tau}\,\int_z \eta^2_{z,\tau}\right]\,
\int_x\, \left( - \frac{\delta S}{\delta \phi_{x, \tau}} +2 \int_y \, r_{\Lambda}(\Delta_x)\, \delta^{(d)}(x-y) \, \frac{\delta}{\delta \eta_{y,\tau}} \right)\, \frac{\delta F[\phi]}{\delta \phi_{x, \tau}}  \nonumber \\
 &= \left\langle \int_x\, \left( - \frac{\delta S}{\delta \phi_{x, \tau}} 
  + 2 \int_y \, r_{\Lambda}(\Delta_x)\, \delta^{(d)}(x-y) \, \frac{\delta}{\delta \eta_{y,\tau}}\right) \, 
 \frac{\delta F[\phi]}{\delta \phi_{x, \tau}} \right \rangle \nonumber \\
 &= \left\langle \int_x \left(- \frac{\delta S}{\delta \phi_{x, \tau}} + 2 \int_y\, r_{\Lambda}
 (\Delta_x)\, \delta^{(d)}(x-y) \, \int_w\, \frac{\delta \phi_{w, \tau}}{\delta \eta_{y,\tau}} 
 \frac{\delta}{\delta \phi_{w, \tau}} \right) \frac{\delta F[\phi]}{\delta \phi_{x, \tau}} \right\rangle \nonumber \\
 &= \left\langle  \int_x \left(- \frac{\delta S}{\delta \phi_{x, \tau}} 
 + r^2_{\Lambda}(\Delta_x)\, 
 \frac{\delta}{\delta \phi_{x, \tau}} \right) \frac{\delta F[\phi]}{\delta \phi_{x, \tau}} \right\rangle.
 \label{eq:app-time_deriv_obs}
\end{align}
\end{widetext}
Here, the first equation follows from an 
integration by parts with respect to $\eta$. 
The third equation uses the chain rule
to calculate the functional derivative of $F$ 
with respect to $\eta$. The last equation 
is obtained by using the identity for the
functional derivative of the field $\phi$ 
with respect to $\eta$ from (\ref{eq:app-phi-eta-deriv}).
Moreover, it follows
\begin{widetext}
 \begin{align}
  \frac{d}{d\tau} \langle F[\phi] \rangle
  &= \int \mathcal{D} \phi\, F[\phi] \frac{\partial P(\phi, \tau)}{\partial \tau}
 = \int \mathcal{D} \phi\, 
  \left[\int_x \left(- \frac{\delta S}{\delta \phi_x} 
  + r^2_{\Lambda}(\Delta_x)\, 
  \frac{\delta}{\delta \phi_x} \right) \frac{\delta F[\phi]}{\delta \phi_x}\right] P(\phi, \tau) 
  \nonumber \\
  &= \int \mathcal{D} \phi\, F[\phi] \int_x \, \frac{\delta}{\delta \phi_x} \left(
  \frac{\delta S}{\delta \phi_x} + r^2_{\Lambda}(\Delta_x)\, 
  \frac{\delta}{\delta \phi_x} \right) P(\phi, \tau)\, .
  \label{eq:app-Fokker-Planck-rhs}
 \end{align}
\end{widetext}
The last equation is obtained by functional 
integration by parts with respect to $\phi$.
Thus, we arrive at the Fokker-Planck equation 
for the stochastic process with colored noise
\begin{equation}
 \frac{\partial}{\partial \tau} P[\phi, \tau] =
 \int \mathrm{d}^dx \frac{\delta}{\delta \phi_x} \left(\frac{\delta S}{\delta \phi_x} 
 + r^2_{\Lambda}(\Delta_x) \frac{\delta}{\delta \phi_x}\right) P(\phi, \tau)\, .
 \label{eq:app-FPE_colored}
\end{equation}
\begin{figure*}[t]
\includegraphics[width=0.8\columnwidth]{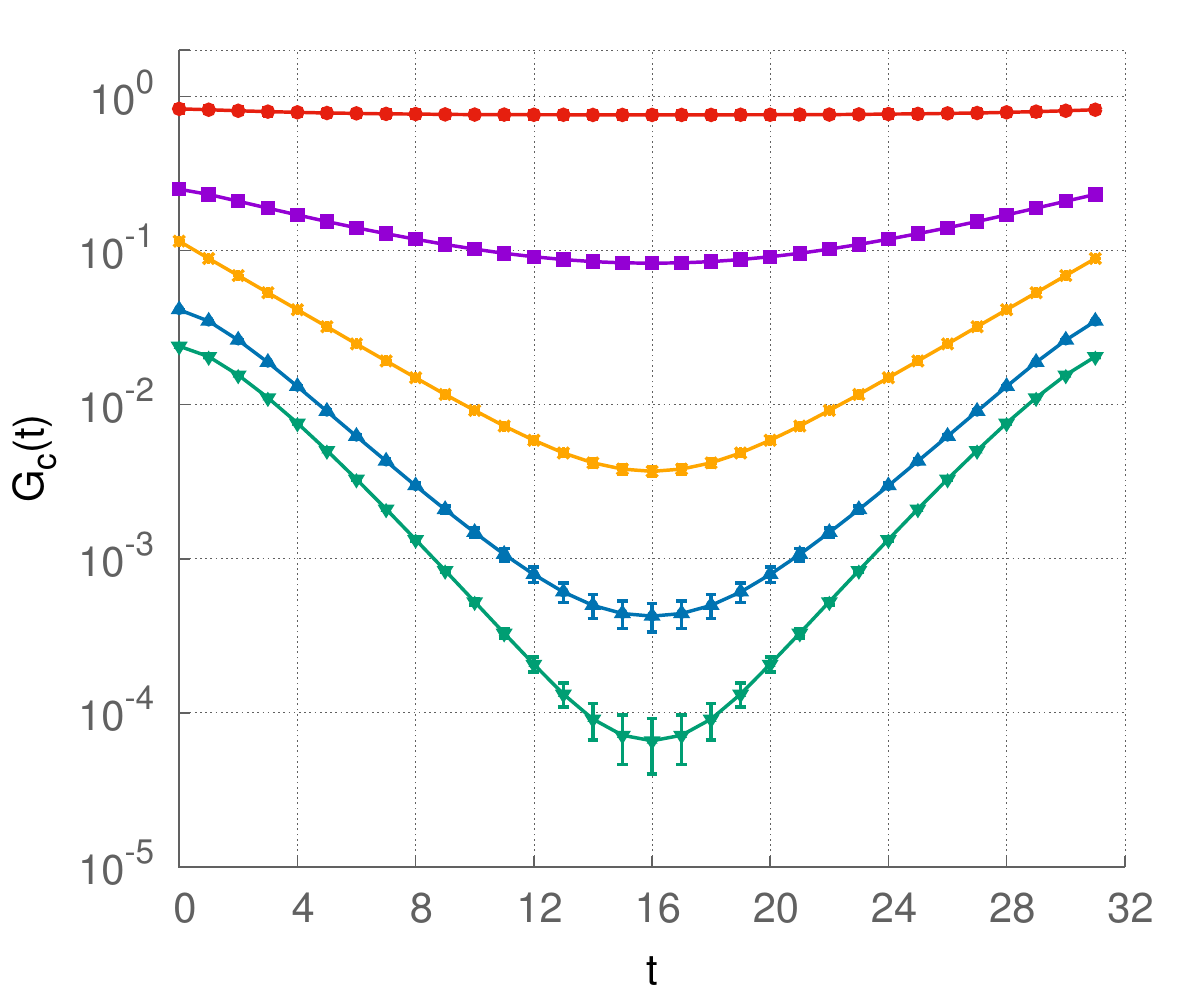}
~
\includegraphics[width=\columnwidth]{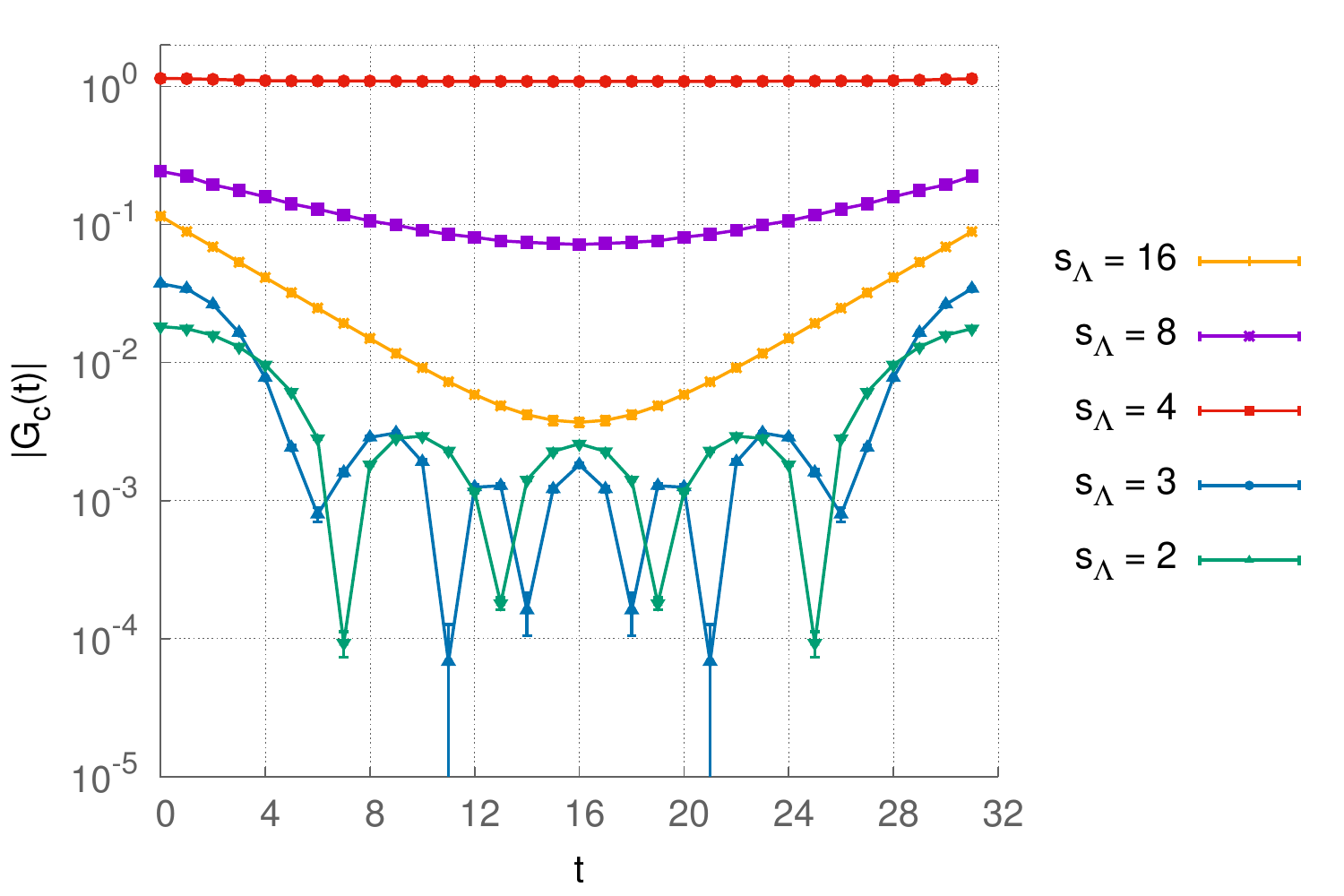}

\caption{Time slice correlation functions on a $32 \times 32$ lattice for fixed 
         parameters $\kappa = 0.26$ and $\lambda = 0.02$.
         The orange curve shown in both graphs is the result obtained 
         from a Langevin simulation with white noise for cutoff $\latcutoff = 16$.
         The full quantum theory is in the symmetric phase for the given choice 
         of parameters.
         (Left) The plot shows the correlation function of
         time slices using a Pauli-Villars regularization function
         for different cutoffs. 
         Halving the maximum lattice momentum cutoff to $\latcutoff = 8$ yields 
         the violet curve showing that the mass decreases. It becomes minimal 
         for $\latcutoff = 4$ were the theory is close to the critical point
         and the correlator is flat, see the red curve. 
         Decreasing the cutoff further to $\latcutoff=3$ and $\latcutoff = 2$ respectively 
         shows that the correlator decays stronger again meaning that the 
         mass in lattice units grows, but now with the theory being in the broken phase. 
         (Right) The same setup as on the left is shown but here 
         the sharp cutoff (\ref{eq:sharp-cutoff-continuum}) is used. 
         Note that for $\latcutoff = 2,3$ the correlator 
         resembles a Bessel function which is an artifact of the noise correlation 
         (\ref{eq:app-col-noise-correlation}).}
\label{fig:Corr-TS-N-32-2D-PV-sharp}
\end{figure*}
\subsection{Alternative regularization functions}
\label{sec:app-reg-function}

The regularization scheme used in \cite{Bern:1985he} is a
Pauli-Villars regularization with cutoff parameter $\Lambda$.
Explicitly, the regularization function is defined as
\begin{equation}
 r_{\Lambda}(\Delta_x) = \left(1 - \frac{\Delta_x}{\Lambda^2}\right)^{-1}\, .
 \label{eq:app-PV-cutoff-continuum}
\end{equation}
For most purposes it suffices to use the sharp regulator introduced in
(\ref{eq:sharp-cutoff-continuum}). However for certain cases, 
smooth regularization functions such as the Pauli-Villars 
type cutoff (\ref{eq:app-PV-cutoff-continuum}) or smooth
approximations to the sharp cutoff may be required. A crucial
disadvantage of the sharp cutoff is that it gives rise to artifacts
appearing in the noise correlation function, as discussed for the
continuum in Appendix \ref{sec:app-spacetime-correlator-CN}. 
Those are clearly visible in the
correlation functions of time slices and pose 
difficulties, for instance to the determination of masses because 
standard exponential fit techniques are not applicable. 
The lattice version of the Pauli-Villars regularization function
reads
\begin{equation}
r_{\tilde{s}_{\Lambda}}(\tilde{p}^2):= 
\left(1 - \frac{\tilde{p}^2}{\tilde{s}^2_{\Lambda}}\right)^{-m}\, \delta_{x, y}\, ,
\label{eq:PV-reg-lat-space-time}
\end{equation}
where $m \in \mathbb{N}$ as introduced in \cite{Bern:1985he}. 
Here $"\sim"$ refers to the physical momenta introduced in 
(\ref{eq:app-phys-momenta}).
Alternatively, a smooth approximation to
the sharp regulator on the lattice reads
\begin{equation}
r_{\tilde{s}_{\Lambda}}(\tilde{p}^2):= 
\frac{1}{2} \left(1 - \tanh\left[\alpha \left(\frac{\tilde{p}^2}{\tilde{s}^2_{\Lambda}}
        - 1\right)\right]\right)\,,
\label{eq:smooth-appr-reg-lat-space-time}
\end{equation}
where the parameter $\alpha \in \mathbb{R}_{+}$ can be tuned to vary
the steepness around the cutoff momentum. Both of the regulator
functions mentioned here are currently under study. They
may reduce the above mentioned artifacts arising from the use of the
sharp regulator. However, throughout the course of this work, we use
the sharp regulator for all quantitative studies. Smooth regulators
are only used in this section to illustrate a qualitatively different behaviour 
visible in the observables.

We discuss the effects of different choices of the 
regularization function by means of 
the two-point correlation function of time slices shown
in Fig.\ \ref{fig:Corr-TS-N-32-2D-PV-sharp}. 
The correlators were computed for parameters $\kappa = 0.26$, $\lambda = 0.02$ on
a $32 \times 32$ lattice, that is for the same choice as for Fig.\
\ref{fig:WF-CN-WN-Benchmark}.  Note, that the curves are represented
in logarithmic scaling.  The orange curve visible in both plots was
computed in a simulation with Gaussian white noise ($\latcutoff=16$) and
reproduces the hyperbolic cosine behaviour, typical for lattice
correlators.  The remaining correlators were produced in simulations
with colored noise. The left plot in Fig.~\ref{fig:Corr-TS-N-32-2D-PV-sharp} 
stems from a simulation with a
smooth Pauli-Villars regularization function. 
The external parameters $\kappa$ and
$\lambda$ are chosen such that by cutting off ultraviolet modes we
interpolate between the phases of the theory.  Close to the phase
transition the mass in lattice units approaches zero. This is
consistent with the flattening of the correlation functions, see the
violet ($\latcutoff=8$) and the red curve ($\latcutoff=4$).  For $\latcutoff = 2, 3$, see
the green and blue curve, the theory is in the broken phase and the
correlator bends for small Euclidean times loosing its typical
exponential shape.  This is a sign of an imprint of the regularization
function in the correlator well visible for
small $\latcutoff$.  This observation is also in agreement with the fact that
the colored noise is correlated in Euclidean space-time.
Moreover, consistently in the broken phase the mass grows again. For
large Euclidean times the correlator also seems to retain the
exponential behaviour. This might allow for the application of fits to
extract mass values or the calculation of effective masses.

The right hand side of Fig.\ \ref{fig:Corr-TS-N-32-2D-PV-sharp} shows
the same setup as on the left but for the sharp regularization
function (\ref{eq:sharp-cutoff-continuum}).  For intermediate
$\latcutoff = 4, 8$, see the red and violet curve, the results qualitatively
agree with the corresponding results obtained with the Pauli-Villars
regularization. For small $\latcutoff = 2, 3$ however, see the green and
blue curve, the correlator shapes differ.  Although at small
Euclidean times the correlator bends similarly, at larger times it
oscillates.  For illustrative reasons we show the modulus of
the correlator $|G_c(t)|$.  The sharp
regularization function leaves an artifact imprinting a Bessel-like
shape on the correlator, see also the discussion in Appendix
\@{\ref{sec:app-spacetime-correlator-CN}}. The qualitative behaviour
of the mass or correlation length agrees for both regularization
functions used here.  In Fig.~\ref{fig:Corr-TS-N-32-2D-PV-sharp} we do not show
the classical correlation function since it is trivially zero. This is
due to the gradient flow driving the field values into the classical
minimum approaching a constant value as $\tau \to \infty$. 


\section{Relation between stochastic regularization and the FRG}
\label{sec:app-SR-FRG}
\begin{figure}[t]
\includegraphics[width=.87\columnwidth]{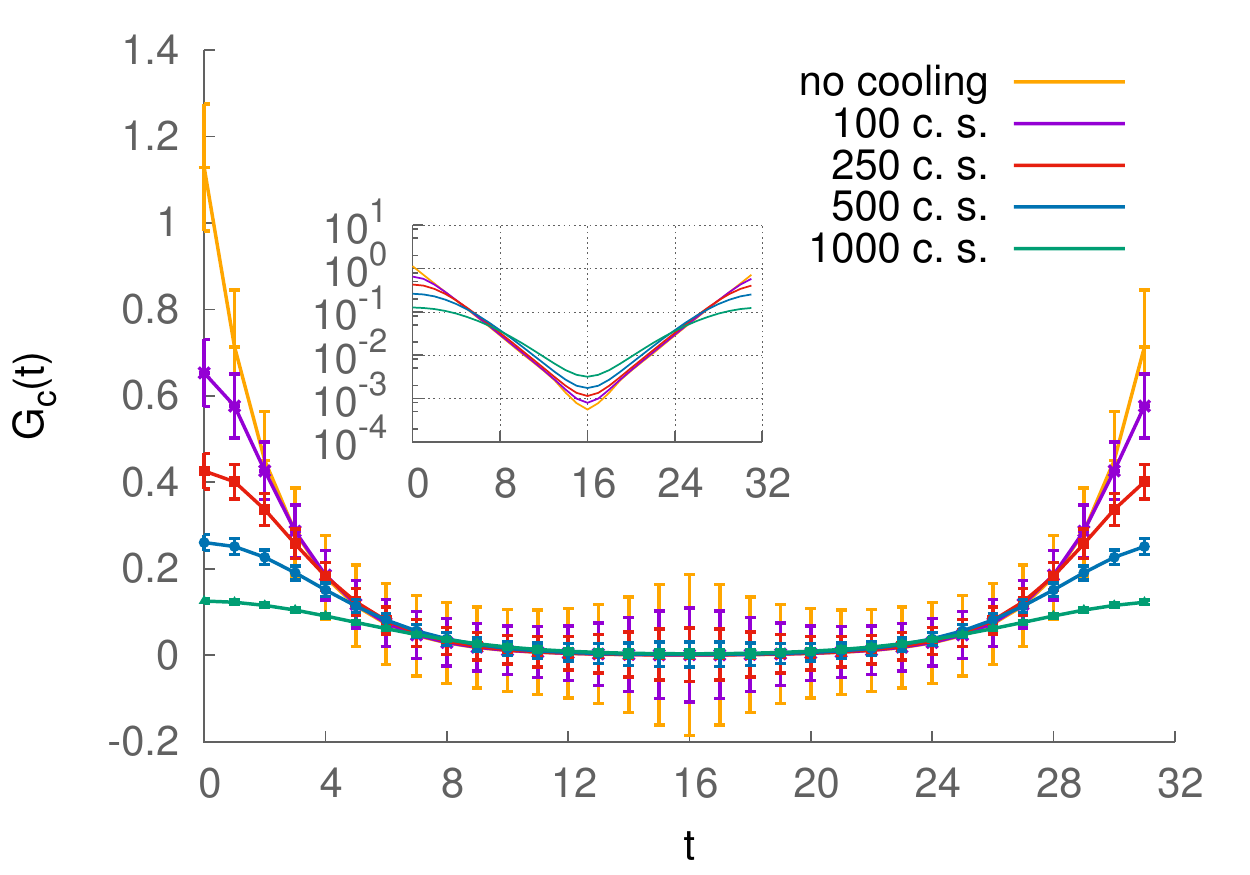}
\caption{Equilibrium configurations from Langevin simulations 
  with white noise are smoothed using the gradient flow. The correlation function of time
  slices $G_c(t)$ shows a similar behaviour as the colored noise result in 
  Fig.~\ref{fig:Corr-TS-N-32-2D-PV-sharp}.}
\label{fig:Correlator_WF}
\end{figure}

Using the sharp momentum cutoff 
\begin{align}
  r_{\Lambda}(p^2) = \theta(\Lambda^2 - p^2)\, ,
  \label{eq:sharp-cutoff-continuumApp}
\end{align}
in the Fokker-Planck equation
\eq{eq:kerneled-FPE} allows for a simple relation of Stochastic
Quantization with colored noise with functional renormalization group
equations. To that end we write the probability
distribution $P(\phi,\tau)$ in \eq{eq:kerneled-FPE} for
$\tau\to\infty$ as
\begin{align}\label{eq:app-Seff}
P_\Lambda(\phi,\tau)= \exp\left(-S -\Delta S_\Lambda\right)\,, 
\end{align}
where $\Delta S_{\Lambda}$ is defined in (\ref{eq:delta-S-sharpcutoff}).
Inserting (\ref{eq:app-Seff}) with (\ref{eq:delta-S-sharpcutoff})
into (\ref{eq:kerneled-FPE}) leads to
the fixed point equation $\partial_\tau  P_\Lambda =0$ in
momentum space with
\begin{align}\label{eq:app-FPL}
\left[ \Bigl(1 - r_\Lambda(p)\Bigr) \0{\delta S}{\delta \phi_p} - 
 r_\Lambda(p) \0{\delta \Delta S_\Lambda}{\delta \phi_p} \right] P_\Lambda(\phi,\tau)=0\,.
\end{align}
With $(1 -r_\Lambda) r_\Lambda\equiv 0$ the two parts on the left hand
side of \eq{eq:app-FPL} have to vanish separately. Now we use that 
\begin{align}\label{eq:app-UVcutoff} 
1 - r_\Lambda(p) = \theta(p^2-\Lambda^2)\, , 
\end{align}
only applies to UV modes. Accordingly we have
\begin{align}\label{eq:app-UVcutoff1} 
  \Bigl(1- r_{\Lambda}(p)\Bigr) \0{\delta S}{\delta \phi_p}P_\Lambda(\phi,\tau)=0\,. 
\end{align}
The prefactor in \eq{eq:app-UVcutoff1} does not vanish on the
ultraviolet modes that do not satisfy the equations of motion,
$\delta S/\delta \phi_p\neq 0$. For these modes \eq{eq:app-UVcutoff1} entails that
the measure has to vanish,
\begin{align}\label{eq:app-sharpP}
\left.\vphantom{\frac{1}{1}} P_\Lambda(\phi_p,\tau)\right|_{|p|>\Lambda}\stackrel{!}{=}0\,, 
\end{align} 
hence the name sharp (UV) cutoff. (\ref{eq:app-sharpP}) requires a
diverging $\Delta S_\Lambda $ for the ultraviolet modes with
$p^2 > \Lambda^2$. In turn, the cutoff term is also constrained for
$p^2 <\Lambda^2$ by \eq{eq:app-FPL} with
\begin{align}\label{eq:app-UVcutoff2} 
  r_\Lambda(p) \0{\delta \Delta S_\Lambda}{\delta \phi_p}=0\,,
\end{align}
and $\Delta S_\Lambda $ has to vanish for the infrared modes. A simple choice for
$\Delta S_\Lambda$ with these properties is given by 
\begin{equation}\label{eq:app-sharpcutoff}
  \Delta S_\Lambda[\phi]=\012 \int_p \phi_p \,\Lambda^2 
  \left(\0{1}{ r_{\Lambda}(p) }-1 \right)    \phi_{-p}\,.
\end{equation}
This cutoff term vanishes for momentum modes with $p^2< \Lambda^2$ and
is infinite for $p^2> \Lambda^2$ leading to
$P_\Lambda(\phi,\tau)=0$. This entails that the UV modes satisfy the
classical equation of motions and no quantum effects are taken into
account. 

We close this section with the remark that smooth cutoff functions
$r_\Lambda(p)$ for the noise do not lead to a measure of the type
\eq{eq:app-Seff}, as the related integrability relations are
violated. This has been already observed in \cite{Bern:1985he} in a
different context.


\section{A qualitative comparison of the gradient flow with colored noise}
\label{sec:comp-WF-CN}

In this section, we briefly and qualitatively focus on the analogous
behaviour of the gradient flow and the Langevin evolution with colored
noise. Thereto, we consider a one-dimensional real scalar field theory
and measure field configurations from a Langevin evolution with
white noise. The configurations are stored and smoothed by means of
the gradient flow. At each cooling step observables and corresponding
errors are calculated. In Fig.~\ref{fig:Correlator_WF} the two-point
correlation function of time slices is depicted. The number of
configurations is of $N_{\text{cf}} \approx O(10^4)$ for
$\kappa = 0.47, \lambda = 0.01$ and lattice size $N = 32$. 
As configurations are smoothed by the gradient flow the
errorbars shrink. To visualize this here, the errorbars are magnified 
by a factor $\sqrt{N_{\text{\tiny{cf}}}}$.
In comparison with the
result obtained using the sharp regulator in 
Fig.~\ref{fig:Corr-TS-N-32-2D-PV-sharp} we find the same behaviour of the
correlator at small Euclidean times.
As the configurations are cooled
the correlator bends. This signalizes the effect of a heat diffusion
equation which has been investigated in \cite{Bonati:2014tqa} in the
context of a massless scalar theory in $d$ dimensions. Note that the gradient
flow for a scalar theory has exactly the form of a heat diffusion equation.


\bibliography{literature}

\end{document}